%% file: paper.tex
\documentclass[10pt,epsf,amssymb
]{article}

\usepackage{graphicx}
\usepackage{amsfonts}
\usepackage{amsmath}
\usepackage[usenames]{color}
\usepackage{amssymb}
\usepackage[mathscr]{eucal} 
\usepackage{url}
\usepackage{cite}
\usepackage{hyperref}

\def\Z{\mathbb{Z}}

\def\R{\mathbb{R}}
\def\C{\mathbb{C}}
\def\P{\mathbb{P}}

\newcommand{\vev}[1]{ \left\langle {#1} \right\rangle }

\newcommand{\be}{\begin{equation}}
 
 \newcommand{\ee}{\end{equation}}               

\newcommand{\nn}{\nonumber}

\begin{document}

\begin{titlepage}
  
 \begin{flushright}
 KCL-mth-13-01 \\
   IPMU12-0235
 \end{flushright}
   
 \vskip 1cm
 \begin{center}
   
 {\large \bf On Singular Fibres in F-Theory}

 \vskip 1.2cm
   
 Andreas P. Braun$^{1}$ and Taizan Watari$^{2}$

 \vskip 0.4cm
  {\it
   $^1$Department of Mathematics, King`s College, London WC2R 2LS, UK
   \\[2mm]
   
  $^2$Institute for the Physics and Mathematics of the Universe, University of Tokyo, Kashiwano-ha 5-1-5, 277-8583, Japan  
   }
 \vskip 1.5cm
   
\abstract{

In this paper, we propose a connection between the field theory local
model (Katz--Vafa field theory) and the type of singular fibre in flat
crepant resolutions of elliptic Calabi-Yau fourfolds, a class of
fourfolds considered by Esole and Yau. We review the
analysis of degenerate fibres for models with gauge groups $SU(5)$ and
$SO(10)$ in detail, and observe that the naively expected fibre type is
realized if and only if the Higgs vev in the field theory local model is
unramified. To test this idea, we implement a linear (unramified) Higgs vev for the 
``$E_6$'' Yukawa point in a model with gauge group $SU(5)$ and verify
that this indeed leads to a fibre of Kodaira type IV$^*$.
Based on this observation, we argue i) that the singular fibre types 
appearing in the fourfolds studied by Esole--Yau are not puzzling at
all, (so that this class of fourfolds does not have to be excluded 
from the candidate of input data of some yet-unknown formulation of F-theory)
and ii) that such fourfold geometries also contain more information than
just the eigenvalues of the Higgs field vev configuration in the field
theory local models.  
} 
  
 \end{center}

\end{titlepage}



\section{Introduction}

Relatively to the world-sheet based formulations of string theories,
Type I, Type IIA, IIB, and the Heterotic string theories, the microscopic
formulation (i.e. the theoretical foundation) is less understood in
11-dimensional supergravity and F-theory. In the last
years, we have seen efforts of using F-theory for a better understanding of 
effective low-energy physics. Despite the fact that singularities 
in the internal geometry play an essential role and the lack of microscopic 
theoretical foundations in F-theory, string dualities have been used to
overcome this problem. In this paper we revisit the issue of which geometries
are appropriate in F-Theory. We do not take a top-down deductive approach,
but rather employ more of a try-and-error experimental technique to learn about
singularities in F-theory. We feel that this might teach us important lessons about 
an underlying fundamental description, as one of the best test arenas for the 
microscopic aspects of a theory of quantum gravity (such as F-theory) must be singular 
geometry. 

Part of the data defining a compactification of F-theory is an elliptically 
fibred Calabi--Yau $n$-fold $X_n$. To be more precise, there is a projection 
morphism which maps $X_n$ to the base manifold $B_{n-1}$,
\begin{equation}
 \pi_X: X_n \longrightarrow B_{n-1}, 
\end{equation}
and a section $\sigma: B_{n-1} \longrightarrow X_n$. By definition, 
the fibre of a generic point in the base is a non-singular curve of 
genus 1. In the context of Type IIB string theory, this fibre geometry 
plays a role similar to a principal bundle for vector bundles; the
``structure group'' is ${\rm SL}(2; \Z)$, and various fields on 
$B_{n-1}$ are in various representations of the structure group. 
Type IIB string theory specifies the ${\rm SL}(2; \Z)$ monodromy 
around a divisor $\{\Delta = 0 \}$ -``7-brane'' in $B_{n-1}$.
In this sense, we know
the theoretical constraints imposed on the fibred geometry over 
$B^\circ := B_{n-1} \backslash \{ \Delta = 0\}$, even in the absence 
of a top-down theoretical principle of F-theory. How should $X_n$ behave
right on top of the ``7-branes'' then?

The answer to this question is not unique in mathematics, but depends
on how one defines the relevant geometry. 
Let $X^\circ_n := \pi_X^{-1}(B^{\circ}_{n-1})$. 
To define $X_n$ whose behaviour on top of ``7-branes'' we examine,
one can require various different sets of conditions on the model $X_n$
that fit into the following commuting diagram:
\begin{equation}
\begin{array}{ccc}
  X_n & \longleftarrow & X^\circ_n \\
  \downarrow & & \downarrow \\
  B_{n-1} & \hookleftarrow & B_{n-1}^{\circ}
\end{array}.
\label{eq:comm-diagram}
\end{equation}
\begin{itemize}
 \item [(a)] One might think of a subvariety of a projective space
       $X_n^{\rm Weierstrass}$ given by a Weierstrass equation, along
       with a projection $\pi^w_{X}: X_n^{\rm Weierstrass} \rightarrow
       B_{n-1}$, and declare to take $X_n^{\rm Weierstrass}$ and 
       $\pi^w_{X}$ as the $X_n$ and $\pi_X$ in the diagram above.
       In this case, $X_n = X_n^{\rm Weierstrass}$ is not necessarily 
       a non-singular variety. One can still study the geometry of 
       the fibre curve at various points on the discriminant locus 
       $\left\{ \Delta = 0 \right\}$, if one is interested.
 \item [(b)] One might find a resolution $(\tilde{X}_n, \rho)$ 
       of\footnote{Here, by ``a resolution'', we meant that 
       $\tilde{X}_n$ is a non-singular variety, and there exists a 
       regular morphism 
       $\rho: \tilde{X}_n \rightarrow X_n^{{\rm Weierstrass}}$ 
       such that $\rho$ restricted on the inverse image of 
       $X_n^{{\rm Weierstrass}} \backslash {\rm (singular~locus)}$ 
       is an isomorphism. } $X_n^{\rm Weierstrass}$, and declare that 
       $\tilde{X}_n$ and $(\pi^w_X \circ \rho)$ are the $X_n$ and 
       $\pi_X$ in the diagram above.
       There always exists such a resolution for 
       $X_n^{\rm Weierstrass}$ (because we consider algebraic varieties 
       over the field $\C$), but such an $X_n = \tilde{X}_n$ is not 
       necessarily unique for a given $X_n^{\rm Weierstrass}$. 
       Thus, one has to ask first how many different choices are
       available for $(\tilde{X}_n, \rho)$, before studying the 
       geometry of the fibre of 
       $\pi_X: \tilde{X}_n \rightarrow B_{n-1}$ over 
       the discriminant locus $\left\{ \Delta = 0 \right\}$. 
\end{itemize}
There are several variations that are located in between those two extreme cases.
\begin{itemize}
 \item [(c)] One might consider a partial resolution\footnote{\label{footnote:partres}By a partial resolution, 
       we mean a pair $(X'_n, \rho)$ 
       where $\rho: X'_n \rightarrow X_n^{\rm Weierstrass}$ is a regular
       morphism, and further an isomorphism when restricted on the
       inverse image of 
       $X_n^{{\rm Weierstrass}} \backslash ({\rm singular~locus})$. 
       $X'_n$ is not necessarily required to be non-singular, however.} 
       $(X'_n, \rho)$ of $X_n^{{\rm Weierstrass}}$ where $\rho$ is
       crepant, and all the complex codimension-2 singularities in 
       $X_n^{{\rm Weierstrass}}$ are resolved in $X'_n$. 
 \item [(d)] One might also think of a crepant resolution 
       $(\tilde{X}_n, \rho)$ of $X_n^{\rm Weierstrass}$ where, for 
       any non-singular irreducible curve $C$ in $B_{n-1}$ which intersects the
       irreducible components of $\left\{ \Delta = 0 \right\}$ transversely, the fibre 
       $(\pi^w_X \circ \rho)^{-1}(C)$ is a non-singular surface.
\end{itemize}
If we are to take $X'_n$ under the condition (c) or 
$\tilde{X}_n$ in (d) as the $X_n$ in the diagram (\ref{eq:comm-diagram}), 
uniqueness or/and existence of such $X_n$ will be the primary questions  
before studying the geometry of the fibre over the discriminant locus.
As mathematics, none of these problems is wrong, and all of them have 
their own answers. Recently, Esole and Yau \cite{Esole:2011sm} introduced 
another set of conditions on $X_n$:
\begin{itemize}
\item [(e)] $(\tilde{X}_n, \rho)$ is a crepant resolution of 
      $X_n^{\rm Weierstrass}$, 
      and $(\pi^w_X \circ \rho): \tilde{X}_n \rightarrow B_n$ 
      remains to be a flat family of curves,\footnote{
      A fibration is flat if all fibres have the same
      dimension.}\raisebox{4pt}{,}\footnote{In \cite{Esole:2011sm} a 
      crepant resolution preserving flatness was obtained by using a small resolution.
      A resolution $(X, \nu)$
      of $Y$ is said to be a small resolution, if for every $r > 0$, 
      the space of points of $Y$ where the inverse image of $\nu$ 
      has dimension $r$ is of codimension greater than $2r$ \cite{Huebsch:1991}. 
      In the case $Y$ and $X$ are fourfolds, the inverse image of any point
      in $Y$ is either a curve or point(s), but not a surface. This is
      why, for any small resolution $(\tilde{X}_4, \rho'')$ of $X'_4$
      that is obtained as a crepant resolution $(X'_4, \rho')$ 
      of codimension-2 singularities in $X_4^{\rm Weierstrass}$, 
      $(\pi^w_X \circ \rho' \circ \rho''): \tilde{X}_4 \rightarrow B_3$ 
      often becomes a flat fibration.
      For $X_{n}$ with $n=5$ or higher, however, the following two 
      conditions on a resolution $(\tilde{X}_n, \rho)$ of 
      $X_n^{\rm Weierstrass}$ are clearly different: i) 
      $(\pi^w_X \circ \rho): \tilde{X}_n \rightarrow B_{n-1}$ is a flat 
      fibration and ii) $\rho: \tilde{X}_n \rightarrow X_n^{\rm
      Weierstrass}$ is a combination of a small resolution and the 
      minimal crepant resolution of codimension-2 singularities. 
      The authors of \cite{Esole:2011sm} did not make a clear bet 
      on either i) or ii), but we take i) as an interpretation of 
      their proposal.}. 
\end{itemize}
and considered $(\tilde{X}_n, \pi^w_X \circ \rho)$ satisfying this
condition as the $(X_n, \pi_X)$ in the diagram (\ref{eq:comm-diagram}). 
The authors of \cite{Esole:2011sm} further studied possible choices of 
$(\tilde{X}_4, \rho)$ for a class of Weierstrass-model Calabi--Yau fourfold 
$X_{n=4}^{{\rm Weierstrass}}$ to be 
used for $SU(5)$ models in F-theory compactifications. 

As physics, on the other hand, those $X_n$ are meant to be the input data 
of a theoretical formulation, both of which are used together to
calculate physical observables.
The combination of the set of input data and the theoretical formulation 
should be self-consistent
and the physics output should be reasonable.  
Based on such criteria, one can proceed in constructing and refining 
theories of physics, while abandoning those that are not functioning well.
However, it is clear at least that it is total non-sense to argue 
which conditions are to be imposed on $X_n$ without referring to 
a theoretical formulation.
In the absence of a microscopic formulation of F-theory, can we ever 
discuss such an issue?

Let us go back to the classic literature on F-theory, and remind 
ourselves of how the geometry--physics dictionary has originally been developed.
Group and matter representation of an F-theory compactification were
determined in the 90's by essentially relying on the Heterotic--F-theory
duality.\footnote{Here, we are talking about determining physics
consequences of F-theory where a straightforward application of weakly
coupled Type IIB string theory alone is not sufficient.}
Counting of light degrees of freedom (both charged and neutral) and 
identification of special loci in the moduli space on both 
sides of the duality are the primary weapons in this game. Under the 
strategy of relying on the duality, one does not have to argue whether the 
geometry data for F-theory remains singular or is resolved. This ties in
with the fact that we do not need a microscopic formulation of F-theory
when we are using dualities (see e.g., \cite{Morrison:1996pp} and
section 4 of \cite{Bershadsky:1996nh}\footnote{See, in particular, how
the dictionary between geometry and hypermultiplets in the ${\bf 56}$
representation of $E_7$ is determined.}).  

There is a beautiful correspondence between i) the A-D-E classification 
of surface singularities, ii) Kodaira's classification of singular fibres 
in non-singular elliptic surfaces and iii) non-Abelian gauge groups appearing 
on 7-branes, first presented in a table in \cite{Morrison:1996pp,
Bershadsky:1996nh}.  
The relation between i) and ii) is a mathematical fact and a priori has nothing 
to do with a theoretical formulation in which we use this data for physics.
Correspondingly, the authors of \cite{Morrison:1996pp,
Bershadsky:1996nh} do not argue that the F-theory geometry be
non-singular rather than singular (or vice versa), and 
do not use the configuration of singular fibres in the non-singular
geometry (ii)) to derive the gauge group on the 7-branes (see e.g. the discussion 
in \cite{Marsano:2011hv}).
What is special for the A-D-E surface singularities is that the correspondence 
between singular and resolved geometry is so unique and automatic that 
there is no need to make a distinction between them, especially when 
a microscopic theoretical formulation of F-theory is absent. In a sense, 
string duality and 16 supercharges are so powerful that we do not have
to argue about the difference.

Kodaira's study \cite{Kodaira1,Kodaira2}, see also \cite{peters}, is
based on the assumption that the base manifold $B_{n-1}$ is a complex 
curve (so that irreducible components of singular fibres are divisors of
$X_n$) and $X_{n=2}$ is non-singular. Tate's algorithm \cite{Tate} 
specifies the situation in terms of orders of vanishing when each type 
of singular fibre in Kodaira's list is realized; ``order of vanishing'' 
is a physics translation of a notion associated with discrete valuation 
rings in mathematics, and the definition of discrete valuation rings is 
satisfied by the ring of formal power series of 1 variable, $\C[[x]]$, not by 
the one with two variables, $\C[[x,y]]$. Hence Kodaira's classification of 
singular fibres (and its criterion specified in Tate's
algorithm)\footnote{The local choice of a gauge group at a generic point
on 7-branes (in field theory in 8-dimensions) is essentially an issue of
codimension-one in the base, and hence Tate's algorithm is applicable
without modification, even when the base $B_{n-1}$ is a surface or
3-fold. When one is interested in a symmetry left unbroken by $X_n$
($n>2$) in the effective theory in $(12-2n)$-dimensions, however, this
is not an issue that Tate's paper was concerned about. When a discrete
valuation (order of vanishing) $\nu$ is introduced in association with
the smallest power of $x$, any power series with $\nu=0$ is invertible
in $\C[[x]]$, but that is not true in $\C[[x,y]]$; consider a power
series beginning with $y \times (\cdots ) + x \times (\cdots)$; it is
not invertible, even though $\nu = 0$ holds. The $y=0$ locus within the
7-brane characterised by $x =0$ corresponds to the enhanced singularity
point. For these reasons, Tate's algorithm has been used with some
modification for symmetry groups left unbroken by $X_n$ for cases with 
$n > 2$ \cite{Bershadsky:1996nh,Katz:2011qp}.} 
cannot be used when the base manifold is of dimension higher than 1.

One might still think that the ``adiabatic argument'' may be used to
infer the physics consequences (such as matter and interactions) from the
singular fibres in a non-singular $X_n$, but it is not more than a 
(not too rigorous) guiding idea based on physics 
intuition. Furthermore, there is no theoretical top-down principle 
telling us how to read out matter content or Yukawa interactions (corresponding
to effects respecting 8 and 4 supercharges, respectively) from singular
fibres over higher codimension loci in the base. At best, we can hope to
read out a dictionary between the geometry and its physics consequences
after the physics is determined in a more convincing way. 

Following the successful tradition of studies on F-theory in the 90's, 
a combination of Heterotic--F-theory duality, effective field theory 
descriptions of Katz--Vafa type \cite{Katz:1996xe} and a bit of adiabatic 
argument, has recently been used to determine the physics consequences associated with 
codimension-3 loci of the base manifold $B_{n-1}$ \cite{Donagi:2008ca,
Beasley:2008dc, Hayashi:2008ba, Hayashi:2009ge}.\footnote{
See also the appendix C of \cite{Hayashi:2009bt}, \cite{Cecotti:2010bp} 
and \cite{Kawano:2011aa}.} 
Just like in the 90's, this was achieved without getting involved in 
such issues as (or assuming) whether to (or how to) resolve
singularities. We are thus ready to ask if there is any dictionary
between the singular fibre of $X_4$ and some physics information, 
if we are to assume that some sort of resolved $X_4$ is relevant to the 
description of F-theory. One may even hope to proceed further, and try 
to infer which sort of resolved geometry is (not) suitable as the input 
data for (yet unknown microscopic formulations of) F-theory, based on
whether the dictionary looks reasonable or not.

An alternative approach to elucidate the necessary input data of $X_4$ is to focus on 
the formulation of a topological observable. A canonical example is given by the 
D3-brane tadpole (see e.g. \cite{Andreas:1999ng}), which receives a contribution proportional to the Euler characteristic 
if $X_4$ is smooth. 
For a singular space, however, we can come up with different notions which all agree
with the Euler characteristic in the smooth case. Again, the underlying physics is what is
responsible for selecting one of the (mathematically) possible choices. Interestingly, even
though a singular Calabi-Yau manifold can have different crepant
resolutions, the Betti numbers (and hence the Euler characteristic) are
the same for any resolution respecting the Calabi-Yau condition \cite{1997alg.geom.10020B}. Even though this
reinforces the idea of using crepant resolutions, it also hints at the
existence of a way to define a physically sensible notion of the Euler
characteristic without considering a resolved geometry first (as in the
condition (a) in the introduction). We do not try to include the test of
conditions (d, e) in this second approach in this article, however.
      
Reference \cite{Esole:2011sm} (see also \cite{Marsano:2011hv, Lawrie:2012gg}) reported that, 
for $X_4^{{\rm Weierstrass}}$ for $SU(5)$ models, the singular fibre of $\tilde{X}_n$ specified under 
the condition (e) has only six irreducible components over codimension-3 
points in $B_3$ characterized as ``$E_6$'' and ``$A_6$'' points, 
whereas the singular fibre over the points characterized as ``$D_6$'' has seven irreducible components. 
Hence the number of the irreducible components in the singular fibre is NOT the same 
as the number of nodes of the ``corresponding'' extended Dynkin diagrams on some 
of the codimension-3 points, while it is still the same for others. 

This does not jeopardise the generation of the Yukawa couplings which are expected from heterotic duality and
the field theory local model, as has been shown in \cite{Marsano:2011hv}. Here, the fibre components over matter curves
were shown to recombine such that they give rise to the expected Yukawa couplings between charged states coming from wrapped
M2-branes for both the ``$E_6$'' and the ``$D_6$'' points.

Even though there is a priori no reason to believe\footnote{This should be clear already from the discussion so far, and we will add more
discussion in the middle of section \ref{ssect:fibreSO(10)} and also in 
\ref{ssect:SO10-discussion}. For these reasons, we maintain
quotation marks.} that the number of irreducible components in a singular fibre has anything to 
do with extended Dynkin diagrams of A-D-E type also for degenerations occurring over loci of higher codimension, 
one may still be puzzled by this result. 
Why does the beautiful fact that the fibre shows the extended Dynkin diagram of the gauge group
not have a higher-dimensional analogue? What discriminates between cases where the fibre has the
``expected'' number of components and those where it has not? One may even be tempted to take this 
as an indication that the condition (e) on $X_4$ is not the right one for the input data of F-theory, 
no matter how F-theory is eventually formulated.

In this article, however, we study the resolution $\tilde{X}_4$ under the condition (e) and examine the geometry 
of singular fibres more extensively than in \cite{Esole:2011sm}. We conclude that what seemed puzzling 
was not actually puzzling at all, but there is a clear rule controlling the number of irreducible components of the 
singular fibre. The key idea is to use the link to the effective field theory description and exploit the connection
between the Higgs vev and the Weierstrass equation.

The outline of this paper is as follows: 
In sections \ref{ssect:SO10-resolution} and \ref{ssect:fibreSO(10)}, 
we discuss the resolution of fourfolds under the condition (e) for 
$X_4^{{\rm Weierstrass}}$ with generic choice of complex structure for
$SO(10)$ models \cite{Tatar:2012tm,Lawrie:2012gg} \footnote{Here, the choice of conditions such as 
(b)--(e) is not specified clearly. We will make it clear in
footnotes \ref{fn:tatar-comm-1}, \ref{fn:tatar-comm-2} and \ref{fn:tatar-comm-3} how the resolution in 
\cite{Tatar:2012tm} is related to our presentation in section
\ref{ssect:SO10-resolution} and \ref{ssect:fibreSO(10)}.}  
for the purpose of accumulating more data beyond the analysis of \cite{Esole:2011sm}. Two resolutions $\tilde{X}_4$ 
are available, and they are related by a flop. We study the geometry of singular fibres over higher codimension loci in the base. 
Our results are described in terms of algebraic families of curves in $\tilde{X}_4$ and their intersections, 
so that we can maintain algebraic information such as multiplicity rather than just set-theoretical information.
 
All the concrete studies of resolutions and singular fibres for $SU(5)$ 
and $SO(10)$ models (see \cite{Esole:2011sm,Marsano:2011hv,Krause:2011xj,Tatar:2012tm,Lawrie:2012gg} and
sections \ref{ssect:SO10-resolution} and \ref{ssect:fibreSO(10)} in this
article) 
show clearly that the ``puzzling'' behaviour of the number of irreducible 
components in the singular fibre corresponds precisely to the ramification 
behaviour of the Higgs field. In section \ref{ssect:SO10-discussion}, we
will explain why this is natural and claim that this correspondence  
(empirical rule) holds true not just for the cases that have been
studied. If this natural dictionary holds true,  
then what has been considered a puzzling behaviour no longer has to be taken as 
negative evidence for the condition (e) on $X_n$ as input data for F-theory. 
Furthermore, this also implies that the fibre structure of the resolved Weierstrass 
model, and hence also the fourfold $X_4^{{\rm Weierstrass}}$ contains more information 
on the Higgs vev in the  than just its eigenvalues.

In order to accumulate further evidence for the natural dictionary, a
further experiment is carried out in section \ref{sect:Hvev}.
Linear Higgs field vev configurations around codimension-3 loci in the base 
were considered even for F-theory $SU(5)$ models and $SO(10)$ models 
in \cite{Beasley:2008dc}, and the corresponding local geometry of 
$X_n^{\rm Weierstrass}$ is also known \cite{Hayashi:2009ge}; the complex structure 
of $X_n^{\rm Weierstrass}$ just has to be chosen in a special way.  
In section \ref{ssect:resA6}, we pick up a local fourfold geometry for 
an ``$A_6$'' point of $SU(5)$ models corresponding to the linear Higgs vev, 
and show that the natural dictionary holds true; now 7 irreducible components 
are in the singular fibre over the ``$A_6$'' type point, in contrast with just 6 
components for $X_4^{\rm Weierstrass}$ with generic choice of complex structure 
(ramified Higgs vev) studied in \cite{Esole:2011sm}. 
Section \ref{ssect:resE6} is devoted to the local geometry of $X_4^{\rm Weierstrass}$
for an ``$E_6$'' point of $SU(5)$ models corresponding to the linear Higgs vev.
We show that there exist 24 resolutions satisfying the condition (e), 
which are all related by flops (there are just six resolutions related by 
flops if we consider the local geometry for an ``$E_6$'' point with generic complex 
structure \cite{Esole:2011sm}). We will see that in all of these 24 
resolutions, the singular fibre over the ``$E_6$'' point has 7 irreducible 
components. Those two experiments confirm the natural dictionary we claim above. 
Finally, section \ref{ssect:resE7} provides a brief sketch of how to 
carry out similar experiments for the local geometry of ``$E_7$'' type points 
in $SO(10)$ models corresponding to the linear Higgs vev configuration.

It is an option for busy readers to skip sections \ref{ssect:SO10-resolution} 
and \ref{ssect:fibreSO(10)}, which are rather technical in nature, and 
proceed directly to \ref{ssect:SO10-discussion}. 
Section \ref{ssect:SO10-discussion} also plays the role of a summary 
in this article. 

We are aware that there are works addressing singularity
resolutions of elliptic Calabi--Yau fourfolds in toric language,
see e.g. \cite{Blumenhagen:2009yv,Grimm:2009yu,Collinucci:2010gz,Knapp:2011wk,Krause:2011xj,Collinucci:2012as}. 
The toric language is used for the study of singularity resolution
also in this article in section \ref{sectresE6lH}, and the flatness of the fibration is examined.
That is done for a local geometry (or a very special global geometry) only.
Studying this for a more general class of compact fourfolds while keeping the flatness of the 
fibration, however, is beyond the scope of this article.

\section{$SO(10)$ models}
\label{sect:so10}

Let us consider an F-theory compactification on an elliptically fibred 
Calabi-Yau fourfold $X_4^{\rm Weierstrass}$ given as a Weierstrass equation:
\begin{equation}
 Y^2 + z \beta'_5 XYW + z^2 \beta_3 YW^3 = X^3 + z \beta_4 X^2 W^2
  + z^3 \beta_2 X W^4 + z^5 \beta_0 W^6\, .  
\label{SO10X4}
\end{equation}
Let $S_{\rm GUT}$ be an effective divisor in the base 3-fold $B_3$, and 
$z \in \Gamma(B_3; {\cal O}_{B_3}(S_{\rm GUT}))$ so that the zero locus
of $z$ is $S_{\rm GUT}$. 
$\beta_{i=4,3,2,0} \in \Gamma(B_3; 
 {\cal O}_{B_3}(-(6-i)K_{B_3}-(5-i)S_{\rm GUT}) ) )$, and 
$\beta'_5 \in \Gamma(B_3; {\cal O}_{B_3}(-K_{B_3}-S_{\rm GUT}))$ 
so that the massless $SO(10)$ gauge field is on 
the $S_{\rm GUT} \times \R^{3,1}$ 
7-brane~\cite{Bershadsky:1996nh,Marsano:2009ym}.  
$[X:Y:W]$ are the homogeneous coordinates of $\P^2_{1,2,3}$ in which the
elliptic fibre is embedded.  

The discriminant of this elliptic fibration is given by 
\begin{align}
\Delta & = z^7 \left[ \beta_{3}^2 \beta_{4}^3 + z \left(
    \frac{27}{16} \beta_{3}^4
  - \frac{9}{4}  \beta_{3}^3 \beta_{4} \beta'_{5}
  + \frac{1}{2} \beta_{3}^2 \left(
          -9 \beta_{2} \beta_{4}
          + \beta_{4}^2 \beta_{5}^{'2} \right)
  - \beta_{3} \beta_{2} \beta_{4}^2 \beta'_{5} 
  - (\beta_{2}^2 - 4 \beta_{4} \beta_{0}) \beta_{4}^2
\right)  \right.
 \nn \\
 &  \qquad \left. + {\cal O}(z^2) \right] \, .
\end{align}
The $z=0$ locus in $B_3$ corresponds to the GUT divisor $S_{\rm GUT}$ and the $\beta_{4}|_{S_{\rm GUT}}=0$ and 
$\beta_{3}|_{S_{\rm GUT}}=0$ loci are the matter curves of 
$SO(10)$-${\bf 16}+\overline{\bf 16}$ and -${\bf 10}$ representations, 
respectively \cite{Bershadsky:1996nh}.
If we are to take a local neighbourhood (in the complex analytic sense) 
of these matter curves within $S_{\rm GUT}$, then the local fourfold 
geometry is known to be approximately a fibred geometry,  
with the fibre space being the ALE space of $E_6$ type and $D_6$ type 
deformed by one parameter, respectively \cite{Katz:1996xe}. 
Even the coefficient of the $z^8$ term vanishes at points in $B_3$ 
specified 
by $\beta_{4}|_{S_{\rm GUT}} = \beta_{3}|_{S_{\rm GUT}} = 0$, and 
by $\beta_{3}|_{S_{\rm GUT}} = 
  (\beta_{2}^2 - 4 \beta_{4} \beta_{0})|_{S_{\rm GUT}} = 0$ 
\cite{Andreas:1999ng, Hayashi:2008ba}.
The local geometry of the fourfold is approximately a fibration 
over a local open patch in $S_{\rm GUT}$ containing such a point, 
with the fibre geometry being an ALE space of $E_7$ and $D_7$ type,
respectively, with two deformation parameters \cite{Hayashi:2009ge}.
An ansatz that the physics of F-theory associated with these local 
geometries of the Weierstrass-model $X_4^{\rm Weierstrass}$ are 
described (approximately) by supersymmetric 
$E_6$, $SO(12)$, $E_7$ and $SO(14)$ gauge theories on 7+1-dimensions, 
respectively, with appropriately chosen background field configuration
\cite{Hayashi:2008ba, Hayashi:2009ge, Hayashi:2009bt, 
Cecotti:2010bp}, is now known to be consistent with the
Heterotic--F-theory duality.
Therefore, for the study of low-energy physics, we can simply use 
these gauge theories. There is nothing more to add about that 
in this article. 

As we have already stated in Introduction, however, there still remains 
a theoretical (rather than a practical) issue that is related to 
the formulation of F-theory itself. We might think of a resolution 
\begin{equation}
 \rho: \tilde{X}_4 \longrightarrow X_4^{\rm Weierstrass} \,  
\end{equation}
satisfying the property (e) in the Introduction, and 
formulate a theory based on $\tilde{X}_4$ rather than 
$X_4^{\rm Weierstrass}$.
We will construct such resolutions in section \ref{ssect:SO10-resolution}, 
and study the geometry of singular fibre in section
\ref{ssect:fibreSO(10)} purely as a problem in mathematics. From the physics
perspective, a discussion is given in section \ref{ssect:SO10-discussion}.
\newpage
\subsection{Resolution}
\label{ssect:SO10-resolution}

{\bf Crepant resolution of the $D_5$ singularity}
\\
\\
We construct resolutions of $X_4^{\rm Weierstrass}$ in two steps. 
First, the Weierstrass model Calabi-Yau fourfold $X_4^{\rm Weierstrass}$ 
is singular along its codimension-2 subvariety specified by 
$X = Y = z = 0$. This singular locus in $X_4^{\rm Weierstrass}$ 
is\footnote{Because this singularity locus stays away from the
zero-section of the elliptic fibration $\pi^w_X : X_4^{\rm Weierstrass}
\rightarrow B_3$, that is, the $W = 0$ locus, we can use $x = X/W^2$ and
$y = Y/W^3$ as the inhomogeneous coordinates for the study of
singularity resolutions.} 
mapped to $S_{\rm GUT}$ in $B_{3}$ under $\pi^w_X$. 
At a generic point in this singular locus, 
which is codimension-2 in $X_4^{\rm Weierstrass}$, 
the fourfold geometry forms a surface singularity of $D_5$ type 
in the two directions transverse to the singular locus.
In the first step, therefore, this $D_5$ singularity along the  
complex 2-dimensional subvariety is resolved. The new fourfold geometry 
is denoted by $X'_4$, and the birational morphism between $X'_4$ and 
$X_4^{\rm Weierstrass}$ by 
\begin{equation}
 \rho': X'_4 \longrightarrow X_4^{\rm Weierstrass} \, .
  \label{eq:SO10-first-step}
\end{equation}

The construction of $X'_4$ and $\rho'$ is done by a sequence of blow-ups. 
As this is a rather standard procedure, we will only highlight
the crucial steps and give the result, while 
explaining our notations and clarifying subtleties that are sometimes 
ignored in the physics literatures.

The first blow-up is centred at the two dimensional subvariety 
specified above: $x=y=z=0$. If we are to take an open patch ${\cal U}$ 
of $X_4^{\rm Weierstrass}$ around a generic point in the centre of the blow up, 
then the blown-up ambient space is covered by three patches 
${\cal U}_{1,2,3}$. In the third patch ${\cal U}_3$, 
where the coordinates $(x,y,z)$ in ${\cal U}$ and a new set of
coordinates $(x_3,y_3,z_3)$ in ${\cal U}_3$ are related by 
\begin{align}
 (x_3, y_3, z_3) \mapsto (x,y,z) = (z_3 x_3, z_3 y_3, z_3 ) \, ,
\label{eq:SO10-1st-3patch-map}
\end{align}
the expression for the proper transform (new fourfold) becomes
\begin{equation}\label{X4U3}
y_3\left(y_3 + z_3x_3\beta'_5+ z_3\beta_3\right)  =
 z_3\left(x_3^3 +x_3^2\beta_4  + x_3 z_3\beta_2 +z_3^2\beta_0 \right)\, .
\end{equation}
The exceptional locus $D_B$ of this blow-up is located at $z_3=0$ 
(from which $y_3=0$ also follows).

Let us denote this proper transform as $X'_{4,1}$, and the birational 
map (given by (\ref{eq:SO10-1st-3patch-map}) in ${\cal U}_3$) as 
$\rho'_1: X'_{4,1} \rightarrow X_4^{\rm Weierstrass}$. 
$D_B$ is a three-dimensional subvariety (divisor) in $X'_{4,1}$.

The fibre over $S_{\rm GUT}$ in 
$(\pi^w_X \circ \rho'_{4,1}): X'_{4,1} \rightarrow B_3$ has a second 
irreducible component besides $D_B$. This extra component, which is 
denoted as $D_{\infty}$, meets the zero section of the elliptic fibration. 
$D_\infty$ cannot be seen in the patch ${\cal U}_{3}$, but it is found 
in ${\cal U}_1$.
In the ${\cal U}_1$ patch, the fourfold $X'_{4,1}$ is defined by 
\begin{equation}
y_1\left(y_1 + z_1x_1\beta'_5 + z_1^2 x_1\beta_3\right) = x_1 +  z_1 x_1 \beta_4 +z_1^3 x_1^2  \beta_2  +z_1^5 x_1^3 \beta_0 \, , 
\end{equation}
and the map $\rho'_1$ is given by 
$(x_1,y_1,z_1)\mapsto (x,y,z)=(x_1,y_1 x_1, z_1 x_1)$. 
The inverse image of 
$S_{\rm GUT}$---$0=\rho^{' \, -1}_1(z)=(z_1 x_1)$---consists of 
two components; 
$D_B$ corresponds to the $x_1=0$ locus (from which $y_1 = 0$ also
follows), and the other irreducible component $D_{\infty}$ to $z_1=0$ 
(from which $x_1 = y_1^2$ follows). These two components intersect along 
a codimension-two subvariety 
$\left\{x_1 = y_1 = z_1 = 0 \right\} \subset {\cal U}_1$. 
It should be clear that we follow most of the notation of 
Ref.~\cite{Esole:2011sm}.

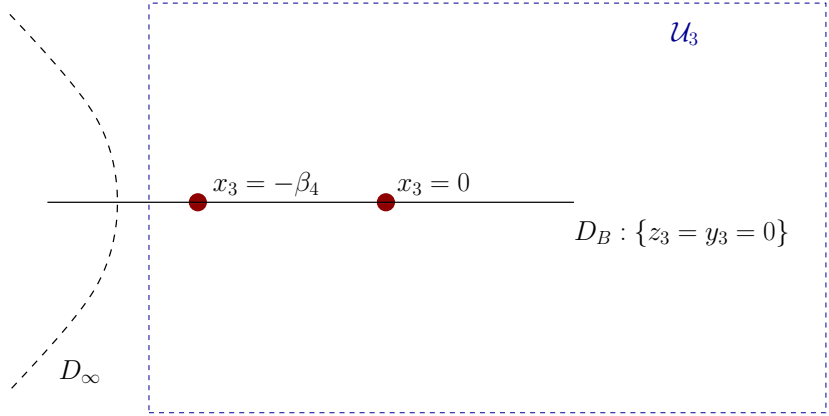
\begin{figure}[h!]
\begin{center}
\scalebox{.5}{ \input{SO10bu1.pspdftex}} 
\caption{The situation after the first blow-up. There are two remaining
  singularities sitting on the exceptional divisor $D_B$. (see also 
 the caption of fig. \ref{SO10bu4}, and the comment just before the ``step
  2''.) \label{SO10bu1}} 
\end{center}
\end{figure}

Note that $X'_{4,1}$ still has two singular codimension-two loci: 
besides the remaining singularity at $x_3=y_3=z_3=0$, there is another 
singularity of type $A_1$ at $z_3=y_3=(x_3+\beta_4)=0$. 
We have depicted the situation after the first blow-up in fig. \ref{SO10bu1}.

To completely resolve the codimension-two singularities of $X'_{4,1}$ 
(which is over the codimension one locus $S_{\rm GUT}$ in the base), 
we successively blow up along the codimension-two subvariety $x_3=y_3=z_3=0$ to obtain 
$X'_{4,2}$, then along $x_{31}=y_{31}=z_{31}=0$ to
obtain\footnote{\label{fn:tatar-comm-1} The first three steps
of blow-ups of the ambient space (and the corresponding proper 
transformations starting from $X_4^{\rm Weierstrass}$) in this article 
are the same as the first three blow-ups in \cite{Tatar:2012tm}.} 
$X'_{4,3}$, and finally along\footnote{\label{fn:tatar-comm-2}
A subvariety specified by 
$\left\{z_{311} = y_{311} = 0 \right\} \subset {\cal U}_{311}$ and 
$\left\{y_1 = x_1 = 0 \right\} \subset {\cal U}_1$ in the ambient space 
is chosen as the centre of the fourth blow-up in \cite{Tatar:2012tm} instead. 
Even though the ambient space is blown-up differently at this stage, and
is not isomorphic to ours, we confirmed that there is an isomorphism
between the proper transform after the fourth blow-up in
\cite{Tatar:2012tm} and $X'_{4,4}$ in this article.} 
$z_{311}=y_{311}=(x_{311}+\beta_4)=:\chi$ to obtain $X'_{4,4}$. 
The proper transform obtained in this way, $X'_{4,4}$, is the $X'_4$ we aimed to find, 
and the product of the birational maps associated
with these four blow-ups is $\rho'$ in (\ref{eq:SO10-first-step}).
This procedure---the well-known crepant resolution of a surface singularity of
type $D_5$---resolves all singularities of 
$X_4^{\rm Weierstrass}$ \eqref{SO10X4} 
which occur over the codimension one locus $S_{\rm GUT}$ in the base
$B_3$. This birational map $\rho'$ between the fourfolds is crepant. 

The fibre of any points $p \in B_3$, $\rho^{' -1}(p)$, is of
dimension one because each blow-up at most replaces a point 
by a curve, and only a finite number of such points are found in the 
fibre of any point in $S_{\rm GUT}$. 
Thus the fibration $(\pi^w_X \circ \rho'): X'_4 \rightarrow B_3$
still defines a flat family. 

The singular fibre of $(\pi_X^w \circ \rho'): X'_4 \rightarrow B_3$
corresponds to the inverse image of $S_{\rm GUT}$, and this consists of 
multiple irreducible components; two among them correspond to the 
proper transforms in $X'_4$ obtained by starting from $D_B$ and 
$D_{\infty}$ in $X'_{4,1}$, and are also denoted by $D_B$ and $D_\infty$ 
(in a slight abuse of notation). 
The other components are denoted by $D_{A, C}$ and $D_{\pm}$.
Those divisors intersect along codimension-two subvarieties in $X'_4$; 
to take an example, $D_{\infty} \cdot D_B$ corresponds to 
$\left\{ x_1 = y_1 = z_1 = 0 \right\} \subset {\cal U}_1$.

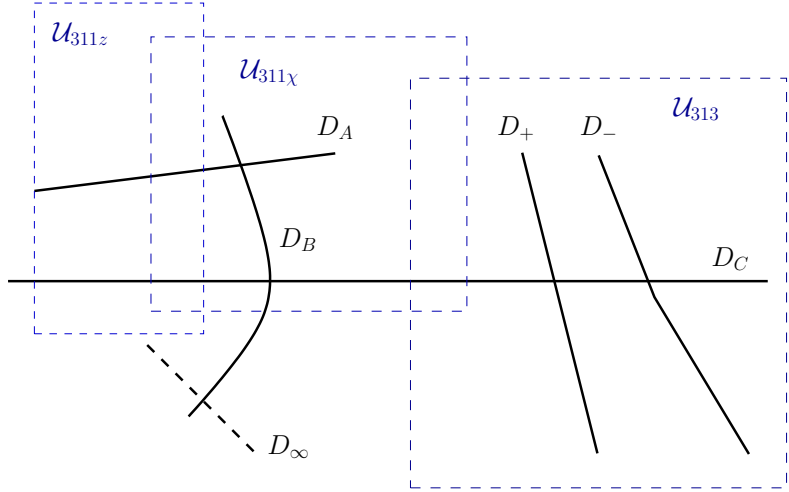
\begin{figure}[h!]
\begin{center}
\scalebox{.5}{ \input{SO10bu4.pspdftex}} 
\caption{The configuration of exceptional curves over a generic point 
on $S_{\rm GUT}$, after the last blow-up for the crepant resolution of a $D_5$
singularity. We have labelled the fibre components by the
exceptional divisor $D_i$ they originate from. Irreducible curves are
 drawn by lines, and when two curves share a point, they are drawn so
 that they intersect in this figure (just like in Kodaira's paper).
As the present discussion requires using multiple coordinate charts, 
we have furthermore sketched the location of the fibre components 
in the relevant patches.
\label{SO10bu4}} 
\end{center}
\end{figure}

We have depicted the configuration of curves in the fibre 
$\rho'^{\, -1}(p)$ of a generic point $p \in S_{\rm GUT}$ 
in fig. \ref{SO10bu4}, along with information about which charts cover 
which irreducible components.
We abuse the notation further in the figure  
by using the labels $D_A$, $D_B$ etc. also for 
the curves $D_A \cap (\pi^w_X \circ \rho')^{-1}(p)$, 
$D_B \cap (\pi^w_X \circ \rho')^{-1}(p)$, etc. 
As expected from the procedure of resolution we have employed, their 
intersection pattern (see also table \ref{D5curves}) agrees with that 
of the Dynkin diagram $D_5$---the intersection form of exceptional
curves appearing in the crepant resolution of a surface singularity of
type $D_5$. 
Here, the role of the extended node is played by the fibre component originating from
$D_\infty$. We avoided to draw a picture that looks like an extended Dynkin
diagram (and used one that looks like $I_1^*$ type instead), because the
intersection pairing of divisors $D_{A,B,C,\pm,\infty}$ does not provide
intersection ``numbers'' anymore.\footnote{The Cartan matrix 
(which equals the intersection form for resolved ADE surface
singularities) determines the (extended) Dynkin diagram. 
Although intersection {\it numbers} cannot be defined for a pair of 
two parameter families of curves in a fourfold, it is still
possible to define something similar as follows. First, $D_i \cdot D_j$
can be regarded not just as a codimension-two subvariety in $X'_4$, but
also as a divisor in $D_j$. Since $D_j$ can be
regarded as a fibred space over the GUT divisor $S_{\rm GUT}$, the fibre 
of a generic point in $S_{\rm GUT}$ defines a codimension-two subvariety
in $D_j$. Taking the intersection between this divisor and codimension-two
subvariety in $D_j$ we obtain a number that plays a role analogous to 
$D_i \cdot D_j$ in the case of surfaces. This definition of ``$D_i \cdot D_j$'' for the
two-parameter families of curves, however, cannot be extended
immediately for one-parameter families of curves \cite{Marsano:2011hv} (i.e. matter surfaces), which we discuss later. 
We will not try to formulate such numbers in this article, and draw
pictures of singular fibre geometry based only on the information of 
whether irreducible components share points set-theoretically or not.} 

\noindent
{\bf Step 2: Remaining small resolution}
\\
\\
Although the codimension-two singularity of $X_4^{\rm Weierstrass}$ 
(which was in the fibre of $S_{\rm GUT}$) was resolved in $X'_4$, 
there is still a codimension-three singularity in $X'_4$, and we seek 
for a resolution,
\begin{equation}
 \rho'': \tilde{X}_4 \rightarrow X'_4\, , \qquad 
 \rho=(\rho' \circ \rho''): \tilde{X}_4 \rightarrow 
 X_4^{\rm Weierstrass} \, ,
\end{equation}  
so that $\tilde{X}_4$ is smooth, $\rho$ is crepant, and 
$(\pi^w_X \circ \rho): \tilde{X}_4 \rightarrow B_3$ remains 
a flat fibration.

The remaining codimension-three singularity of $X'_4$ is found in the 
${\cal U}_{313}$ patch. The ${\cal U}_{311\chi}$ patch and 
${\cal U}_{311z}$ patch do not contain such singular loci, because 
the defining equation of $X'_4$ in these patches is given by 
\begin{align}\label{U311zchi}
 {\cal U}_{311z}:\quad  &  y_z (y_z+(\chi_z z_z-\beta_4)\beta'_5+\beta_3)=\chi_z + (\chi_z z_z-\beta_4)\beta_2+ z_z (\chi_z z_z-\beta_4)^2\beta_0  \\
 {\cal U}_{311\chi}:\quad & y_\chi (y_\chi+z_\chi(\chi_\chi-\beta_4)\beta'_5+ \beta_3 z_\chi) = z_{\chi} \left[ 1+z_\chi(\chi_\chi-\beta_4)\beta_2+ 
z_\chi^2 \chi_\chi (\chi_\chi - \beta_4)^2\beta_0 \right]\, .
\end{align}
The defining equation in the patch ${\cal U}_{313}$ is, on the other
hand,\footnote{Note also that $X'_4$ is already smooth even in the
fibre $(\pi^w_X \circ \rho')^{-1}(p)$ of a generic point $p$ in the 
matter curve for the $SO(10)$-${\bf 16}+\overline{\bf
16}$-representation; TW thanks Radu Tatar for discussion on this. } 
\begin{equation}
\label{U313} y_{313} (y_{313}+z_{313}x_{313}\beta'_5+\beta_3) = x_{313} \left[ x_{313} z_{313} + \beta_4 + \beta_2 z_{313}+\beta_0z_{313}^2 \right]\, .
\end{equation}
$X'_4$ is singular along a codimension-three locus (curve) specified by 
$y_{313}=x_{313}=\beta_3=(\beta_4 + \beta_2 z_{313}+\beta_0
z_{313}^2)=0$; $X'_4$ forms a conifold singularity in the three dimensions 
transverse to this curve. 

This curve in $X'_4$---the locus of the codimension-three singularity---is 
found in the fibre of the matter curve for the 
$SO(10)$-${\bf 10}$-representation, because $\beta_3 = 0$. 
It is a double cover over the matter curve, because for a given point
$p$ in the matter curve, there are two roots in the equation 
$\beta_4(p) + \beta_2(p) z_{313} + \beta_0(p) z_{313}^2 = 0$.
This double covering is ramified over points in the matter curve 
characterized by $z=0$, $\beta_3 = 0$ and $(\beta_2^2 - 4 \beta_4 \beta_0)=0$, 
which is known as ``$D_7$'' type points.
We are already familiar with such a double covering curve over the 
matter curve for the ${\bf 10}$-representation in $SO(10)$ models 
in the analysis in \cite{Hayashi:2008ba, Hayashi:2009ge}; those double
covering curves found before are in the dual Heterotic Calabi--Yau 
threefold \cite{Hayashi:2008ba}, or in the total space of canonical
bundle over $S_{\rm GUT}$ \cite{Hayashi:2009ge}, not in the fourfold 
geometry for the F-theory compactification. It is interesting that 
a similar object can be identified in the process of constructing 
a resolved geometry $\tilde{X}_4$ from a singular Weierstrass model 
$X_4^{\rm Weierstrass}$ in F-theory. 

Just like in the conifold resolution, we introduce a $\P^1$ with 
homogeneous coordinates $[a:b]$ and replace \eqref{U313} by the two
equations\footnote{\label{fn:tatar-comm-3}The
last blow-up in \cite{Tatar:2012tm} actually corresponds to this small resolution.}\raisebox{4pt}{,}\footnote{There 
are two different ways to resolve a conifold
singularity, and this is one of the two.}
\begin{align}
y_{313} a & = x_{313} b  \nn \\ 
\left(x_{313}z_{313}+\beta_4 + z_{313}\beta_2+z_{313}^2\beta_0\right) a
 & = \left(y_{313}+z_{313}x_{313}\beta'_5+\beta_3\right) b \, .
\end{align}
In terms of affine coordinates, we may write $y_{313}=y_a x_a$ and $x_{313}=x_a$ (that is, $y_a = (b/a)$) in 
${\cal U}_a$ as well as $y_{313}=y_b$ and 
$x_{313}=x_b y_b$ (that is, $x_b = (a/b)$) in ${\cal U}_b$. 

The proper transform of \eqref{U313}---$\tilde{X}_4$---in the patches 
${\cal U}_a$ and ${\cal U}_b$ is 
\begin{align}\label{Uab}
{\cal U}_a : \quad &y_a\left(y_a x_a + z_{313} x_a \beta'_5 + \beta_3\right)  
 = x_a z_{313} + \beta_4 + z_{313}\beta_2+z_{313}^2\beta_0 \nn \\
{\cal U}_b :\quad &y_b + z_{313} x_b y_b \beta'_5 + \beta_3  
 = x_b \left(x_b y_b z_{313} + \beta_4 +z_{313}\beta_2 + z_{313}^2 \beta_0\right)  \, .
\end{align}
For generic sections $\beta_i$, this completely resolves the singularity
in the fourfold $X_4^{\rm Weierstrass}$. In particular, there are no
point-like singularities left over the ramification point 
$(\beta_2^2-4\beta_4 \beta_0)=0$, $z_{313} = -\beta_2/(2\beta_0)$, and 
over the points $\beta_4 = 0$ (i.e., ``$E_7$'' point), as long as 
the sections $\beta_i$'s are generic.

The birational map $\rho''$ is crepant, and therefore the resolution map 
$\rho: \tilde{X}_4 \rightarrow X_4^{\rm Weierstrass}$ is also crepant. 
This is because $\rho''$ is a small resolution, and only introduces an
exceptional curve over the curve of conifold singularities in $X'_4$. 
Thus, the exceptional locus of $\rho''$ is a surface, not a divisor 
and there is no way there can be a dis{\it crepancy} between the two divisors 
$K_{\tilde{X}_4}$ and $\rho^{'' *}(K_{X'_4})$ in $\tilde{X}_4$.

Although the combination of the resolution map 
$\rho: \tilde{X}_4 \rightarrow X_4^{\rm Weierstrass}$ and the original 
fibration map $\pi^w_X : X_4^{\rm Weierstrass} \rightarrow B_3$ defines 
a new fibration map $(\pi^w_X \circ \rho)$, for which the generic fibre 
is still an elliptic curve, some singular fibres may not necessarily be 
of the same dimension as the generic fibre (i.e. may not be a flat fibration). 
The resolved manifold $\tilde{X}_4$ we constructed above, however, 
is still a flat fibration; this is because the conifold resolution
$\rho''$ only introduces at most one-dimensional object for a point 
in $X'_4$, and there are at most a finite number of (two) isolated points 
in the fibre of any points in $S_{\rm GUT}$. Thus, the fibre geometry 
of $(\pi^w_X \circ \rho)$ at any point in $B_3$ is of dimension one,
so that the fibration is flat. The other conifold resolution leads to 
a similar result. Hereafter, we denote $\pi^w_X \circ \rho$ by $\tilde{\pi}_X$.

\subsection{Fibre structure}
\label{ssect:fibreSO(10)}

Let us study the geometry of the fibre of the smooth manifold
$\tilde{X}_4$ with the projection map $\tilde{\pi}_X = (\pi^w_X \circ \rho)$ 
constructed above. This is primarily a question of mathematics; we
postpone the discussion  
from the physics perspective to section \ref{ssect:SO10-discussion}.

\subsubsection{Singular fibres over the GUT divisor $S_{\rm GUT} \subset B_3$}

The fibre geometry is an elliptic curve over a generic point in $B_3$,
and the fibre degenerates over subvarieties with various codimensions in
the base. Over the codimension-one subvariety $S_{\rm GUT} \subset B_3$, 
the fibre geometry of a generic point in $S_{\rm GUT}$ is the $I^*_1$ type 
in the Kodaira classification. 

In order to talk about multiplicity of irreducible components of
singular fibres, and about various ``limits'' of the fibre geometry 
over the loci of higher codimensions in the base, we should deal with 
this singular fibre of (generically) $I^*_1$ type as an {\it algebraic family}, 
rather than as a singular fibre over each point in the base individually. 
In order to track multiplicities, a set-theoretic description is not
sufficient, but we need the algebraic information as well. 
\begin{table}[h!]
\centering
\begin{tabular}{l|l|l|l|l}
& ${\cal U}_{a}$ & ${\cal U}_{b}$ & ${\cal U}_{311z}$ & ${\cal U}_{311\chi}$ \\
\hline
$D_+$ & $x_a=0$  & $y_b=0$ &  &  \\
& $[y_a\beta_3=\beta_4+z_{313}\beta_2$  & $[\beta_3=x_b(\beta_4+z_{313}\beta_2$ &  &  \\
& $\qquad \qquad \quad +z_{313}^2\beta_0)]$ & $\qquad \qquad \quad +z_{313}^2\beta_0]$ & & \\
\hline
$D_-$ &  & $x_b=0$  &  &  \\
&  & $[y_b=-\beta_3]$  &  &  \\
\hline
$D_C$ & $z_{313}=0$ & $z_{313}=0$ & $\chi_z z_z=\beta_4$ &  $\chi_\chi=\beta_4$ \\
& $[\beta_4=y_a(y_ax_a+\beta_3)]$  & $[x_b\beta_4=y_b+\beta_3]$  & $[y_z(y_z+\beta_3)=\chi_z]$ &  $[y_\chi(y_\chi+z_\chi\beta_3)=z_\chi]$ \\
\hline
$D_B$ & & &  &  $z_\chi=0$ \\
& &  & &  $[y_\chi=0]$ \\
\hline
$D_A$ &  &  & $z_z=0$ &  $\chi_\chi=0$ \\
& & & $\left[y_z(y_z-\beta_4\beta'_5+\beta_3) \right.$ &
      $\left[y_\chi(y_\chi-z_\chi(\beta_4\beta_5-\beta_3)) \right.$ \\
& & & $\left. =\chi_z-\beta_4\beta_2 \right]$ &
      $\left. =z_\chi(1-\beta_4\beta_2z_\chi) \right]$
\end{tabular}
\caption{This table shows the expressions (in the patches relevant for our discussion) for the exceptional divisors in $\tilde{X}_4$.
Here, empty fields indicate that the corresponding components cannot be seen
in this patch. The expressions for the other irreducible component 
$D_{\infty}$, which is contained in the ${\cal U}_1$ patch, is omitted 
from this table (the information is found in the text).  
For all the irreducible components in all the patches in this table, 
the equations in the first line can be taken as the defining equations 
of these components, and the equations in the second line (in 
[ $\cdots$ ] ) follow from the combination of the equations in the first
line for that component and the defining equations of $\tilde{X}_4$ in
 that patch. This does not mean, however, that the equations in
 the second line are not important. In fact, we define $D_B$ in the
 ${\cal U}_{311\chi}$ patch, not as ${\rm div} (z_\chi) \cdot
 \tilde{X}_4$, but as 
${\rm div} (z_\chi) \cdot {\rm div} (y_\chi)$ in the ambient space 
${\cal U}_{311\chi}$. 
These two are the same set theoretically, but different by 
multiplicity. }
\label{D5curves}
\end{table}
The codimension-one locus $S_{\rm GUT}$ in the base is characterized by 
a divisor $\{ z = 0 \} = {\rm div} \; (z)$ in the base, and the
pull-back of the divisor under $\tilde{\pi}_X$ also defines a divisor 
in $\tilde{X}_4$. This three-dimensional subvariety of $\tilde{X}_4$ 
corresponds to the algebraic family of singular fibres in $\tilde{X}_4$. 
This three-dimensional subvariety is not irreducible, however, and 
it turns out, as a divisor, that 
\begin{equation}
 {\rm div} \; \left( \tilde{\pi}^*_X (z) \right)
  = D_{\infty} + D_A + D_+ + D_- + 2(D_B + D_C) \, ,
  \label{eq:SO10-singfib-total-Sgut}
\end{equation} 
in terms of the irreducible divisors $D_i$. In order to read out the
multiplicity of $D_A$, $D_B$ and $D_C$, for example, the ${\cal
U}_{311\chi}$ patch containing all these three divisors can be used. 
\begin{equation}
 {\rm div} \; \left( \tilde{\pi}^*_X (z) \right) |_{{\cal U}_{311\chi}}
 = {\rm div} \; \left( z_{\chi} \chi_\chi (\chi_\chi - \beta_4)^2
		\right)
 = {\rm div} \; (z_\chi) + {\rm div} \; (\chi_\chi) + 2 {\rm div} \; (\chi_\chi-\beta_4)\, ;
\end{equation}
the three terms in the right-hand side correspond to $D_B$, $D_A$ and
$D_C$, as will be clear from the ${\cal U}_{311\chi}$ column of 
table \ref{D5curves}, and the multiplicity of $D_C$ comes from the 
coefficient 2 here. The multiplicity of $D_B$ is $2$, not $1$,
because $z_\chi \sim y_\chi^2$ for $z_\chi \simeq 0$ in this patch,  
and hence ${\rm div} \; (z_\chi)$ corresponds to 
$2 {\rm div} \; (y_\chi) = 2 D_B$.
The defining equations of the irreducible components of the algebraic 
family, $D_A$, $D_B$, $D_C$ and $D_{\pm}$, are
summarised in table \ref{D5curves}.
Note that we use $D_A$, $D_B$, etc. for 
the three-dimensional subvarieties in $\tilde{X}_4$, not just for those~in~$X'_4$.

In order to determine the fibre over a point $p$ in $B_3$ one might be tempted 
to consider $D_i \cap \tilde{\pi}_X^{-1}(p)$. Again, we wish to stress that
while this gives the correct set-theoretic information it fails to capture
any information regarding multiplicities. Instead, we may specify a point 
on $B_3$ by intersecting three appropriate divisors of the base. 
Under the map $\tilde{\pi}^*_X$ these will give rise to divisors on
$\tilde{X}_4$, the 
intersection of which will describe the fibre over $p$ in an algebraic way. 
Technically, such an intersection is performed by simply setting
the sections $\beta_i$ to their value at the point $p$, so that they become
constants in the defining equations of the divisors $D_i$. 
This construction applies to all points in the base including those on 
$S_{\rm GUT}$, matter curves (codimension-two loci) and 
codimension-three loci (Yukawa points), as long as the
$\beta_i$s are generic. 

So far, we have not done anything more than just reproducing classic 
results on the singular fibre geometry that have been known since the 
days of Kodaira. 
We now move on to study the geometry of singular fibres 
in the smooth geometry $\tilde{X}_4$ over a subvariety in $B_3$
with codimension higher than one. At the beginning of this section, 
we have already identified codimension-two and codimension-three loci 
of the base $B_3$ where further degeneration of the $I^*_1$ fibre 
may take place. We will turn our attention to those higher codimension 
loci one by one.

\subsubsection{Singular fibres over the ${\bf 16}$ matter curve 
$\beta_4|_{S_{\rm GUT}}=0$}

In the fibre over a generic point in the matter curve 
$\beta_4|_{S_{\rm GUT}} = 0$ for the 
$SO(10)$-${\bf 16}+\overline{\bf 16}$ representation, the fourfold 
geometry $X'_4$ obtained after the crepant resolution of 
codimension-two $D_5$ singularity is already smooth. 
This fact, however, does not say anything a priori about whether the singular 
fibre over the surface $S_{\rm GUT}$ degenerates further or not over 
this matter curve. We will see in the following that it does. 

The singular fibre over the matter curve $\beta_4|_{S_{\rm GUT}} = 0$ 
forms an algebraic family; it is the two-dimensional subvariety 
${\rm div}(\tilde{\pi}_X^*(z)) \cdot {\rm div}(\tilde{\pi}_X^* (\beta_4))$ 
in $\tilde{X}_4$.  
This two-dimensional subvariety has seven irreducible components, 
${S}_{1,2,3,4,5,6}$ and $S_{\infty}$.
The defining equations of those irreducible components are summarised 
in table \ref{tab:SO10-singfib-16curve}.

The family of singular fibres over $S_{\rm GUT}$ (as a three-dimensional
subvariety in $\tilde{X}_4$) has six irreducible components, whereas 
the family of singular fibres over the curve $\beta_4|_{S_{\rm GUT}}$ 
has seven components.\footnote{This phenomenon itself is nothing
surprising. Consider a family of curves 
$C = \{ (x,y,t) \in \C^3 | xy = t\} $ parametrized by 
$S = \{ t \in \C \}$. 
This two-dimensional variety $C$ is irreducible, but the fibre 
of $\{ t=0\} \in S$, $C_{t=0} := \{(x,y) \in \C^2 | xy = 0 \}$  
is not.} This is not counterintuitive, because one can readily see 
that the defining equations for the three-dimensional family $D_C$
in table \ref{D5curves} becomes factorizable in the $\beta_4=0$ limit. 
One can see after a bit of analysis that the families of irreducible 
components of the singular fibre over $S_{\rm GUT}$ correspond to the 
following combinations of the families over the $\beta_4|_{S_{\rm GUT}}=0$
matter curve:
\begin{eqnarray}
 D_+ \cdot {\rm div}(\tilde{\pi}_X^* (\beta_4)) = S_1\, , & \quad & 
 D_A \cdot {\rm div}(\tilde{\pi}_X^* (\beta_4)) = S_3\, , \quad
 D_B \cdot {\rm div}(\tilde{\pi}_X^* (\beta_4)) = S_6\, , \nn \\
 D_- \cdot {\rm div}(\tilde{\pi}_X^* (\beta_4)) = S_5\, , & \quad &  
 D_C \cdot {\rm div}(\tilde{\pi}_X^* (\beta_4)) =
    S_2 + S_4 + S_3\, .  
\label{eq:limit-SO10-Sgut-16curve}
\end{eqnarray}

The 2-dimensional subvarieties $S_{1,2,3,4,5,6}$ (and $S_{\infty}$) 
generate a lattice of algebraic cycles in 
$H_4(\tilde{X}_4 \; \Z) \cap [H^{2,2}(\tilde{X}_4)]^*$. The class 
defined by 
\begin{equation}
{\rm Span}_{\Z} \left\{ S_{1,2,3,4,5,6} \right\} / 
{\rm Span}_{\Z} \left\{ D_{A,B,C,\pm} \cdot {\rm div}(\tilde{\pi}_X^* (\beta_4)) \right\}
\end{equation}
is non-trivial. This class corresponds to one possible (mathematically 
precise) formulation of the 4-cycle over which the four-form flux is 
integrated in the net chirality formula for the chiral matter 
in the $SO(10)$--${\bf 16}$-$\overline{\bf 16}$ representation \cite{Hayashi:2008ba}. 

The singular fibre over this matter curve as a whole is decomposed as 
\begin{equation}
  {\rm div}(\tilde{\pi}_X^*(z)) \cdot {\rm div}(\tilde{\pi}_X^*(\beta_4)) = 
   S_{\infty} + S_1 + S_5 + 2(S_2+S_4+S_6) + 3 S_3\, ;
\end{equation}
this result is obtained either by directly calculating the left-hand side 
in various patches, or by combining (\ref{eq:SO10-singfib-total-Sgut}) 
and (\ref{eq:limit-SO10-Sgut-16curve}).

\begin{figure}[!h]
\begin{center}
\scalebox{.5}{ \input{curves16matter.pspdftex} }
\caption{A schematic picture of irreducible components of the singular fibre 
over a generic point in the matter curve $\beta_4|_{S_{\rm GUT}}=0$. 
We have labelled the fibre components by the complex surfaces $S_i$ 
they originate from. 
\label{Dynkin16}}
\end{center} 
\end{figure}
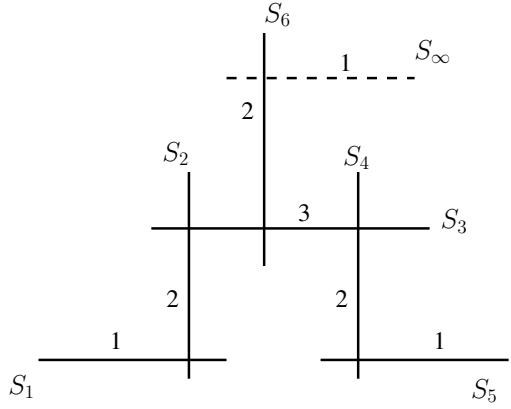

Over a generic point in the matter curve of the 
$SO(10)$-${\bf 16}+\overline{\bf 16}$ representation, 
the family of singular fibres $S_{1,2,3,4,5,6}$ and $S_{\infty}$ leaves 
seven corresponding irreducible curves. Information of whether those 
irreducible curve components share a point or not is 
schematically drawn as in figure \ref{Dynkin16}.

%
\begin{table}[h!]
\centering
\begin{tabular}{l|l|l|l|l}
& ${\cal U}_{a}$ & ${\cal U}_{b}$ & ${\cal U}_{311z}$ & ${\cal U}_{311\chi}$ \\
\hline
$S_1$ & $[x_a] \cdot $ & $[y_b] \cdot $ &  &  \\
&  $[z_{313}\beta_2+z_{313}^2\beta_0-y_a \beta_3]$  &  $[x_b(z_{313}\beta_2+z_{313}^2\beta_0)-\beta_3]$ &  &  \\
\hline
$S_5$ & &  $[x_b] \cdot [y_b+\beta_3]$   &  &  \\
\hline
$S_4$ & $[z_{313}] \cdot [y_ax_a+\beta_3]$
            & $[z_{313}] \cdot [y_b+\beta_3]$
            & $[\chi_z] \cdot [y_z + \beta_3]$
            & \\
\hline
$S_2$ &  $[z_{313}] \cdot [y_a]$ &
            & $[\chi_z] \cdot [y_z]$ & \\
\hline
$S_3$ & & & $[z_z] \cdot $ 
            & $[z_\chi] \cdot $ \\
           & & & $ [y_z(y_z+\beta_3)-\chi_z]$ 
            & $[y_\chi(y_\chi + \beta_3 z_\chi)-y_\chi]$ \\ 
\hline
$S_6$ & & &  &  $[z_\chi] \cdot [y_\chi]$ \\
\end{tabular}
\caption{
This table shows how the two-dimensional subvarieties $S_{1,2,3,4,5,6}$ of 
$\tilde{X}_4$ are specified in the ambient space of $\tilde{X}_4$; 
$\tilde{X}_4$ is defined by a single equation, (\ref{Uab}, \ref{U311zchi}), 
in any one of the patches ${\cal U}_a$, ${\cal U}_b$, ${\cal U}_{311z}$ and 
${\cal U}_{311\chi}$, and $S_{1,2,3,4,5,6}$ by intersection of three divisors. 
For example, $S_4$ is given by 
${\rm div}(z_{313}) \cdot {\rm div}(y_b+\beta_3) \cdot {\rm div}(\beta_4)$ in 
the ${\cal U}_b$ patch. The intersection with ${\rm div}(\beta_4)$ is omitted 
in this table, because it is common to all the entries of this table. 
$\beta_2$, $\beta_3$ and $\beta_4$ in this table should actually be 
$\tilde{\pi}_X^*(\beta_{2,3,4})$; we omitted $\tilde{\pi}_X^*$ in order 
to save space. }
\label{tab:SO10-singfib-16curve}
\end{table}

\subsubsection{Singular fibres over the ${\bf 10}$ matter curve 
$\beta_3|_{S_{\rm GUT}}=0$}\label{sssect:10matter}

Let us now study how the singular fibres over 
$\tilde{\pi}_X: \tilde{X}_4 \rightarrow B_3$ over the codimension-one locus 
$S_{\rm GUT} \subset B_3$ degenerate further over the codimension-two locus 
$\beta_3|_{S_{\rm GUT}}=0$. This is the matter curve for the $SO(10)$-${\bf 10}$ 
representation. 

After a bit of analysis, we find that the fibre over this matter curve, 
${\rm div}(\tilde{\pi}_X^*(z)) \cdot {\rm div}(\tilde{\pi}_X^*(\beta_3))$, 
forms a two-dimensional subvariety of $\tilde{X}_4$ with six irreducible 
components. Those irreducible components are denoted by 
$S_{\rm i, ii, iii, iv, v}$ and $S_{\infty}$. Their defining equations 
are summarised in table \ref{tab:SO10-singfib-10curve}. Although the number 
of irreducible components remains the same as that of the $I^*_1$ type 
in the Kodaira classification, the components are not in one to one 
correspondence. In fact, we found that 
\begin{eqnarray}
  D_A \cdot {\rm div}(\tilde{\pi}_X^*(\beta_3)) = S_{\rm i}, & \quad &
  D_B \cdot {\rm div}(\tilde{\pi}_X^*(\beta_3)) = S_{\rm ii}, \quad 
  D_C \cdot {\rm div}(\tilde{\pi}_X^*(\beta_3)) = S_{\rm iii}, \nn \\
  D_- \cdot {\rm div}(\tilde{\pi}_X^*(\beta_3)) = S_{\rm iv}, & \quad &
  D_+ \cdot {\rm div}(\tilde{\pi}_X^*(\beta_3)) =
    S_{\rm iv} + S_{\rm v}\, . 
\label{eq:limit-SO10-Sgut-10curve}
\end{eqnarray}
As the singular fibre over this matter curve as a whole we have 
\begin{equation}
{\rm div}(\tilde{\pi}_X^*(z)) \cdot {\rm div}(\tilde{\pi}_X^*(\beta_3))
 = S_{\infty} + S_{\rm i} + 2(S_{\rm ii} + S_{\rm iii} + S_{\rm iv}) + S_{\rm v}\, , 
\end{equation}
which is obtained either in a direct computation or by combining 
(\ref{eq:SO10-singfib-total-Sgut}) and (\ref{eq:limit-SO10-Sgut-10curve}).

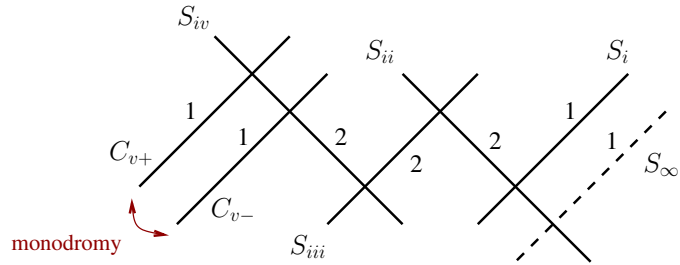
\begin{figure}[h!]
\begin{center}
\scalebox{.5}{ \input{curves10matter.pspdftex}} 
\caption{The fibre components over a generic point $p$ in the matter curve 
$(\beta_3|_{S_{\rm GUT}})=0$, including information of their multiplicities. For the sake of simplicity, we 
are using the labels $S_{\rm i, ii, iii,iv}$ and $S_{\infty}$, which were given to the surfaces 
introduced in table \ref{tab:SO10-singfib-10curve}, also for the irreducible curve components 
they give rise to. Since the restriction of $S_{\rm v}$ on the fibre is not irreducible (as 
explained in the text), its irreducible components are denoted by 
$C_{ {\rm v}\pm}$. Upon encircling 
the locus $(\beta_2^2-4\beta_4\beta_0)|_{S_{\rm GUT}}=0$, the two fibre
 components  
$C_{{\rm v}+}$ and $C_{{\rm v}-}$ are interchanged. \label{Dynkin10}}
\end{center} 
\end{figure}

Over a generic point $p$ in the matter curve of $SO(10)$-${\bf 10}$ representation, the fibres of $S_{\rm i, ii, iii, iv}$  
are all irreducible, but that of $S_{v}$ is not. 
It consists of two disjoint $\P^1$'s, because there are two 
solutions in $\beta_4(p) + \beta_2(p) z_{313} + \beta_0(p) z_{313}^2=0$. 
Therefore, the singular fibre of a given generic point in this matter curve 
has seven irreducible components and looks like figure \ref{Dynkin10}.

The two disjoint $\P^1$'s, $C_{{\rm v}+}$ and 
$C_{{\rm v}-}$,
form the single irreducible algebraic surface $S_{\rm v}$ 
in $\tilde{X}_4$ when fibred over the matter curve. This is simply because the two roots of 
$\beta_4(p) + \beta_2(p) z_{313} + \beta_0(p) z_{313}^2 = 0$ are interchanged 
when encircling the locus $(\beta_2^2-4\beta_4\beta_0)|_{S_{\rm GUT}}=0$.
Those fibre $\P^1$'s are non-split as a codimension-two subvariety 
in $\tilde{X}_4$, just like the non-split cases  
of \cite{Aspinwall:1996nk,Bershadsky:1996nh} occur for divisors. 

The ramification of spectral surfaces (7-brane monodromy) is a
notion understood in terms of the canonical bundle of the non-Abelian
7-branes (GUT divisor) \cite{Hayashi:2009ge}, not the fourfold. 
Our analysis here shows that the ramification of the spectral surface 
(i.e. the 7-brane monodromy) also has a corresponding phenomenon in 
terms of algebraic cycles in the fourfold.\footnote{A similar correspondence 
between the monodromy of spectral surface and discriminant loci is observed 
in \cite{Hayashi:2010zp}, although fibre components are not studied there. }

In such a generic choice of the complex structure of $X_4^{\rm Weierstrass}$ 
for an $SO(10)$ model, 
\begin{equation}
 {\rm Span}_{\Z} \left\{ S_{{\rm i,ii,iii,iv,v}} \right\} / 
 {\rm Span}_{\Z} \left\{ D_{A,B,C,\pm} \cdot {\rm div}(\tilde{\pi}_X^*(\beta_3))
                \right\}\, 
\end{equation}
is trivial. This guarantees that the net chirality cannot be generated 
in the $SO(10)$-${\bf 10}$ vector representation without breaking 
the $SO(10)$ symmetry, which is a known fact in physics.

%
\begin{table}[h!]
\centering
\begin{tabular}{l|l|l|l|l}
& ${\cal U}_{a}$ & ${\cal U}_{b}$ & ${\cal U}_{311z}$ & ${\cal U}_{311\chi}$ \\
\hline
$S_{\rm iv}$ & & $[y_b] \cdot [x_b]$  &  &  \\
\hline
$S_{\rm v}$ & $ [x_a] \cdot $ & 
                  $[y_b] \cdot $ &  &  \\
& $[\beta_4+\beta_2 z_{313} + \beta_0 z_{313}^2]$ &
  $[\beta_4+\beta_2 z_{313} + \beta_0 z_{313}^2]$ & & \\ 
\hline
$S_{\rm iii}$ & $[z_{313}] \cdot [x_a y_a^2 - \beta_4]$
            & $[z_{313}] \cdot [y_b-x_b \beta_4]$
            & $[\chi_z z_z - \beta_4] \cdot [\chi_z - y_z^2]$
            & $[\chi_\chi - \beta_4] \cdot [y_\chi^2-z_\chi]$ \\
\hline
$S_{\rm ii}$ & & & & $[z_\chi] \cdot [y_\chi]$ \\
\hline
$S_{\rm i}$ & & & $[z_z] \cdot $ 
            & $[\chi_\chi] \cdot $ \\
           & & & $ \left[ y_z(y_z-\beta_4\beta'_5) \right.$  
            & $\left[y_\chi(y_\chi -\beta_4\beta'_5 z_\chi) \right.$ \\ 
 & & & $\left. -\chi_z+\beta_4 \beta_2 \right]$
     & $\left. -z_\chi(1-\beta_4 \beta_2 z_\chi)\right]$ 
\end{tabular}
\caption{
This table shows how the two-dimensional subvarieties 
$S_{\rm i,ii,iii,iv,v}$ of $\tilde{X}_4$ are specified in the ambient 
space of $\tilde{X}_4$; 
For example, $S_{\rm iii}$ is given by 
${\rm div}(z_{313}) \cdot {\rm div}(y_b-x_b \beta_4) \cdot {\rm div}(\beta_3)$ in 
the ${\cal U}_b$ patch. The intersection with ${\rm div}(\beta_3)$ is omitted 
in this table, because it is common to all the entries of this table.  
}
\label{tab:SO10-singfib-10curve}
\end{table}

\subsubsection{Singular fibres over the codimension-three points of 
type ``$D_7$''}
\label{sssec:SO10-D7pt}

There are isolated points in $B_3$ specified by $\beta_3|_{S_{\rm GUT}}=0$ 
and $(\beta_2^2-4\beta_4\beta_0)|_{S_{\rm GUT}}=0$ on $S_{\rm GUT} \subset B_3$.
The singular fibre of $\tilde{X}_4 \rightarrow B_3$ degenerates further over 
these points. This is a special case of the study in section \ref{sssect:10matter}. 
Intersecting the surfaces $S_i$ of section \ref{sssect:10matter} with
${\rm div}(\tilde{\pi}_X^*(\beta_2^2-4\beta_4\beta_0))$, 
we will obtain the same copy for every such codimension-three point of 
type ``$D_7$''. The resulting formulae should thus involve an intersection number 
${\rm div}((\beta_3|_{S_{\rm GUT}}) \cdot 
 {\rm div}((\beta_2^2 - 4 \beta_4 \beta_0)|_{S_{\rm GUT}})$ on 
$S_{\rm GUT}$, but for the sake of brevity, we will present results on
multiplicity information from which this trivial multiplicative factor is removed.

\begin{figure}[!h]
\begin{center}
\scalebox{.5}{ \input{curvesD7.pspdftex} }
\caption{\label{DynkinD7} 
The fibres over codimension-three loci of type ``$D_7$'', which are characterized by the conditions 
$\beta_3|_{S_{\rm GUT}} = (\beta_2^2-4\beta_4\beta_0)|_{S_{\rm GUT}}=0$.
This picture is not like the $I_3^*$ type fibre.
If we see the local geometry of $\tilde{X}_4$ as an ALE-space fibration, 
then the fibre surfaces over such points of ``$D_7$'' type have two 
$A_1$ singularity points; they are on the $C_v$ component and are drawn as
the two blobs in this figure. If they were replaced by $\P^1$'s, this figure 
would look like the $I_3^*$ type of Kodaira classification. }
\end{center} 
\end{figure}
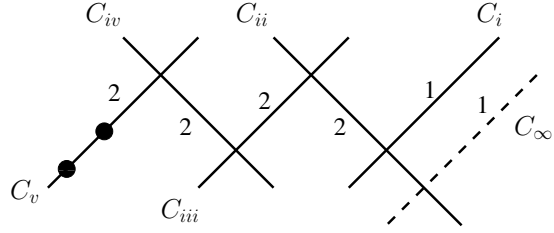

Intersecting the surfaces $S_{\rm i,ii,iii,iv}$ over the matter curve for the ${\bf 10}$ 
representation with the divisor ${\rm
div}(\tilde{\pi}^*_X(\beta_2^2-4\beta_4\beta_0))$, we obtain the
irreducible curves $C_{\rm i,ii,iii,iv}$.
However, for $S_{\rm v}$, 
\begin{equation}
  S_{\rm v} \cdot {\rm div}(\tilde{\pi}_X^*(\beta_2^2 - 4 \beta_4 \beta_0)) =
      2 C_{\rm v},
\end{equation}
as we define a component $C_{\rm v}$ by 
${\rm div}(\beta_3)\cdot {\rm div}(y_b) \cdot {\rm div}(\beta_2^2-4\beta_4\beta_0) \cdot {\rm div}(\beta_0 z+\beta_2/2)$ 
in the ${\cal U}_b$ patch. In the ${\cal U}_a$ patch, ${\rm div}(y_b)$ is replaced by ${\rm div}(x_a)$. 
Therefore, the singular fibre over the codimension-three points of ``$D_7$'' 
type is decomposed as 
\begin{equation}
 {\rm div}(\tilde{\pi}_X^*(z)) \cdot 
 {\rm div}(\tilde{\pi}_X^*(\beta_3)) \cdot 
 {\rm div}(\tilde{\pi}_X^*(\beta_2^2-4\beta_4 \beta_0)) 
 = C_{\infty} + C_{\rm i} +
   2(  C_{\rm ii} +  C_{\rm iii} +  C_{\rm iv}
      +  C_{\rm v}).
\end{equation}
A schematic picture of the singular fibre over this type of codimension-three 
points is shown in fig. \ref{DynkinD7}. 

If we are to take a local neighbourhood (in the complex analytic sense, 
not in the Zariski topology) of a point of this type, then a local geometry 
of $X_4^{\rm Weierstrass}$, and therefore that of $\tilde{X}_4$, can be 
regarded as an ALE-space fibration over the local patch of $S_{\rm GUT}$.  
In this context, $\beta_3|_{S_{\rm GUT}}$ and 
$(\beta_2^2-4\beta_4\beta_0)|_{S_{\rm GUT}}$ can be used as a 
set of local coordinates on $S_{\rm GUT}$ in the local neighbourhood, 
if the complex structure of $X_4^{\rm Weierstrass}$ is that of a generic 
$SO(10)$ model. 
Similarly to the fact that the non-singular nature of $\tilde{X}_4$ as 
a fourfold does not necessarily imply that the fibre curve geometry over 
all points in the base $B_3$ is non-singular, the non-singular nature 
of $\tilde{X}_4$ does not guarantee that the ALE fibre geometry here is 
non-singular everywhere. 
Indeed, the ALE fibre geometry over the point given by 
$(\beta_3|_{S_{\rm GUT}}, (\beta_2^2-4\beta_4\beta_0)|_{S_{\rm GUT}})=(0,0)$
is singular. There are two $A_1$ singularities at $(x_b,y_b,z)=(x_{*\pm},0,z_*)$,
where $\beta_0 z_* + \beta_2/2=0$ and $x_{*\pm}$ are the two roots of 
$1+z_* \beta'_5 x_b - z_*^2 x_b^2=0$. If we were interested in a fourfold 
geometry in which all such singularities of the ALE fibre are resolved 
(that is, $X_n$'s satisfying the condition (d) in Introduction), 
those surface singularities of $A_1$ type would also have to be resolved.
Consequently, two more irreducible components would appear and all the exceptional curves 
of the $D_7$ singularity resolution would be obtained 
(see fig. \ref{DynkinD7}). We are not studying such a fourfold geometry in this article, however,
as we are only resolving the singularities of the fourfold. In section \ref{ssect:SO10-resolution}, 
we constructed a fourfold so that the elliptic fibration (and the 
ALE fibration) as a family (i.e., as a fourfold) becomes non-singular, 
without trying to make all the elliptic (or ALE) fibre geometries non-singular. 
Thus, under such a mathematical construction (condition (e)), there is
nothing wrong in obtaining the singular fibre curve configuration
fig. \ref{DynkinD7} which is different from the naively expected 
$I_3^*$ type fibre in the Kodaira classification.

\subsubsection{Singular fibres over the codimension-three points of type ``$E_7$''}
\label{sssec:SO10-E7pt}

Finally, we summarise the results on the singular fibre over the
codimension-three points in $B_3$ satisfying 
$\beta_3|_{S_{\rm GUT}} = \beta_4|_{S_{\rm GUT}} = 0$. The two different
kinds of matter curves intersect transversely at these points, if the complex 
structure of $\tilde{X}_4$ is that of a generic $SO(10)$ model. Of course, 
there will be many points in the base for which the above conditions are 
satisfied; there are 
${\rm div}(\beta_3|_{S_{\rm GUT}}) \cdot {\rm div}(\beta_4|_{S_{\rm
GUT}})$ 
of them. As in the last section, we omit this intersection number 
from all expressions.

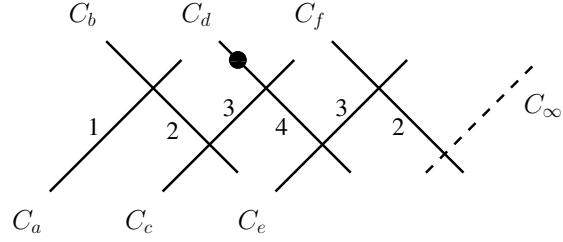
\begin{figure}[!h]
\begin{center}
\scalebox{.5}{ \input{curvesE7.pspdftex} }
\caption{The fibre over $\beta_3|_{S_{\rm GUT}}=\beta_4|_{S_{\rm GUT}}=0$.
This picture is not like the III$^*$ type fibre.
The blob in this figure represents the point which looks like an $A_1$ singularity 
when an ALE fibre space is sliced out (see the text). As a fourfold, 
though, this is not a singularity of $\tilde{X}_4$. 
\label{DynkinE7}}
\end{center} 
\end{figure}

The defining equations of the irreducible components of the
singular fibre over a point of this type are summarised 
in table \ref{tab:SO10-singfib-E7pt}.
%
\begin{table}[h!]
\centering
\begin{tabular}{l|l|l|l|l}
& ${\cal U}_{a}$ & ${\cal U}_{b}$ & ${\cal U}_{311z}$ & ${\cal U}_{311\chi}$ \\
\hline
$C_{b}$ &  & $[y_b] \cdot [x_b]$ &  &  \\
\hline
$C_{c}$ & $[x_a] \cdot [z_{313}]$ & $[y_b] \cdot [z_{313}]$ & & \\
\hline
$C_{a}$ & $[x_a] \cdot [\beta_2 + \beta_0 z_{313}]$ & 
                      $[y_b] \cdot [\beta_2 + \beta_0 z_{313}]$ & &  \\
\hline
$C_{d}$ & $[z_{313}] \cdot [y_a]$ & & $[\chi_z] \cdot [y_z]$ & \\
\hline
$C_{e}$ & & & $[z_z] \cdot [\chi_z - y_z^2]$ &
                          $[\chi_\chi] \cdot [z_\chi - y_\chi^2]$ \\
\hline
$C_{f}$ & & & & $[z_\chi] \cdot [y_\chi]$ 
\end{tabular}
\caption{
This table shows how the one-dimensional subvarieties 
$C_{a,b,c,d,e,f}$ of $\tilde{X}_4$ are specified in the ambient 
space of $\tilde{X}_4$. 
For example, $C_{a}$ is given by 
${\rm div}(y_b) \cdot {\rm div}(\beta_2 + \beta_0 z) \cdot {\rm
 div}(\beta_3) \cdot {\rm div}(\beta_4)$ in the ${\cal U}_b$ patch. 
The intersection with ${\rm div}(\beta_3) \cdot {\rm div}(\beta_4)$ 
is omitted from this table because it is common to all entries.  
}
\label{tab:SO10-singfib-E7pt}
\end{table}
A schematic picture of the fibre geometry is shown in fig. \ref{DynkinE7}.

We have already worked out the irreducible decomposition of the two parameter 
algebraic family over $S_{\rm GUT} \subset B_3$ and one parameter 
families over the matter curves $\beta_4|_{S_{\rm GUT}} = 0$ and 
$\beta_3|_{S_{\rm GUT}} = 0$. The fibres in these one-parameter families 
over the two different matter curves both degenerate further 
at this type of codimension-three point in the base.
The irreducible components in the family over the matter curve for 
$SO(10)$-${\bf 16}+\overline{\bf 16}$ representation degenerate in the
following way:
\begin{eqnarray}
 S_1 \cdot {\rm div}(\tilde{\pi}_X^*(\beta_3)) =
    C_c +  C_b +  C_a, & 
 S_2 \cdot {\rm div}(\tilde{\pi}_X^*(\beta_3)) =
   C_d, & 
 S_3 \cdot {\rm div}(\tilde{\pi}_X^*(\beta_3)) =
   C_e, \\
 S_4 \cdot {\rm div}(\tilde{\pi}_X^*(\beta_3)) =
   C_d + C_c, & 
 S_5 \cdot {\rm div}(\tilde{\pi}_X^*(\beta_3)) =
   C_b, & 
 S_1 \cdot {\rm div}(\tilde{\pi}_X^*(\beta_3)) =
   C_f,
 \label{eq:limit-SO10-16curve-E7pt}
\end{eqnarray}
while the irreducible components\footnote{The two irreducible components
of $S_{\rm v}$ restricted on the fibre of a generic point in the matter
curve, $C_{{\rm v}\pm}$, are algebraically equivalent to $C_c$ and
$C_a$, respectively.} over the matter curve for 
$SO(10)$-${\bf 10}$ representation degenerate as in 
\begin{eqnarray}
 S_{\rm i} \cdot {\rm div}(\tilde{\pi}_X^*(\beta_4)) = 
    C_e, & \quad & 
 S_{\rm iii} \cdot {\rm div}(\tilde{\pi}_X^*(\beta_4)) = 
     C_c +  C_e +  2 \; C_d, 
   \label{eq:limit-SO10-10curve-E7pt} \\
 S_{\rm ii} \cdot {\rm div}(\tilde{\pi}_X^*(\beta_4)) = 
     C_f, & \quad & 
 S_{\rm iv} \cdot {\rm div}(\tilde{\pi}_X^*(\beta_4)) = 
      C_b, \qquad 
 S_{\rm v} \cdot {\rm div}(\tilde{\pi}_X^*(\beta_4)) = 
       C_c +  C_a.  \nn 
\end{eqnarray}
The degeneration of the irreducible components of the two parameter family 
over $S_{\rm GUT}$ can be worked out by combining 
(\ref{eq:limit-SO10-Sgut-16curve}) and
(\ref{eq:limit-SO10-16curve-E7pt}), or by combining 
(\ref{eq:limit-SO10-Sgut-10curve}) and 
(\ref{eq:limit-SO10-10curve-E7pt}). The result is 
\begin{eqnarray}
 D_- \cdot
{\rm div}(\tilde{\pi}_X^*(\beta_4)) \cdot {\rm div}(\tilde{\pi}_X^*(\beta_3))
 =  C_b, & & 
 D_+ \cdot
{\rm div}(\tilde{\pi}_X^*(\beta_4)) \cdot {\rm div}(\tilde{\pi}_X^*(\beta_3))
 =  C_c + C_b + C_a, \\
 D_A \cdot
{\rm div}(\tilde{\pi}_X^*(\beta_4)) \cdot {\rm div}(\tilde{\pi}_X^*(\beta_3))
 = C_e, & \quad & 
 D_C \cdot
{\rm div}(\tilde{\pi}_X^*(\beta_4)) \cdot {\rm div}(\tilde{\pi}_X^*(\beta_3))
 =  C_c + 2 C_d + C_e, \\
 D_B \cdot
{\rm div}(\tilde{\pi}_X^*(\beta_4)) \cdot {\rm div}(\tilde{\pi}_X^*(\beta_3))
 =  C_f.
\end{eqnarray}
All of the irreducible components over the codimension-three points of this
type combined become
\begin{equation}
 {\rm div}(\tilde{\pi}_X^*(z)) \cdot 
 {\rm div}(\tilde{\pi}_X^*(\beta_4)) \cdot
 {\rm div}(\tilde{\pi}_X^*(\beta_3))
 = 
 C_{\infty} + 2 C_f + 3 C_e + 
4 C_d + 3 C_c + 2 C_b + C_a.
\end{equation}

If we focus on a local patch (in the complex analytic sense) of a 
such a point of ``$E_7$'' type in $S_{\rm GUT}$, then some local geometry 
of $\tilde{X}_4$ can be regarded as an ALE fibration over the 
local patch of $S_{\rm GUT}$.
Just like in section \ref{sssec:SO10-D7pt}, the fibre ALE geometry 
is not non-singular when we are on top of this codimension-three point. 
The singularity is found on the $C_d$ component; set
$\beta_3 = \beta_4=0$ in the defining equation in the ${\cal U}_{311z}$ 
patch (\ref{U311zchi}), and treat $\beta'_5, \beta_2$ and $\beta_0$
just as parameters; the singularity is at $\chi_z = y_z = 0$, 
$1+\beta_2 z_z = 0$. If this particular ALE fibre were to be sliced out, 
and resolved, then the additional exceptional curve would fill the 
missing irreducible piece of the ${\rm III}^*$ type fibre in the Kodaira 
classification. In this article, however, we do not try to resolve
singularities in the ALE fibre (or elliptic fibre) individually or to find 
$X_4$ satisfying the condition (d), but
construct a resolution that is non-singular as a fourfold. Thus, the singular 
fibre over the codimension-three locus in the base can be different 
from any one in the list of Kodaira. 

The observation of the last paragraph also shows the limited applicability of
the adiabatic argument in the context of F-Theory on Calabi-Yau fourfolds. 
The adiabatic argument was used as a powerful guiding idea in extending 
string duality in higher dimensions to those in lower dimensions, because it 
expresses a belief that the physics of a compactification on a fibred 
geometry changes only gradually in the base as the moduli parameter 
of the fibre geometry changes gradually over the base. 
Such things as symmetry restoration or further degeneration of singular 
fibres, however, are non-adiabatic changes, and the pioneers in the 90's did 
not rush to cross such non-trivial gaps without accumulating supporting 
evidence.  
If we are to adopt the condition (e) for F-theory compactification, 
then the mathematical object of our interest is the family of 
$X_n$'s satisfying the condition (e): the family, which we denote 
${\cal X}_n^{(e)}$, is a fibration over the moduli space 
${\cal M}_n^{(e)}$ of such geometry, with each fibre 
$(X_n^{(e)})_m$ for $m \in {\cal M}_n^{(e)}$ being 
an $X_n$ satisfying the condition (e). We are facing a question whether
the $(n+d)$-dimensional compactification---$X_{n+d}$---to be used in the
adiabatic argument always has a local geometry given by ${\cal
X}_n^{(e)} \times_{{\cal M}_n^{(e)}} \C^d$ for some map from $\C^d$ to 
${\cal M}_n^{(e)}$. The answer is no, and hence the adiabatic argument 
should not be used even to ``guess'' the further degeneration of 
singular fibres at higher codimension loci as long as we adopt the
condition (e).

\subsection{Discussion}
\label{ssect:SO10-discussion}

It is a well-defined problem in mathematics to look for $\tilde{X}_4$
satisfying condition (e), and study the geometry of singular fibres, 
as we have done so far. 
From the physics perspective, however, we should say that the condition
(e) is not derived or justified from anywhere as a property required
for the input data $X_n$ of some formulation of F-theory,  
although it certainly looks like one of the most promising among the conditions 
presented as (a)--(e).\footnote{The condition (a) may be just as 
promising as (e). Different conditions on the singularities of $X_4$ can also contain equivalent information
if the corresponding fourfolds can be uniquely constructed from one another. It is then a merely a question of 
formulation which one is chosen. Even though the short list of conditions (a) to (e) seemed rather natural to us,
we can by no means exclude others.} Floating in the air as a ``puzzle'' after the work of
\cite{Esole:2011sm} was what to think of the geometry of singular fibres 
over codimension-3 loci in the base $B_3$. 

In the absence of theoretical formulations relating the geometry of
singular fibres to some physical consequences or predictions, 
mathematical results on the singular fibre geometry do not pose any
puzzle to begin with. Codimension-three loci (in the base) of various types have been referred
to in the recent literatures on F-theory phenomenology by using the A-D-E 
classification. This naming, however, originates from the fact 
that the physics associated with the local geometries around those
codimension-3 loci is described approximately by a gauge theory with the
gauge group of that A-D-E type \cite{Donagi:2008ca,
Beasley:2008dc, Hayashi:2008ba, Donagi:2008kj, Hayashi:2009ge}, and 
also from the fact that the local geometry is approximately regarded 
as a fibration of a deformed ALE space of that A-D-E type 
(cf. \cite{Tatar:2006dc}). We have used a notation like ``$D_7$'' point or 
``$E_7$'' point with quotations in this article, but there is nothing
wrong with  
using the A-D-E type for classification of codimension-3 loci 
in the sense stated above. This A-D-E type classification of
codimension-3 loci does not refer to the type of singular fibre, 
type of singularity\footnote{
In the classification theory of singularity, exists a class of 
singularities called ``simple singularities''. Simple singularities 
are given a label which is one of $A_n$, $D_n$ and $E_{6,7,8}$.
They are given by the following equations:
$\sum_i (y_i)^2 + z^{n+1} = 0$ ($A_n$ type), 
$\sum_i (y_i)^2 + x^2 z + z^{n-1} = 0$ ($D_n$ type), 
$\sum_i (y_i)^2 + x^3 + z^4 = 0$ ($E_6$ type),
$\sum_i (y_i)^2 + x^3 + x z^2 = 0$ ($E_7$ type) and 
$\sum_i (y_i)^2 + x^3 + z^5 = 0$ ($E_8$ type), respectively. 
These are stabilizations of the ADE surface singularities.
Thus, one could think of simple singularities of A-D-E type in a
fourfold, but they are isolated point singularities, and are not 
the type of singularities we address in this article. 
The A-D-E labels for the codimension-three loci in the base did not 
originate from such a classification in singularity theory.} 
of $X_4^{\rm Weierstrass}$, or even to whether the 4-fold $X_4$ is $X_4^{{\rm Weierstrass}}$ or one of 
its resolutions. Having made this point clear, we will drop the
quotation marks for the A-D-E labels referring to the codimension-three 
loci in the base in the rest of this article. 

As shown in \cite{Marsano:2011hv}, the resolved geometry of \cite{Esole:2011sm}
does indeed lead to the expected Yukawa couplings via recombination of wrapped
M2-branes. If we focus our attention on the types of singular fibres, however,
there still remains a question. If the condition (e) proposed by 
\cite{Esole:2011sm} does not seem so bad (in the sense of physics), 
isn't there a rule or dictionary relating the type of singular fibres  
and physics at / around the codimension-3 loci?

Having worked out the geometry of singular fibres also for $SO(10)$
models, we now know the singular fibre for five different types of 
codimension-3 loci: three ones in $SU(5)$ models as in \cite{Esole:2011sm}, 
and two ones in $SO(10)$ models as discussed in \cite{Tatar:2012tm} and here. 
It is now evident what the rule is. 
The number of irreducible components in the singular fibre is the 
same as the number of nodes in the extended Dynkin diagram associated 
with the A-D-E label of the codimension-3 loci, if and only if 
the Higgs field in the field theory local model (Katz--Vafa field 
theory \cite{Katz:1996xe}) begins with a linear term in the local 
coordinates on the non-Abelian 7-branes.\footnote{The $D_6$ points 
in $SU(5)$ models correspond to this case.}
Whenever the spectral cover for the Higgs field configuration is 
ramified at the codimension-3 loci,\footnote{This happens for the $E_6$ 
points and $A_6$ points in $SU(5)$ models, and the $D_7$ points and
$E_7$ points in $SO(10)$ models.} the number of irreducible components in the singular
fibre is less than the number of nodes in the extended Dynkin diagram, see
e.g. sections \ref{sssec:SO10-D7pt} and \ref{sssec:SO10-E7pt} 
(and also \ref{sssect:10matter}). 
It is evident that the reduction in the number of irreducible components 
is related to the ramification behaviour. 

In a patch where the spectral surface is ramified at a codimension-3
locus, let the defining equation for the spectral surface approximately be 
\begin{equation}
 \xi^2 - u \xi - v = 0 \, , 
\label{eq:spec-surf-2}
\end{equation}
where $(u,v)$ is a set of local coordinates on the 
non-Abelian 7-branes (GUT divisor), and $\xi$ is the fibre coordinate 
of the total space of the canonical bundle on the GUT divisor. We assume
that the origin $(u,v)=(0,0)$ is one of the codimension-3 points of $B_3$, 
and $\{ v=0 \}$ is the matter curve in the GUT divisor. The corresponding 
Higgs field configuration is given by (see appendix C of
\cite{Hayashi:2009bt})
\begin{equation}
 \widetilde{\varphi} = \left( \begin{array}{cc}
			0 & v \\ 1 & u 
			      \end{array} \right).
\end{equation}
There is a non-vanishing Higgs field vev remaining at the origin $(u,v) =
(0,0)$, even though its two eigenvalues vanish there. This double cover 
spectral surface is what we encounter 
at the $E_6$ and $A_6$ type points in $SU(5)$ models, and 
at the $E_7$ type points in $SO(10)$ models. Although 
a four-fold spectral cover is necessary for the $D_7$-type points in 
$SO(10)$ models, it remains true that a non-vanishing Higgs vev remains 
even at the codimension-3 loci; see \cite{Hayashi:2008ba, Hayashi:2009ge}.
On the other hand, the situation is quite different in the $D_6$ type 
points in $SU(5)$ models. Here, the spectral surface is given by 
$\xi^2 + (u+v) \xi + uv = 0$, and the Higgs field configuration is 
given by \cite{Hayashi:2009ge}
\begin{equation}
 \widetilde{\varphi} = \left(\begin{array}{cc}
			-u & \\ & -v
			     \end{array}\right).
\end{equation}
Hence the Higgs field vev vanishes at the $D_6$ type points of $SU(5)$ models. 

The symmetry unbroken by the Higgs field vev is a well-defined notion 
captured in the language of the field theory local model (Katz--Vafa  
field theory) on 7+1 dimensions for each point in the base. Given
a patch which contains only a single singularity of codimension three,
there furthermore is a unique minimal choice for the gauge group of
the field theory local model. The rule that we discovered empirically 
is that the number of irreducible components in the singular fibre in $\tilde{X}_4$ 
is the same as the number of nodes in the extend Dynkin diagram if and only if 
the ``symmetry that is not broken by the Higgs field vev'' at the 
codimension three loci is the same as this minimum choice of the gauge group of 
the field theory local model. 
This empirical rule leaves an impression to us that things are sort of 
``going well'' in the geometry satisfying the condition (e). 
If we are convinced that this empirical rule holds true, then 
we will be able to conclude that the ``puzzling'' behaviour of singular 
fibres observed in \cite{Esole:2011sm} is no longer a negative evidence against 
existence of a microscopic formulation of F-theory in which $\tilde{X}_4$ 
under the condition (e) is used as input data $X_4$ in the diagram 
(\ref{eq:comm-diagram}).

Another lesson that we can extract from studying resolutions of 
$X_4^{\rm Weierstrass}$ is this. The Higgs field vev in the field theory 
local models and the coefficients (complex structure information) 
in the defining equation of $X_4^{\rm Weierstrass}$ are related by 
the Hitchin/Katz--Vafa map \cite{Katz:1996xe, Hayashi:2009ge}, but 
this map only extracts eigenvalues of the Higgs field configuration. 
Thus, it has been considered difficult for the fourfold geometry $X_4$ 
to deal with some physics data of compactifications such as i) 
whether the spectral surface is regular\footnote{In 
\cite{Donagi:2000dr}, spectral surfaces are said to be regular if and only if the number 
of Jordan blocks of the Higgs field is the same as the number of 
distinct eigenvalues of the Higgs field at any point.} or 
not \cite{Hayashi:2009bt, Cecotti:2010bp}, and   
ii) the extension structure of the Higgs bundle\footnote{The 
off-diagonal component of the Higgs vev triggers symmetry breaking, but 
it does not change the Higgs field vev eigenvalues in this case.}, which is 
motivated by the spontaneous R-parity violation scenario \cite{Tatar:2006dc, Tatar:2009jk}.
Although the fourfold $X_4^{\rm Weierstrass}$ did not seem to contain 
data on the off-diagonal vev of the Higgs field at all, yet once it is 
resolved, at least a shadow of the information corresponding to i) seems 
to show up in the singular fibre geometry.

\section{Linear higgs vevs and change in the singular fibre}
\label{sect:Hvev}

In order to test the idea that the Higgs vev controls the reduction of
irreducible components in the singular fibre, we do the following
experiment in this section. We know that the two-fold covering spectral 
surface (\ref{eq:spec-surf-2}) becomes factorized for a special
choice of complex structure and the monodromy group is reduced. 
We therefore construct a fourfold $X_4^{(\rm Weierstrass)}$ for 
such a special complex structure, and carry out a resolution analysis similar 
to \cite{Esole:2011sm} (and the one presented in the previous section). 
We will see indeed that the number of irreducible components in the singular
fibre becomes the same as the number of nodes in the extended Dynkin
diagram, even over the codimension-3 loci in the base $B_{3}$. 
This is a non-trivial confirmation of the empirical rule 
(and observations that follow from the rule) above.  

\subsection{The codimension-three loci of $A_6$ type in $SU(5)$ models}
\label{ssect:resA6}

Let us begin with the analysis of a simple configuration which
already clearly shows the influence of the Higgs vev on the fibre
structure. For this, we consider a geometry $X_4^{(\rm Weierstrass)}$ 
given by 
\begin{equation}
 y^2 = x^2 + z^N (s_1(u,v) + s_2(u,v) z + z^2),  \qquad N=5, 
\end{equation}
where $(z,u,v)$ are local coordinates on $B_3$ and $(x, y)$ those 
for the elliptic fibre. The above fourfold also has the structure
of an ALE fibration. Here, $(x,y,z)$ are coordinates along the fibre directions
and $(u,v)$ parametrise the base. For $N=5$, this geometry is then regarded as a 
fibration of an ALE space of $A_6$ type deformed by two parameters
$s_{1,2}(u,v)$ so that an $A_4$ singularity remains at $(x,y,z)=(0,0,0)$. 
This is meant to be a local geometry of a fourfold 
$X_4^{\rm Weierstrass}$ containing a codimension-three point of $A_6$ 
type in the base. 
If the complex structure of $s_1(u,v)$ and $s_2(u,v)$ is generic, 
then the spectral surface $\xi^2 + s_2 \xi + s_1 = 0$ is essentially 
of the form (\ref{eq:spec-surf-2}). This is the case studied in 
\cite{Esole:2011sm}.

When we choose $s_1$ and $s_2$ in a very special way, namely, 
\begin{equation}
 s_1(u,v) = s_+(u,v) s_- (u,v) \quad {\rm and} \quad 
 s_2 (u,v) = - (s_+ + s_-) 
\label{eq:linear-cpxstr-A4}
\end{equation}
for some $s_\pm(u,v)$, then there is no 7-brane monodromy. 
As we will see below, the geometry of singular fibres also becomes 
different in this case. 

\begin{figure}[!h]
\begin{center}
\scalebox{.5}{ \input{A4.pspdftex} }
\caption{The fibre components over a generic point meet according to the Dynkin diagram $A_4$. Note that we do not
include the fibre component at infinity as it cannot be seen in the ALE space \eqref{a6nonfact}.
\label{A4figure}}
\end{center} 
\end{figure}
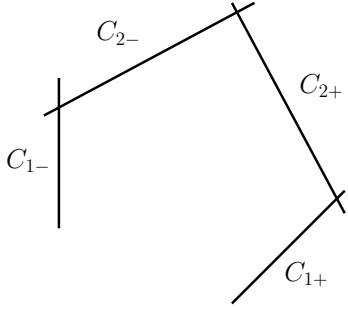

As a first step, the crepant resolution of the $A_4$ singularity 
along $(x,y,z) = (0,0,0)$ is carried out. Let the proper transform 
be $X'_4$. In the ${\cal U}_{33}$ patch, $X'_4$ is given by 
\begin{equation}\label{a6nonfact}
 (y_{33})^2 =  (x_{33})^2 + z_{33} (z_{33}^2 + s_2 z_{33} + s_1 ).
\end{equation}
Two exceptional divisor $D_{1\pm}$ appear in the first blow-up, and 
two more $D_{2\pm}$ in the second one. $D_{1\pm}$ do not appear in the 
${\cal U}_{33}$ patch, and $D_{2\pm}$ are characterized by 
$z_{33} = 0$ and $y_{33} \pm x_{33} = 0$. The exceptional divisors
meet according to figure \ref{A4figure}.
Up until now, it does not matter whether $s_1(u,v)$ and $s_2(u,v)$ 
have a special dependence on $(u,v)$. In the following we will discuss the resolutions
of the remaining singularity both for the case of generic $s_1(u,v)$ and $s_2(u,v)$
and for the case in which the condition \eqref{eq:linear-cpxstr-A4} holds. This discussion will 
entirely take place in the patch ${\cal U}_{33}$, so that we may drop the label
$_{33}$ from now on. Furthermore, the exceptional divisors $D_{1\pm}$
do not suffer any transmutation when moving around on the complex $u,v$ plane, so that we
may limit the following discussion to the study of the fate of $D_{2\pm}$.

Let us first describe the resolution of the singularity remaining in
$X'_4$ \eqref{a6nonfact} for the case of generic $s_1(u,v)$ 
and $s_2(u,v)$; this is already described in \cite{Esole:2011sm}, but 
we will discuss it in detail here so that one can easily see how the generic and 
non-generic choices of the complex structure of $s_{1,2}(u,v)$ make a
difference in the geometry of singular fibre. 
Here, we can resolve \eqref{a6nonfact} by introducing a new ambient
space $\left\{(x,y,z,u,v,[\xi_+:\xi_-]) \right\} = \C^5 \times \P^1$,
where the proper transform $\tilde{X}_4$ is given by two equations:
\begin{align}
\xi_- (y+x) & =\xi_+ z \nn\\
\xi_+ (y-x) & = \xi_- (z^2 + s_2 z + s_1 ) \, .
\end{align}
We now have achieved to realize condition (e): $\tilde{X}_4$ is smooth, and this process is a small
resolution.

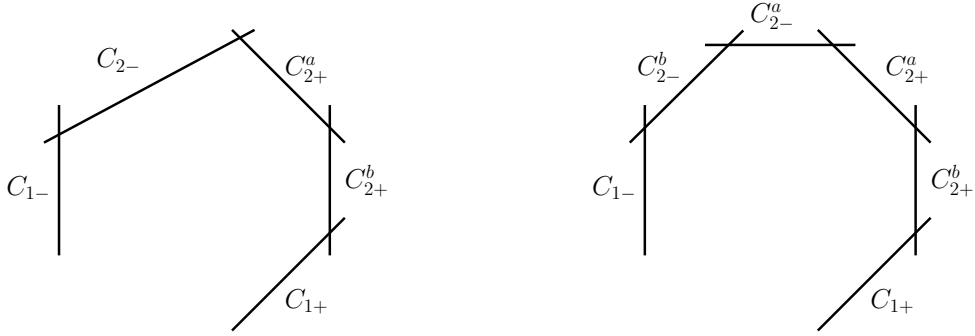
\begin{figure}[!h]
\begin{center}
\scalebox{.5}{ \input{A6nlH.pspdftex} }\hspace{3cm}
\scalebox{.5}{ \input{A6lH.pspdftex} }
\caption{The fibre components over the point $s_1=s_2=0$. In case $s_1$ and $s_2$ are generic,
the exceptional divisors $D_{2\pm}$ and $D_{1\pm}$ split up into $5$ irreducible curves, as shown
on the left. If we choose $s_1$ and $s_2$ to be of the form \eqref{eq:linear-cpxstr-A4}, we obtain
$6$ irreducible curves, as shown on the right. Together with the
 component $C_\infty$ that originates from $D_\infty$, these singular
 fibre configurations are like $I_6$ type and $I_7$ type, respectively. 
\label{A6nlHvslH}}
\end{center} 
\end{figure}

The exceptional divisors $D_{2\pm}$ in $X'_4$ now have a non-trivial 
proper transform in $\tilde{X}_4$, which we denote by $D_{2\pm}$ as before. 
It can be described in terms of an intersection of three divisors in
the ambient space $\C^5 \times \P^1$:
\begin{align}
D_{2-}:&\quad {\rm div}(z)\cdot{\rm div}(y-x)\cdot{\rm div}(\xi_-) \nn\\
D_{2+}:&\quad {\rm div}(z)\cdot{\rm div}(y+x)\cdot{\rm div}\left(\xi_+ (y-x)  - \xi_-s_1 \right) \, .
\end{align}
In order to find the geometry of the fibre over the
codimension-three locus at $(z,u,v)=(0,0,0)$ in $B_3$, we only need to 
take the intersections among $D_{2\pm}$ (or $D_{1\pm}$), 
$\left\{s_1 = 0 \right\}$ and $\left\{s_2 = 0 \right\}$.
While $D_{2-}$ and $D_{1\pm}$ lead to irreducible curves when intersected with ${\rm div}(s_1)$ and ${\rm div}(s_2)$, $D_{2+}$ 
splits into the two irreducible components:
\begin{align}
C_{2+}^a:&\quad {\rm div}(s_1)\cdot{\rm div}(s_2)\cdot{\rm div}(x)\cdot{\rm div}(y)\cdot{\rm div}(z) \nn\\
C_{2+}^b:&\quad {\rm div}(s_1)\cdot{\rm div}(s_2)\cdot{\rm div}(x+y)\cdot{\rm div}(z)\cdot{\rm div}(\xi_+) \, .
\end{align}
Note that only $s_1$ plays a non-trivial role and this splitting occurs already
over the $s_1=0$ matter curve. Nothing extra happens when $s_2$ is also
set to zero. Together with $C_{2-}$ and $C_{1\pm}$, the fibre 
components over the point $s_1=s_2=0$ ($A_6$ type point) make up a
configuration that looks like the $I_6$ type of Kodaira classification 
(rather than the $I_7$ type), as shown in fig. \ref{A6nlHvslH}.

Let us now discuss the case where \eqref{eq:linear-cpxstr-A4} is 
satisfied, i.e. we take the polynomials $s_1$ and $s_2$ to reflect 
a Higgs vev which is linear in the vicinity of the $A_6$ point.
In this case, we can write \eqref{a6nonfact} as
\begin{equation}\label{a6fact}
 (y-x)(y+x) =  z (z-s+)(z-s_-).
\end{equation}
The crucial change is that there is now a further factorization,
changing the structure of the singularity at $s_1=s_2=0$. Note that this
type of singularity is precisely the one discussed in
\cite{Esole:2011sm} (albeit here it occurs in a different context), 
so we can use one of the small resolutions discussed there to resolve it. 
Let us choose the one given by
\begin{align}
(y-x)\xi_1 & =\xi_2 (z-s_+) \nn\\
(x+y)\phi_1 & = \phi_2 (z-s_-) \nn\\
\xi_2\phi_2 & = z\xi_1\phi_1 \, ,
\label{smallresa4}
\end{align}
where the new ambient space is 
$\left\{ (x,y,z,u,v,[\xi_1: \xi_2],[\phi_1: \phi_2]) \right\} =
 \C^5 \times \P^1 \times \P^1$.
$\tilde{X}_4$ obtained in this way is smooth.

The proper transforms of the divisors $D_{2\pm}$ of $X'_4$ can be expressed as intersections
in the ambient space.  They read 
\begin{align}
D_{2+}:\quad & {\rm div}(z)\cdot{\rm div}(x+y)\cdot{\rm div}(\phi_2)\cdot{\rm div}\left( (y-x)\xi_1-s_+\xi_2 \right)\nn\\
D_{2-}:\quad & {\rm div}(z)\cdot{\rm div}(x-y)\cdot{\rm div}(\xi_2)\cdot{\rm div}\left( (y+x)\phi_1+\phi_2s_-\right)\, .
\label{eq:45}
\end{align}
Here, we have used the same notation $D_{2 \pm}$ for the proper 
transforms (\ref{eq:45}) in $\tilde{X}_4$. The situation now presents
itself as very symmetric: 
Intersecting the divisors $D_{2\pm}$ with $\left\{s_+=0 \right\}$, only 
$D_{2+}$ becomes reducible and splits into
\begin{align}
S_{2+}^a:\quad & {\rm div}(s_+)\cdot{\rm div}(x)\cdot{\rm div}(y)\cdot{\rm div}(z)\cdot{\rm div}(\phi_2) \nn\\
S_{2+}^b:\quad & {\rm div}(s_+)\cdot{\rm div}(x+y)\cdot{\rm div}(z)\cdot{\rm div}(\xi_1)\cdot{\rm div}(\phi_2) \, ,
\end{align}
whereas intersecting with $\left\{s_-=0 \right\}$ results only in the
splitting of $D_{2-}$ into the two components:
\begin{align}
S_{2-}^a:\quad & {\rm div}(s_-)\cdot{\rm div}(x)\cdot{\rm div}(y)\cdot{\rm div}(z)\cdot{\rm div}(\xi_2) \nn\\
S_{2-}^b:\quad & {\rm div}(s_-)\cdot{\rm div}(x-y)\cdot{\rm div}(z)\cdot{\rm div}(\phi_1)\cdot{\rm div}(\xi_2) \, . 
\end{align}
Finally, the fibre over the point $(z,s_+,s_-)=(0,0,0)$ can be 
obtained by taking an intersection of 
$\left\{ s_+ = 0 \right\} \cdot \left\{ s_- = 0 \right\}$ and
$D_{2\pm}$. We find that $D_{2\pm}$ split into four curves, which, 
together with $D_{1\pm}$ and $D_\infty$, constitute the seven
irreducible components of the singular fibre over the codimension-three 
point of $A_6$ type. The information of a pair of such curves having a common
point is displayed in figure \ref{A6nlHvslH}; it is  
just like the $I_7$ type fibre of Kodaira classification.\footnote{We
avoid drawing an extended Dynkin diagram of $A_6$ for the reason 
we stated at the end of Step 1 in section \ref{ssect:SO10-resolution}.}
Note that there are six different small resolutions of the
geometry given by equation (\ref{a6fact}) \cite{Esole:2011sm}, so that
there are five others in addition to the one \eqref{smallresa4} 
we have just examined.  
The other five correspond to changing the role among $(y \pm x)$, and 
also among $z$, $(z-s_+)$ and $(z-s_-)$.
The $(z,s_+,s_-)=(0,0,0)$ is the common zero of the last three factors, 
and nothing should change essentially by choosing $(z-s_{\pm})$ instead
of $z$. Thus, the fibre structure over this locus must be the same 
for all of the six small resolutions of \eqref{a6fact}.

\subsection{The codimension-three loci of $E_6$ type in $SU(5)$ models}
\label{ssect:resE6}

Let us turn to the analysis of the resolution of a local geometry
associated with codimension-three loci of $E_6$ type in $SU(5)$
models. For a general choice of complex structure, the spectral surface
is ramified (``7-brane'' monodromy is non-trivial) in this case
\cite{Hayashi:2009ge}, and the number of components of the singular
fibre is less than 7 \cite{Esole:2011sm}. 

Instead of considering $X_4^{(\rm Weierstrass)}$ given by 
\begin{equation}\label{weiersu5sing}
 y^2 + \beta_5 yx + \beta_3 z^2 y = x^3 + \beta_4 z x^2 + \beta_2 z^3 x
  + \beta_0 z^5,
\end{equation}
with $\beta_{i=0,2,3,4,5}$ depending {\it generically} on a set of local
coordinates $(u,v)$ on a local patch of $S_{\rm GUT}$, we take an 
$X_4^{(\rm Weierstrass)}$ with a very special choice of complex structure 
\begin{equation}
 y^2 + (a b) y x + z^2 y = x^3 + (a + b) z x^2, 
\label{eq:weierSU5-E6-linear}
\end{equation}
where $(a,b)$ is a set of local coordinates on $S_{\rm GUT}$.
This choice is such that the two-fold covering spectral surface 
\begin{equation}
 \xi^2 -(a+b) \xi + (a b) = 0
\end{equation}
factorizes completely, so that the 7-brane monodromy is gone.\footnote{ 
Let us motivate this choice and point out its connection to our discussion in 
section \ref{ssect:SO10-discussion}. As explained in \cite{Hayashi:2009ge}, the 
physics associated with local geometry  (in the complex analytic sense) of 
$X_4^{\rm Weierstrass}$ around a type $E_6$ codimension-three point in $B_3$ can
be captured approximately by an $E_6$ gauge theory on a local patch of $S_{\rm GUT}$ (field theory local model), 
with the Higgs field vev given by 
$(\beta_3|_{S_{\rm GUT}}) \xi^2 + (\beta_4|_{S_{\rm GUT}}) \xi + (\beta_5|_{S_{\rm GUT}})=0$.
The condition for a breaking of $E_6$ to $A_4$ with a linear
Higgs vev is that we can write the $\beta_i$'s in terms of some local
coordinates $a,b$ as
\begin{equation}
 \beta_4/\beta_3 = a+b \, ,\hspace{1cm} \beta_5/\beta_3 = ab \, ,
\end{equation}
so that the spectral surface factorizes.
Whereas this analysis uses a two-fold spectral cover ($E_6$ gauge theory)
as an approximate description of the local geometry of $X_4^{\rm Weierstrass}$, 
the field theory local model with $E_6$ gauge group may be regarded as 
an approximation of a similar model with a larger gauge group (such as $E_7$ 
or $E_8$). Factorization conditions should be imposed on the coefficients 
of the spectral surface for the larger gauge group then. Such a description 
is still approximate in nature, however, even when the gauge group is chosen 
to be the maximal one, $E_8$ \cite{Tatar:2009jk, Hayashi:2010zp}. 
Reference \cite{Grimm:2010ez} 
(see also \cite{Choi:2012pr} and \cite{Mayrhofer:2012zy}) 
proposed a prescription of extending the notion of factorized spectral surface 
in field theory local models into a language of global geometry 
$X_4^{\rm Weierstrass}$ (where gravity is also involved). Here one
promotes the conditions on $\beta_i|_{S_{\rm GUT}}$ in a local patch of 
$S_{\rm GUT}$ to those on $\beta_i$ to be satisfied globally on $B_3$.
When the defining equation (\ref{eq:weierSU5-E6-linear}) is regarded as 
that of a global geometry $X_4^{\rm Weierstrass}$, the choice of the
complex structure in (\ref{eq:weierSU5-E6-linear}) is regarded as the
extension of the condition for the factorization of the two-fold
spectral cover under the prescription of \cite{Grimm:2010ez}. Note that
this means that the line bundle 
${\cal O}_{B_3}(-3 K_{B_3}-2S_{\rm GUT})$ admits a global section that
vanishes nowhere in $B_3$, and $a$ and $b$ are global holomorphic sections 
of ${\cal O}_{B_3}(-2K_{B_3}-S_{\rm GUT})$; this is the case, for example, 
if ${\cal O}_{B_3}(-3K_{B_3}-2S_{\rm GUT})$ is trivial, $K_{B_3}+S_{\rm GUT}$
is effective, and hence $K_{S_{\rm GUT}}$ is also effective; 
see \cite{Hayashi:2010zp}.

If one takes a less aggressive view, however, and considers
(\ref{eq:weierSU5-E6-linear}) as the defining equation of a local
geometry $X_4^{(\rm Weierstrass)}$, then $\beta_{0,2}$ are simply
ignored because they are irrelevant in this approximation scheme, and  
the choice $\beta_3 = 1$ only means that $\beta_3|_{S_{\rm GUT}}$ does not 
vanish in a local neighbourhood of an $E_6$ type codimension-three 
point \cite{Hayashi:2009ge}. }

After the crepant resolution of the $A_4$ singularity along 
$(x,y,z) = (0,0,0)$, we obtain $X'_4$, which is given by 
\begin{equation}
 y_{31}(y_{31} + (a b) +  z_{31} ) = x_{31} z_{31} (x_{31} + a + b) \, 
\label{eq:A4-resolved-eq}
\end{equation} 
in the ${\cal U}_{31}$ patch.

\subsubsection{Singularities in higher codimension and their resolution}
\label{sectresE6lH}

After the crepant resolution of the $A_4$ singularity in $X_4^{(\rm Weierstrass)}$, 
there still remain singular loci of codimension three or higher in $X'_4$.
We begin by identifying such singularities and explain how to resolve them 
in section \ref{sectresE6lH}, and study the geometry of singular fibres in 
section \ref{E6lHfibretrans}.

As we will see in the following, $X'_4$ in the ${\cal U}_{31}$ patch 
contains six distinct codimension-three singular loci, which all meet 
at one point. While other patches such as ${\cal U}_1$, ${\cal U}_2$ and 
${\cal U}_{32}$ also contain codimension-three singular loci of 
$X'_4$, these are the same as those already captured 
in the ${\cal U}_{31}$ patch, and we will discuss them in the 
${\cal U}_{31}$ patch in the following. For this reason, we will drop 
subscripts ${}_{31}$ from now on.

There are six codimension-three singular loci in $X'_4 \cap {\cal U}_{31}$. 
They are given by  
\begin{eqnarray}
C_{{\bf 10}a}: & & a=0, \quad x=0, \quad y=0, \quad z=0, \label{eq:6singl-1} \\
C'_{{\bf 10}a}: & & a=0, \quad (x+b)=0, \quad y=0, \quad z=0,\\
C_{{\bf 10}b}: & & b=0, \quad x=0, \quad y=0, \quad z=0, \\
C'_{{\bf 10}b}: & & b=0, \quad (x+a)=0, \quad y=0, \quad z=0, \\
C_{{\bf 5}}: & & (a+b)=0, \quad x=0, \quad y=0, \quad z+ab = 0,
    \label{eq:6singl-5} \\
C_{{\rm bifund}.}: & & a=b, \quad x=-a, \quad -y=z=a^2.  \label{eq:6singl-6}
\end{eqnarray}
The first four are in the fibre of the matter curve 
$\beta_5|_{S_{\rm GUT}}=ab = 0$ for the $SU(5)$-${\bf 10}+\overline{\bf 10}$ representation, 
and the fifth one is in the fibre of the matter curve 
$(\beta_0 \beta^2_5 - \beta_2 \beta_5 \beta_3 + \beta_4 \beta_3^2)|_{S_{\rm GUT}}
  \rightarrow \beta_4|_{S_{\rm GUT}} = (a+b) = 0$ for the 
$SU(5)$-${\bf 5}+\bar{\bf 5}$ matter curve. 
Although all the codimension-three singular loci of $X'_4$ for a generic 
complex structure are in the fibre of either one of these two matter curves, 
now a new singular locus, $C_{{\rm bifund}.}$, appears in the case of a
linear Higgs vev in the $E_6$ field theory local model. It is located
over the curve $a=b$ in $S_{\rm GUT}$, above which the two irreducible
pieces of the spectral surface, $(\xi-a)=0$ and $(\xi-b)=0$, intersect
and an SU(2) symmetry is enhanced in the field theory local
model. These six codimension-three singular loci of $X'_4$ meet 
at $(x,y,z,a,b)=(0,0,0,0,0)$.

Since the singularity structure in $X'_4$ for the linear Higgs vev is 
clearly different from the one for generic complex structure, we need to 
find a new resolution $(\tilde{X}_4, \rho'')$ of $X'_4$ for the linear 
Higgs case. We will first present a resolution which recycles a morphism 
used in \cite{Esole:2011sm} as a partial resolution, and later explain a
systematic way to find other resolutions by using toric language. 

It is natural to try to use the change of the ambient space employed 
for a small resolution of higher codimension singularities 
in \cite{Esole:2011sm}, because the defining equation
(\ref{eq:A4-resolved-eq}) still looks very similar to 
the one studied in \cite{Esole:2011sm}. 

Let us hence replace the ambient space 
$\C^5 = \left\{ (x,y,z,a,b) \right\}={\cal U}_{31}$ of $X'_4$ 
(\ref{eq:A4-resolved-eq}) by 
$\C^5 \times \P^1 \times \P^1 $, on which we have coordinates
$ \left\{ (x, y, z, a, b, 
 [\xi_1: \xi_2], [\zeta_1 : \zeta_2]) \right\}$.
The proper transform of $X'_4$ is given by\footnote{This corresponds to
the resolution ${\cal E}_{xw}$ in \cite{Esole:2011sm}. The coordinate
``w'' is now $z_{(31)}$, though.}
\begin{align}
y\xi_1 & =x\xi_2  \label{eq:A4-E6-apre-EY-1} \\
(y+ab+z)\zeta_1 &=  z\zeta_2 \label{eq:A4-E6-apre-EY-2} \\
\xi_2\zeta_2 & =\xi_1\zeta_1(x+a+b) \, . \label{eq:A4-E6-apre-EY-3}
\end{align}

The first five singular loci (\ref{eq:6singl-1}--\ref{eq:6singl-5}) are 
now resolved, but the last one (\ref{eq:6singl-6}) still remains. 
To see this, note that the two equations (\ref{eq:A4-E6-apre-EY-1}) 
and (\ref{eq:A4-E6-apre-EY-3}) can be solved for $x$ and $y$ in the
patch where $\xi_1\neq 0$ and $\zeta_1\neq 0$.   
The remaining equation (\ref{eq:A4-E6-apre-EY-2}) may then be written as
\begin{equation}
 \left(\zeta_2/\zeta_1-1 \right)\left(z-(\xi_2/\xi_1)^2 \right)= \left(\xi_2/\xi_1-a \right)\left(\xi_2/\xi_1-b \right) \, .
\end{equation}
Hence there is a remaining conifold singularity over
\begin{equation}\label{oneconifoldleft}
\xi_2/\xi_1=a=b\, ,\quad z=a^2\, ,\quad \zeta_2/\zeta_1=1 \, . 
\end{equation}
This is in the fibre over $C_{{\rm bifund}.}$. 

It is thus necessary to carry out a further small morphism to achieve a
small resolution of (\ref{oneconifoldleft}). This completes the process
of constructing a resolution of $X'_4$ given by
(\ref{eq:A4-resolved-eq}). As this is a small resolution of $X'_4$  
(and hence crepant), and the fibre over any points in the local patch 
of $B_3$ is always of dimension one; we obtain a fourfold satisfying 
condition (e).

Clearly this is not the only $\tilde{X}_4$ satisfying the condition (e).
The resolution constructed above consists of two steps, and at least 
there are two different ways to do the conifold resolution in the second step.
One may further speculate that, because the choice of the new ambient space 
in our first step is not more than one of the six different choices in 
\cite{Esole:2011sm}, there may be 12 different resolutions of $X'_4$ 
(\ref{eq:A4-resolved-eq}) in total. 

As we will discuss in the following, however, it turns out that 
there are 24 different resolutions in total, and 
only half of the possible small resolutions of \eqref{eq:A4-resolved-eq} 
can be obtained this way. 

In order to systematically find all the possible small resolutions of the
singularities in \eqref{eq:A4-resolved-eq}, it proves useful to rewrite it as
\begin{align}\label{E6pointeqc}
\left(y+ab-(x+a)(x+b)\right)\,\,\left(y+z+(x+a)(x+b)\right)=-x\,(x+a)\,(x+b)\,(x+a+b) \, .
\end{align}
This equation has the form
\begin{equation}\label{toric5fold}
 y_+y_-= u_1u_2u_3u_4\, 
\end{equation}
if we identify
\begin{align}
y_+& = y+ab-(x+a)(x+b) \nn\\
y_-& = -\left(y+z+(x+a)(x+b)\right)  \nn\\
u_1 &= x+a  \nn\\
u_2 &= x+b \nn\\
u_3 &= -\left(x+a+b\right)  \nn\\
u_4 &= -x  \, .
\end{align}
Thus, $X'_4$ given by (\ref{eq:A4-resolved-eq}) can be regarded as 
a dimension-four subvariety in an ambient space $\C^6 = \left\{ (y_{\pm},
u_{1,2,3,4})\right\}$ determined by the common zero of
(\ref{toric5fold}) and the linear relation 
\begin{equation}
 u_1 + u_2 + u_3 + u_4 = 0 \, .
 \label{eq:lin-hyper}
\end{equation}
This reformulation explicitly shows all of its singular loci in
codimension three (\ref{eq:6singl-1}--\ref{eq:6singl-6}); they are at the common 
zeros of $y_{\pm}$ and two out of the $u_{1,2,3,4}$. 
From equation (\ref{toric5fold}) it is also evident that they 
are all conifold singularities fibred over curves.

The advantage of working with the coordinates $y_\pm$ and $u_k$ is that \eqref{toric5fold} is a toric fivefold.
Its fan consists of a single non-simplicial five-dimensional cone $\Sigma$, generated by the lattice vectors
\begin{equation}\nn
 v_1=\left(\begin{tabular}{l}
            0 \\ 1 \\ 0 \\ 0 \\ 0
           \end{tabular}
 \right) \,
 v_2=\left(\begin{tabular}{l}
            0 \\ 0 \\ 1 \\ 0 \\ 0
           \end{tabular}
 \right) \,
 v_3=\left(\begin{tabular}{l}
             0 \\ 0 \\ 0 \\ 1 \\ 0
           \end{tabular}
 \right) \,
 v_4=\left(\begin{tabular}{l}
             0 \\ 0 \\ 0 \\ 0 \\ 1
           \end{tabular}
 \right) \,
\end{equation}
\begin{equation}\label{gentoric5fold1}
 v_5=\left(\begin{tabular}{l}
            1 \\ 1 \\ 0 \\  0 \\ 0 \\
           \end{tabular}
 \right) \,
 v_6=\left(\begin{tabular}{l}
            1 \\ 0 \\ 1 \\ 0 \\ 0
           \end{tabular}
 \right) \,
 v_7=\left(\begin{tabular}{l}
             1 \\ 0 \\ 0 \\ 1 \\ 0
           \end{tabular}
 \right) \,
 v_8=\left(\begin{tabular}{l}
             1 \\ 0 \\ 0 \\ 0 \\ 1
           \end{tabular}
 \right) \, .
\end{equation}
Associating homogeneous coordinates to the lattice vectors, 
we can construct the invariant monomials\footnote{These monomials
correspond to 
$\tilde{\nu}_{y+} = (-1, 1,1,1,1)$, $\tilde{\nu}_{y-} = (1,0,0,0,0)$, 
$\tilde{\nu}_{u1} = (0,1,0,0,0)$, $\tilde{\nu}_{u2} = (0,0,1,0,0)$, 
$\tilde{\nu}_{u3} = (0,0,0,1,0)$ and $\tilde{\nu}_{u4} = (0,0,0,0,1)$. 
$z_{1,\cdots,8}$ are the homogeneous coordinates corresponding to 
$v_{1,\cdots,8}$.}
\begin{align}
y_+=z_1z_2z_3z_4  \hspace{1cm} y_-=z_5z_6z_7z_8 \\
u_1=z_1z_5  \hspace{.5cm} u_2=z_2z_6  \hspace{.5cm} u_3=z_3z_7  \hspace{.5cm} u_4=z_4z_8  
\end{align}
which satisfy \eqref{toric5fold}, proving that $X'_4$ given by 
(\ref{eq:A4-resolved-eq}) can be regarded as a hyperplane of a toric 
fivefold indeed.

\begin{figure}[!h]
\begin{center}
  \includegraphics[width=5cm]{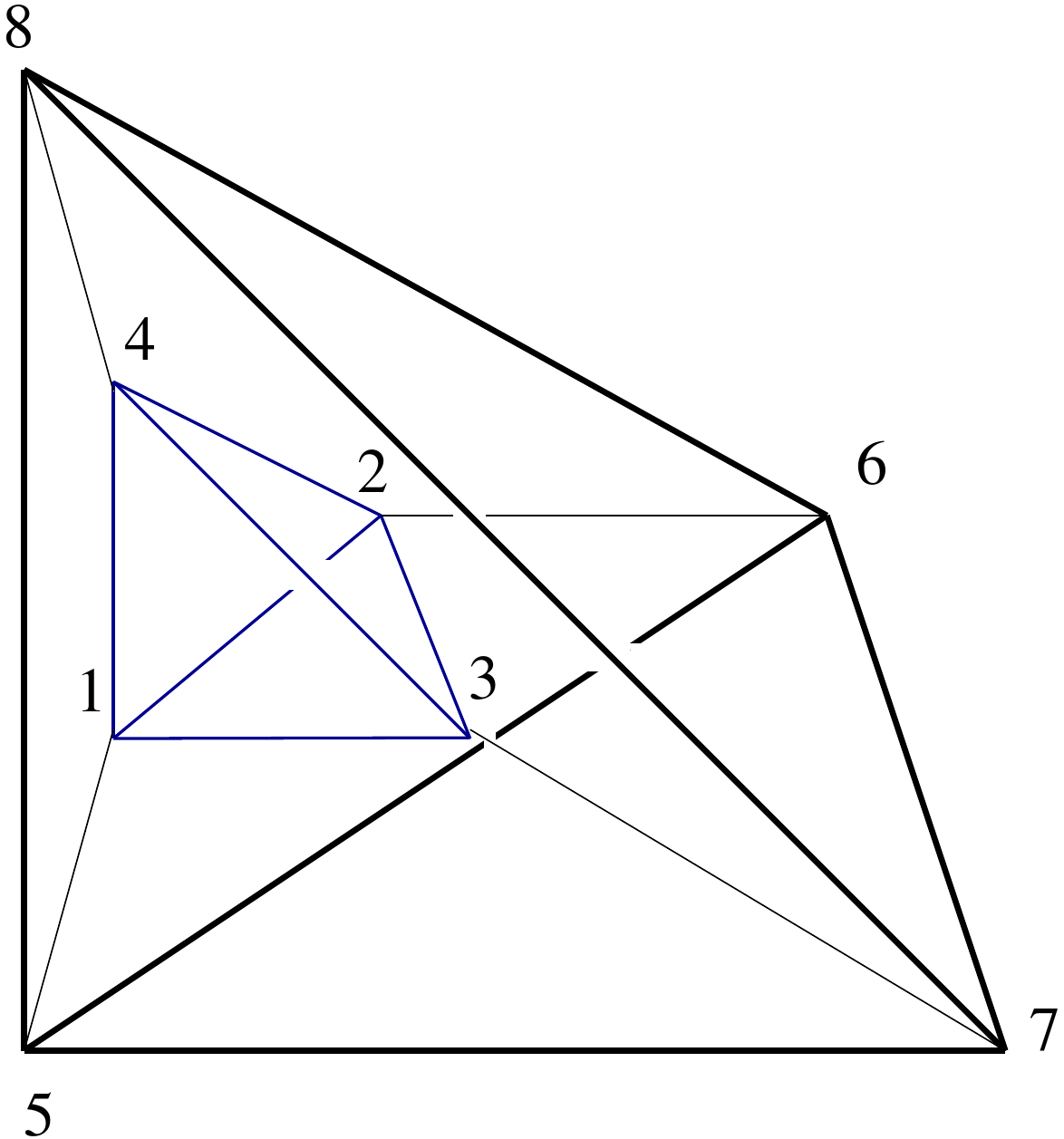} 
\caption{The Schlegel diagram of the tetrahedral prism $\Delta$. This
 picture can be obtained by realizing that the eight vectors $v_i$
 describe two tetrahedrons which are separated in an extra direction. 
 The projection used here identifies this extra direction
 with the radial direction in $\mathbb{R}^3$, i.e. we find a tetrahedron
 sitting inside a tetrahedron. We have labelled the vertices by their
 relation to the toric divisors.   
\label{4tetra}}
\end{center} 
\end{figure}

The five-dimensional cone $\Sigma$ generated by $v_i$'s has 
a four-dimensional tetrahedral prism $\Delta$ as its base. 
We have drawn a visualization of $\Delta$ (called its {\it Schlegel diagram})
in fig. \ref{4tetra}. One can easily rediscover the singularities 
of \eqref{toric5fold} in $\Delta$. Any non-simplicial
two-dimensional face of $\Delta$ corresponds to a conifold singularity 
fibred over a surface. The linear hypersurface equation $u_1+u_2+u_3+u_4=0$ turns 
these into conifold singularities fibred over curves. The polytope $\Delta$ has six 
such faces, which can
be seen as the quadrangles connecting the edges of the two tetrahedrons
in fig. \ref{4tetra}, i.e. they are spanned by the $\binom{4}{2}=6$ 
combinations of generators $v_i,v_k,v_{i+4},v_{k+4}$ for any 
$i,k \in 1 \cdots 4$. These six faces correspond to the six
singularities over $y_+=y_-=u_i=u_k=0$. E.g. the lower face pointing
towards the observer corresponds to $z_1=z_3=z_5=z_7=0$, which is
translated to $y_+=y_-=u_1=u_3=0$.

Crepant resolutions of the toric fivefold \eqref{toric5fold} can be
found by triangulating $\Delta$, leaving the generators $v_i$
untouched. As the $v_i$ all lie on a four-dimensional
subspace,\footnote{The normal vector of the four-dimensional hyperplane
is $(0,1,1,1,1)$.} the canonical divisor of the associated toric variety
is trivial both before and after the resolution. 
Because $u_1+u_2+u_3+u_4$ is a section of a trivial line bundle,  
the crepant resolutions of the ambient fivefold also induce 
crepant resolutions of the linear hyperplane equation given 
by (\ref{eq:lin-hyper}). 
Triangulations of $\Delta$ can be found by hand or by resorting to
computer algebra, such as the package 
TOPCOM \cite{Rambau:TOPCOM-ICMS:2002}. One finds that $\Delta$ admits 24
triangulations, which we have listed in appendix~\ref{24resolutions}. 

The corresponding 24 resolutions can also be found using algebraic
equations. We introduce three $\P^1$s  with homogeneous coordinates 
$[\xi_1:\xi_2]$, $[\zeta_1:\zeta_2]$, $[\phi_1:\phi_2]$ and 
replace \eqref{toric5fold} by the smooth fourfold $(\tilde{X}_4)_p$:
\begin{align}\label{resolutionp(k)}
y_+\xi_1 & = u_{p(1)} \xi_2 \nn\\
y_-\zeta_1 & = u_{p(2)} \zeta_2 \nn\\
\phi_1\xi_2 & =\phi_2\xi_1u_{p(3)}  \nn\\
\phi_2\zeta_2 & =\phi_1\zeta_1u_{p(4)} \, . 
\end{align}
Here $p(k)$ denotes the $k$-th element of a permutation of the numbers
$1,2,3,4$ ($p \in \mathfrak{S}_4$). Hence we recover the fact that
we can do $4!=24$ different resolutions. 

The exceptional set of the these resolutions 
$\rho'': (\tilde{X}_4)_p \rightarrow X'_4$ described above 
at most has a curve in $\P^1 \times \P^1 \times \P^1$ as the fibre 
over any point in the centre of the blowup, 
(\ref{eq:6singl-1}-\ref{eq:6singl-6}).\footnote{In toric language, 
this can be seen as follow. The two-dimensional cones 
$\vev{v_{p(1)} v_{4+p(a)}}|_{a=2,3,4}$, 
$\vev{v_{p(b)} v_{4+p(2)}}|_{b=3,4}$ and $\vev{v_{p(3)} v_{4+p(4)}}$ 
are mapped to the non-simplicial three-dimensional cones 
$\vev{v_{p(1)} v_{p(a)} v_{4+p(1)} v_{4+p(a)}}|_{a=2,3,4}$, 
$\vev{v_{p(2)} v_{p(b)} v_{4+p(2)} v_{4+p(b)}}|_{b=3,4}$ and 
$\vev{v_{p(3)} v_{p(4)} v_{4+p(3)} v_{4+p(4)}}$, respectively. 
The two-dimensional cones correspond to surfaces after imposing the
relation (\ref{eq:lin-hyper}). This map between toric fans induces 
a toric morphism between the corresponding toric varieties.
Since the bulk of the exceptional loci is captured by the algebraic tori
corresponding to those two-dimensional simplicial cones, there is no 
chance that some irreducible component of surfaces in $(\tilde{X}_4)_p$ 
is mapped to a point in $X'_4$.} 
Thus, $\rho'': (\tilde{X}_4)_p \rightarrow X_4'$ is a small resolution. Over all points
of the base $B_3$, the elliptic fibre gains at most (reducible) curves, so that
the fibration $\pi_X: (\tilde{X}_4)_p \rightarrow B_3$ remains flat.

\subsubsection{Fibre structure}\label{E6lHfibretrans}

{\bf Family of singular fibres over $S_{\rm GUT}$} 
\\
\\
The exceptional divisors of the blow-up of the $A_4$ singularity 
(which are divisors in $X'_4$ \eqref{eq:A4-resolved-eq} and are in the
fibre of $S_{\rm GUT}$)  
can be expressed as the following intersections in the ambient space $\C^5$:  
\begin{align}\label{excepdiva4}
D_{1+}:&\quad {\rm div}(y) \cdot {\rm div}(z) \nn\\
D_{1-}:&\quad {\rm div}(y+ab) \cdot {\rm div}(z) \nn\\
D_{2+}:&\quad {\rm div}(y) \cdot {\rm div}(x) \nn\\
D_{2-}:&\quad {\rm div}(y+ab+z) \cdot {\rm div}(x) \, .
\end{align}

To discuss the fibre structure, let us start by examining a particular small resolution given by the permutation $p(1,2,3,4)=(4,3,1,2)$:
\begin{align}\label{oneresoflHE6}
\left(y+ab-(x+a)(x+b)\right)\xi_1 & = - x \xi_2 \nn\\
\left(y+z+(x+a)(x+b)\right)\zeta_1 & = (x+a+b)\zeta_2 \nn\\
\phi_1\xi_2 & =\phi_2\xi_1(x+a) \nn\\
\phi_2\zeta_2 & =\phi_1\zeta_1(x+b)\, . 
\end{align}
We will come back to discuss the geometry of singular fibres 
in $(\tilde{X}_4)_p$ for other $p \in \mathfrak{S}_4$ at the end of 
this section \ref{E6lHfibretrans}.

For the resolution $(\tilde{X}_4)_p$ with the permutation $p$ specified 
above, the algebraic two-parameter family of irreducible components 
of the singular fibre (over $S_{\rm GUT} \subset B_3$) become 
\begin{align}\label{excepdiva4sr}
D_{1+}:\hspace{2cm} & D_{1-}:    \nn \\    
y=z=0   \hspace{2cm} & y+ab=z=0  \nn\\
(x+a+b)\xi_1=\xi_2 \hspace{2cm} & x\zeta_1=\zeta_2 \nn\\
(x+a+b)\zeta_2=(x+a)(x+b)\zeta_1\hspace{2cm} &   (x+a)(x+b)\xi_1=x\xi_2  \nn\\
\phi_1(x+a+b)=\phi_2(x+a)\hspace{2cm} & \phi_1\xi_2=\xi_1\phi_2(x+a)\nn\\
\phi_2\zeta_2=\phi_1\zeta_1(x+b)\hspace{2cm} &  \phi_2x=\phi_1(x+b)\nn\\  
&\nn\\
D_{2+}:\hspace{2cm} & D_{2-}: \nn\\
y=x=0 \hspace{2cm} & y+ab+z=x=0 \nn\\
(z+ab)\zeta_1=(a+b)\zeta_2 \hspace{2cm} & \zeta_2=0 \nn\\
\phi_1\xi_2=\xi_1\phi_2 a \hspace{2cm} & \xi_1=0 \nn\\
\phi_2\zeta_2=\phi_1\zeta_1 b \hspace{2cm} & \phi_1=0 \, .
\end{align}
Those families are defined as subvarieties in the ambient space 
$\C^5 \times \P^1 \times \P^1 \times \P^1$; five independent equations 
for $D_{2\pm}$ leave dimension-three subvarieties of $\tilde{X}_4$; the
six equations for $D_{1\pm}$ are not independent, and they are still 
of dimension-three.

We now turn to the study of the elliptic fibre over the codimension-two
loci ($\approx$ matter curves) and the codimension-three point of $E_6$
type. We have relegated some of the details to appendix \ref{app:lHE6}.
\\
\\
{\bf Families of singular fibres over matter curves}
\\
\\
Let us first investigate the fibre structure over the matter curve 
$\Sigma_{10}: \beta_5|_{S_{\rm GUT}}=0$. As $\beta_5=ab$, this
curve splits up into two irreducible components. If we focus on the
branch $a=0$ we find that the divisor $D_{1+}$ splits into two
four-cycles $S_\alpha$ and $S_\beta$, $D_{1-}$ splits into $S_\beta$ and
$S_\gamma$, $D_{2+}$ splits into $S_\gamma$ and $S_\epsilon$ and
$D_{2-}$ becomes $S_\delta$; see the appendix \ref{app:lHE6} 
for the definition of $S_{\alpha, \beta, \gamma, \delta, \epsilon}$. 
Together with the fibre component at infinity, the fibre components
originating from these algebraic surfaces form an $I_1^*$ fibre over
a generic point in the $a=0$ branch of the matter curve $\Sigma_{10}$.  
The curves $S_\beta$ and $S_\gamma$ appear with multiplicity two as
expected. A similar result is obtained over the branch $b=0$. 

We next turn to the fibre over the matter curve 
$\Sigma_5: \beta_4|_{S_{\rm GUT}}=0$. Here we denote the intersections of the
exceptional divisors $D_{1+}$, $D_{1-}$ and $D_{2-}$ with 
${\rm div}(\tilde{\pi}^*_X(\beta_4))$ in $(\tilde{X}_4)_p$ by
$S_\eta$, $S_\kappa$ and $S_\lambda$, respectively. 
The intersection of the divisor $D_{2+}$ with $\{a+b=0 \}$ splits up into 
the two irreducible surface $S_{\iota}$ and $S_\vartheta$. See appendix
\ref{app:lHE6} for the definitions. Hence each fibre component appears
with multiplicity one. The fibre components make up a fibre of type $I_6$.

There is one more one-parameter algebraic family of fibre curves 
in $\tilde{X}_4$ associated with the singular locus $C_{{\rm bifund}.}$ of 
$X'_4$. It is given by 
\begin{equation}\label{Stau}
S_{\tau}:\quad {\rm div}(x_{31}+a) \cdot {\rm div}(x_{31}+b) \cdot {\rm div}(y_{31}+ab) \cdot {\rm div}(y_{31}+z_{31})
\cdot {\rm div}(\xi_2) \cdot {\rm div}(\zeta_2) 
\end{equation}
in the ambient space $\C^5 \times \P^1 \times \P^1 \times \P^1$ that covers 
the ${\cal U}_{31}$ patch. It is a surface and is regarded as a
$\P^1 = \left\{ [\phi_1 : \phi_2] \right\}$-fibration over a curve 
parametrized by $a = b$. Obviously it is projected down 
to the curve $C_{{\rm bifund}.}$ in $X'_4$, and is
further mapped to a curve 
$(z,a,b) = (z_{31}x_{31},a,b) = (-a^3,a,a) \subset B_3$.

\begin{figure}[!h]
\begin{center}
\scalebox{.5}{ \input{E6ptlH.pspdftex} }
\caption{The geometry of irreducible components in the singular fibre 
over the codimension-three point of $E_6$ type.\label{fig:E6pt}}
\end{center} 
\end{figure}
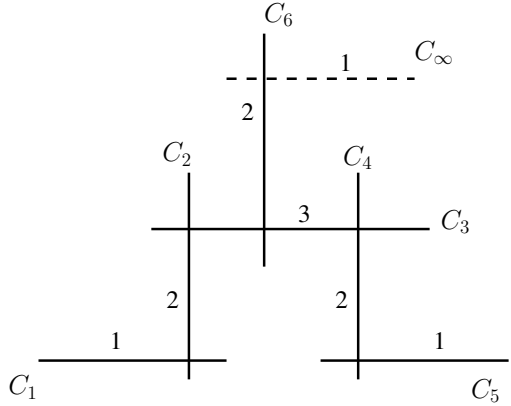

The 1-parameter family of curves $S_\tau$ in $(\tilde{X}_4)_p$ is
associated with a transverse intersection of a pair of 7-branes
(discriminant locus $\{\Delta =0\}$) in $B_3$ in the Weierstrass model 
$X_4^{\rm Weierstrass}$. The $I_1$ type fibre over such single D7-branes
has nothing to do with codimension-two singularity (or with non-Abelian
gauge group with 16 supercharges) and no exceptional divisor is
introduced in the crepant resolution $X'_4$ after the codimension-two
$A_4$ singularity in $X_4^{\rm Weierstrass}$ is resolved. Over the D7-D7
transverse intersection in $B_3$, however, two $I_1$-type fibres
collide, and a conifold singularity is formed. After this singularity in
resolved, an algebraic surface appears in $\tilde{X}_4$
\cite{Braun:2011zm}. Since this D7-D7 intersection is away from the
non-Abelian 7-branes at $z=0$ in $B_3$ for generic $(a,b)$, the
one-parameter family $S_\tau$ of the singular fibre curve component
stands alone, and it is not obtained as a limit of the two parameter 
families $D_{1 \pm}$ and $D_{2\pm}$.
\\
\\
{\bf The singular fibre over the codimension-three locus of $E_6$
type} 
\\
\\
Finally, let us study the singular fibre geometry over the 
codimension-three point of $E_6$ type, characterized by 
$(z,a,b)=(0,0,0) \in B_3$. The fibre geometry at this point 
can be studied by taking a limit from any one of the matter curves 
approaching this point, but also from a generic point on $S_{\rm GUT}$.
Given the discussion in section \ref{ssect:fibreSO(10)}, the equivalence
between these approaches is ensured by the commutativity of the Chow
ring. The most convenient approach is given by starting again from 
the singular fibre over $S_{\rm GUT}$ \eqref{excepdiva4sr} and 
intersecting them with the divisors $\{a=0 \}$ and $\{b=0 \}$ yielding
\begin{align}
 D_{1+}\,&\cdot
{\rm div}(\tilde{\pi}_X^*(a)) \cdot {\rm div}(\tilde{\pi}_X^*(b))=\, C_3 + C_4 + C_6 \nn\\
 D_{1-}\,&\cdot
{\rm div}(\tilde{\pi}_X^*(a)) \cdot {\rm div}(\tilde{\pi}_X^*(b))=\, C_2 + C_3 + C_6 \nn\\
 D_{2+}\,&\cdot
{\rm div}(\tilde{\pi}_X^*(a)) \cdot {\rm div}(\tilde{\pi}_X^*(b))=\, C_2 + C_3 + C_4 + C_5 \nn\\
 D_{2-}\,&\cdot {\rm div}(\tilde{\pi}_X^*(a)) \cdot {\rm div}(\tilde{\pi}_X^*(b))=\, C_1
\end{align}
in terms of the irreducible curves 
\begin{align}
C_1:\quad & 
   {\rm div}(y+z) \cdot {\rm div}(x) \cdot {\rm div}(\zeta_2) 
   \cdot {\rm div}(\xi_1) \cdot {\rm div}(\phi_1)  & \nn\\
C_2:\quad & 
   {\rm div}(y) \cdot {\rm div}(x) \cdot {\rm div}(z) \cdot
   {\rm div}(\phi_1) \cdot {\rm div}(\zeta_2) & \nn\\
C_3:\quad & 
   {\rm div}(y) \cdot {\rm div}(x) \cdot {\rm div}(z) \cdot
   {\rm div}(\xi_2) \cdot {\rm div}(\zeta_2) & \nn\\
C_4:\quad &  
   {\rm div}(y) \cdot {\rm div}(x) \cdot {\rm div}(z) \cdot
   {\rm div}(\xi_2) \cdot {\rm div}(\phi_2) & \nn\\
C_5:\quad & 
   {\rm div}(y) \cdot {\rm div}(x) \cdot {\rm div}(\zeta_1) 
   \cdot {\rm div}(\phi_2) \cdot {\rm div}(\xi_2) \nn\\
C_6:\quad &  
 {\rm div}(z) \cdot {\rm div}(y) \cdot
 {\rm div}(x\xi_1-\xi_2) \cdot {\rm div}(\phi_1-\phi_2) \cdot
 {\rm div}(x\zeta_1-\zeta_2)  
\end{align}
in $\C^5|_{(a,b)=(0,0)} \times \P^1 \times \P^1 \times \P^1$. We have dropped
the factors ${\rm div}(\tilde{\pi}_X^*(a)) \cdot {\rm div}(\tilde{\pi}_X^*(b))$ which are common
to all of the above.
The curves $C_{1,\cdots,6}$ appear with the multiplicity assigned for 
the roots of $E_6$, and share points with one another in the way shown 
in fig. \ref{fig:E6pt}, which looks the same as the IV$^*$ type fibre 
in Kodaira's classification.\footnote{There exists a further fibre
component over every point of the base, $C_{\infty}$, which meets the
zero section of the elliptic fibration.  
This component forms another two parameter family $D_{\infty}$; 
just like $D_\infty$ in the $SO(10)$ models in the previous section, 
it is not seen in the ${\cal U}_{31}$ patch, and this is why it has not 
appeared in the discussion so far in this section. $D_\infty$ is
characterized by $z_1 = 0$ (from which $x_1 = y_1 (y_1 + ab)$ follows)
in the ${\cal U}_1$ patch.
}
\\
\\
{\bf Twenty-three resolutions more} 
\\
\\
Before concluding this section, we would like to comment on the
dependence of our findings on which permutation $p(k)$ is used. One of
the results of \cite{Esole:2011sm} was that the form of the fibre over
the $E_6$ Yukawa point is not the same for all the different small resolutions. 
It is known that any small resolutions of a given singularity have the
same Hodge numbers \cite{1997alg.geom.10020B}. This is not enough, however, to
ensure that the fibre geometry are the same for all the small
resolutions; the study of \cite{Esole:2011sm} shows clearly that such a
phenomenon indeed happens for $SU(5)$ models with generic complex structure. 
In the case of linear Higgs vev, we have so far studied the singular fibre 
geometry only for one of the twenty-four small resolutions. Thus, 
we have to work on twenty-three other resolutions separately.

It is not difficult to see, however, that the singular fibre geometry 
over the $E_6$-type point remains the same for all twenty-four small 
resolutions. An important point is that we can specialize the problem in
the $a = b= 0$ locus first; setting $a=b=0$, $u_{1,2} = x$ and 
$u_{3,4} = -x$, which greatly simplifies the problem.
The fibre geometry for the resolution with $p(1,2,3,4) = (4,3,1,2)$
becomes \eqref{resolutionp(k)} becomes
\begin{align}
y_+\xi_1 & = - x  \xi_2 \nn\\
y_-\zeta_1 & = - x \zeta_2 \nn\\
\phi_1\xi_2 & =\phi_2\xi_1 x  \nn\\
\phi_2\zeta_2 & =\phi_1\zeta_1 x \, .
\end{align}
For other $p \in \mathfrak{S}_4$, the minus signs on the right hand side 
appear in other places. One can construct isomorphisms
among these fibre geometries for different $p \in \mathfrak{S}_4$'s, 
by choosing one of the eight options:
\begin{equation}
 ([\xi_1: \xi_2], [\zeta_1: \zeta_2], [\phi_1 : \phi_2])
\mapsto ([ \pm \xi_1: \xi_2], [ \pm \zeta_1: \zeta_2], [ \pm \phi_1 : \phi_2]).
\end{equation}
Therefore, in the case of linear Higgs vev, the singular fibre geometry 
over the $E_6$ type point remains the same for all the twenty-four
different resolutions $\tilde{X}_4$ satisfying condition (e).

\subsection{The codimension-three locus of $E_7$-type for $SO(10)$ models}
\label{ssect:resE7}

We have already seen that the singular fibre in the resolved geometry 
$\tilde{X}_4$ under the condition (e) follow the empirical rule we proposed 
in section \ref{ssect:SO10-discussion}, when the complex structure of 
$X_4^{\rm Weierstrass}$ is chosen so that the Higgs vev is linear in the local 
coordinates around codimension-three points of $A_4$ type and $E_6$-type 
in $SU(5)$ models. At the end of this article, we briefly mention 
how to carry out similar experiment for the codimension-three points of 
$E_7$ type in $SO(10)$ models. This does not involve extra complications, 
but we will leave the explicit analysis as an open problem. 

As a local geometry of $X_4^{\rm Weierstrass}$ of $SO(10)$ models, 
we can think of $X_4^{({\rm Weierstrass})}$ given by (\ref{SO10X4}) with 
\begin{equation}\label{lHparamso10E7}
\beta_3=a+b\,, \quad \beta_4=ab \,,\quad \beta_0=\beta'_5=0\, ,\quad   \beta_2=1
\end{equation}
in order to realize a linear Higgs vev in the vicinity of an $E_7$-type 
point.\footnote{See \cite{Hayashi:2009ge} and the comments at the beginning 
of section \ref{ssect:resE6}. }
This special choice of complex structure worsens the singularity 
of $X'_4$ for $SO(10)$ models in patch ${\cal U}_{313}$, \eqref{U313}, 
which for a linear Higgs vev reads
\begin{equation}\label{U313simple}
y_{313}(y_{313}+a+b)=x_{313}(x_{313}z_{313}+ab+z_{313})\, .
\end{equation}
Besides the obvious conifold singularity, which we have already encountered 
in section \ref{ssect:SO10-resolution}, there is now also a singularity\footnote{
This locus of singularity also appears in another chart of $X'_4$.
In the ${\cal U}_{311z}$ patch of the ambient space, 
$X'_4$---\eqref{U311zchi}---is now given by 
\begin{equation}
  (y_{z}+a)(y_z+b)=\chi_z (z_z+1)\, ,
\end{equation}
showing that it is singular at 
\begin{equation}
\chi_z=0\,,\quad z=-1\,,\quad -y_z=a=b\,.
\end{equation}
}
 along  
\begin{equation}
 a=b=-y_{313}\,, \quad x_{313}=-1\,, \quad z_{313}=b^2 \, ,
\end{equation}
which is in the fibre of 
$\left\{ (z,a,b)  = (z_{313}^2 x_{313},a,b) = (-b^4,b,b) |{}^\forall b \right\}
\subset B_3$, precisely 
at the place we expect from the locus of intersection of factorized 
spectral surface $(\xi -a)(\xi - b)=0$.

The conifold singularity over the $SO(10)$-${\bf 10}$-representation 
matter curve and this new singular locus of $X'_4$ can be treated 
in a symmetric way, once we rewrite the equation (\ref{U313simple}) 
as
\begin{equation}
 \left(y_{313}+a+b(x_{313}+1)\right)\left(y_{313}-x_{313}b\right)=x_{313}(x_{313}+1)(z_{313}-b^2) \, 
\end{equation}
or
\begin{equation}
 \left(y_{313}+b+a(x_{313}+1)\right)\left(y_{313}-x_{313}a\right)=x_{313}(x_{313}+1)(z_{313}-a^2) \, .
\end{equation}
If we are to denote the two factors in the left-hand sides as $y_\pm$, 
then the two curves of conifold singularity correspond to 
$y_+ = y_- = x_{313} = (z_{313}-b^2) = 0$ and $y_+ = y_- = (x_{313}+1) = (z_{313}-b^2) = 0$, 
respectively.

As we have not worked out possible resolutions under the condition (e) 
for this geometry, we are not in a position to conclude that the empirical 
rule in section \ref{ssect:SO10-discussion} holds true for the case 
of $E_7$-type points in $SO(10)$ models with linear Higgs background.  
A similar analysis can be carried out for the $D_7$ type points of 
$SO(10)$ models or $E_8$-type points of $E_6$ models with linear Higgs 
background as well (see \cite{Hayashi:2008ba, Hayashi:2009ge} for the necessary information).

In between the linear Higgs background (trivial 7-brane monodromy) and 
the full $\mathfrak{S}_r$ monodromy of an $r$-fold spectral cover, one can 
think of many different choices of monodromy groups \cite{Tatar:2006dc,
Bouchard:2009bu, Heckman:2009mn, Marsano:2009gv, Hayashi:2010zp}. 
It will be interesting to carry out the resolution under 
the condition (e) and study the singular fibres for such geometries. The  
results can be used as experimental data for testing the empirical rule 
in section \ref{ssect:SO10-discussion}, and to think more about the 
possibility hinted at the end of section \ref{ssect:SO10-discussion}. 
With such an experimental approach, the empirical rule may be promoted
to a dictionary between the singular fibres of $X_n$ under the condition
(e) and symmetries unbroken at the corresponding points in the base 
(or some other physics concepts); it will even be possible to refine 
the dictionary, based on more data. 
All these things, however, are beyond the scope of this article. 

\subsection*{Acknowledgements}

We have benefited from discussions with Radu Tatar and William Walters, with 
whom we worked together at the earlier stage of this project. TW thanks the 
organizers of CERN theory institute in 2012 summer for providing such an 
opportunity. 
We thank Alexey Bondal and Satoshi Kondo for useful comments. 
A.~P.~B. likes to thank the IPMU Tokyo for kind hospitality during his visit.
The work of A.~P.~B. was supported by a JSPS postdoctoral fellowship 
under grant PE 12530 and by the STFC under grant ST/J002798/1. The work of T.~W. was supported by WPI Initiative, 
MEXT, Japan and a Grant-in-Aid for Scientific Research on Innovative Areas 2303.

\appendix

\section{Triangulations and resolutions of the toric fivefold}
\label{24resolutions}

In this appendix, we record some technical details regarding the possible small resolutions of the fourfold~\eqref{eq:A4-resolved-eq}. 
As discussed in section \ref{sectresE6lH}, this space is realized as the hypersurface $u_1+u_2+u_3+u_4=0$ in the toric fivefold
\begin{equation}
 y_+y_-= u_1u_2u_3u_4\, .
\end{equation}
The toric variety corresponding to this algebraic equation has a fan which just contains a single
five-dimensional non-simplicial cone $\Sigma$ over a polytope $\Delta$, see also fig. \ref{4tetra}. 
This cone is generated by the vectors $v_{1,2,\cdots,8}$ in \eqref{gentoric5fold1}.

\begin{table}[h!]
\centering
\begin{tabular}{l|l|l}
$\#$ & $\Sigma_i$ are generated by & refinement of \\
\hline

$1$
&$\{1,2,3,4,5\},\{2,3,4,5,6\},\{3,4,5,6,7\},\{4,5,6,7,8\}$&${\cal E}_{xw}$\\

$2$
&$\{1,2,3,4,5\},\{2,3,4,5,6\},\{3,4,5,6,8\},\{3,5,6,7,8\}$&${\cal E}_{tw}$\\

$3$
&$\{1,2,3,4,5\},\{2,3,4,5,7\},\{2,4,5,6,7\},\{4,5,6,7,8\}$&\\

$4$
&$\{1,2,3,4,5\},\{2,3,4,5,7\},\{2,4,5,7,8\},\{2,5,6,7,8\}$&\\

$5$
&$\{1,2,3,4,5\},\{2,3,4,5,8\},\{2,3,5,6,8\},\{3,5,6,7,8\}$&\\

$6$
&$\{1,2,3,4,5\},\{2,3,4,5,8\},\{2,3,5,7,8\},\{2,5,6,7,8\}$&\\

$7$
&$\{1,2,3,4,6\},\{1,3,4,5,6\},\{3,4,5,6,7\},\{4,5,6,7,8\}$&${\cal E}_{xw}$\\

$8$
&$\{1,2,3,4,6\},\{1,3,4,5,6\},\{3,4,5,6,8\},\{3,5,6,7,8\}$&${\cal E}_{tw}$\\

$9$
&$\{1,2,3,4,6\},\{1,3,4,6,7\},\{1,4,5,6,7\},\{4,5,6,7,8\}$&\\

$10$
&$\{1,2,3,4,6\},\{1,3,4,6,7\},\{1,4,6,7,8\},\{1,5,6,7,8\}$&\\

$11$
&$\{1,2,3,4,6\},\{1,3,4,6,8\},\{1,3,5,6,8\},\{3,5,6,7,8\}$&\\

$12$
&$\{1,2,3,4,6\},\{1,3,4,6,8\},\{1,3,6,7,8\},\{1,5,6,7,8\}$&\\

$13$
&$\{1,2,3,4,7\},\{1,2,4,5,7\},\{2,4,5,6,7\},\{4,5,6,7,8\}$&${\cal E}_{xt}$ \\

$14$
&$\{1,2,3,4,7\},\{1,2,4,5,7\},\{2,4,5,7,8\},\{2,5,6,7,8\}$&\\

$15$
&$\{1,2,3,4,7\},\{1,2,4,6,7\},\{1,4,5,6,7\},\{4,5,6,7,8\}$&${\cal E}_{xt}$ \\

$16$
&$\{1,2,3,4,7\},\{1,2,4,6,7\},\{1,4,6,7,8\},\{1,5,6,7,8\}$&\\

$17$
&$\{1,2,3,4,7\},\{1,2,4,7,8\},\{1,2,5,7,8\},\{2,5,6,7,8\}$&${\cal E}_{wt}$\\

$18$
&$\{1,2,3,4,7\},\{1,2,4,7,8\},\{1,2,6,7,8\},\{1,5,6,7,8\}$&${\cal E}_{wt}$\\

$19$
&$\{1,2,3,4,8\},\{1,2,3,5,8\},\{2,3,5,6,8\},\{3,5,6,7,8\}$& ${\cal E}_{tx}$\\

$20$
&$\{1,2,3,4,8\},\{1,2,3,5,8\},\{2,3,5,7,8\},\{2,5,6,7,8\}$&\\

$21$
&$\{1,2,3,4,8\},\{1,2,3,6,8\},\{1,3,5,6,8\},\{3,5,6,7,8\}$& ${\cal E}_{tx}$\\

$22$
&$\{1,2,3,4,8\},\{1,2,3,6,8\},\{1,3,6,7,8\},\{1,5,6,7,8\}$&\\

$23$
&$\{1,2,3,4,8\},\{1,2,3,7,8\},\{1,2,5,7,8\},\{2,5,6,7,8\}$& ${\cal E}_{wx}$\\

$24$
&$\{1,2,3,4,8\},\{1,2,3,7,8\},\{1,2,6,7,8\},\{1,5,6,7,8\}$& ${\cal E}_{wx}$
\end{tabular}
\caption{\label{table:24triang}The 24 triangulations of $\Delta$. Each triangulation subdivides the cone
over $\Delta$ into four simplicial cones, each of which is spanned by five vectors
$v_i$. Each row in the table above describes a single triangulation. For each
triangulation, we have collected the generators $v_i$ spanning each four five-dimensional cone
in curly brackets $\{\cdots\}$. For the sake of brevity, we merely write $i$ instead of $v_i$. 
E.g. for the first triangulation, the four different five-dimensional cones are generated by 
$\Sigma_1= \langle v_1,v_2,v_3,v_4,v_5 \rangle $, $\Sigma_2= \langle v_2,v_3,v_4,v_5,v_6  \rangle $, 
$\Sigma_3= \langle v_3,v_4,v_5,v_6,v_7  \rangle $ and $\Sigma_4= \langle v_4,v_5,v_6,v_7,v_8  \rangle $. 
The lower dimensional cones in each triangulation are faces of the $\Sigma_i$. 
Each of these triangulations corresponds to a small morphism. In the right column, we have indicated which of these
morphisms can be obtained by first applying one of the $6$ small morphisms of \cite{Esole:2011sm}, followed by 
a small resolution of the remaining singularity. } 
\end{table}

In order to turn this into a smooth toric variety, we must completely triangulate $\Delta$. Any such triangulation turns $\Delta$ into 
four four-dimensional cells $\hat{\Delta}_i$ and subdivides $\Sigma$ into the four simplicial cones $\Sigma_i$.
In table \ref{table:24triang}, we have listed the $24$ triangulations of $\Delta$ by giving the vectors $v_i$ spanning the simplicial 
cones $\Sigma_i$ over the $\hat{\Delta}_i$ resulting from the triangulation. 

Using the following identification of coordinates:
\begin{align}
y_+& = y+ab-(x+a)(x+b) = z_1z_2z_3z_4 \nn\\
y_-& = -\left(y+z+(x+a)(x+b)\right) =  z_5z_6z_7z_8 \nn\\
u_1 &= x+a =  z_1z_5 \nn\\
u_2 &= x+b =  z_2z_6 \nn\\
u_3 &= -\left(x+a+b\right) =  z_3z_7 \nn\\
u_4 &= -x =  z_4z_8 \, ,
\end{align}
and examining the intersection ring, we can map the $24$ triangulations to the $24$ resolutions that were obtained using algebraic equations. 
Given (\ref{resolutionp(k)}) with $p\in \mathfrak{S}_4$, this corresponds to the triangulation 
$\Sigma_1 = \vev{v_{1,2,3,4},v_{p(2)}}$,  
$\Sigma_2 = \vev{v_{p(1,3,4)},v_{4+p(2,4)}}$, 
$\Sigma_3 = \vev{v_{p(1,3)}, v_{4+p(2,3,4)}}$, 
$\Sigma_4 = \vev{v_{p(1)}, v_{5,6,7,8}}$ 
in table \ref{table:24triang}. 

We can arrive at $12$ of these by starting from \eqref{eq:A4-resolved-eq} and applying one of the six small morphisms ${\cal E}_{ij}$ used 
in \cite{Esole:2011sm}. These morphisms correspond to a partial triangulation of $\Delta$, so that they do not completely 
resolve the singularities of \eqref{eq:A4-resolved-eq}. It turns out that they leave a curve of conifold singularities, which can be resolved
by a further small resolution (for which there are the usual two choices). After performing 
one of the two small resolutions we obtain a smooth fourfold corresponding to one of the $24$ triangulations in table
\ref{table:24triang}. The triangulations which can be obtained this way are listed in the right column.

\section{One-parameter families of fibre components 
for a $SU(5)$ GUT model with linear higgs field}\label{app:lHE6}

In this appendix, we list expressions for the various algebraic submanifolds appearing in section \ref{E6lHfibretrans}.

The matter curve of $SU(5)$-${\bf 10}+\overline{\bf 10}$ representation 
splits into two components, $a=0$ and $b=0$ in the case of linear Higgs 
background (\ref{eq:weierSU5-E6-linear}). 
In the fibre of the $a=0$ curve in $S_{\rm GUT}$, we can construct 
multiple families of fibre components parametrized by the local coordinate 
$b$ of the curve. 
They are given by 
\begin{align}
S_\alpha:\quad & {\rm div}(z) \cdot{\rm div}(y) \cdot {\rm div}(\phi_2)
    \cdot {\rm div}(x+b) \cdot {\rm div}(\xi_2) \nn \\
S_\beta:\quad & {\rm div}(z) \cdot {\rm div}(y) \cdot 
    {\rm div}(\xi_2-\xi_1(x+b)) \cdot {\rm div}(x\zeta_1-\zeta_2) 
    \cdot {\rm div}( \phi_2 x - \phi_1 (x+b)) \nn \\
S_\gamma: \quad & {\rm div}(z) \cdot {\rm div}(y) \cdot {\rm div}(\phi_1) 
    \cdot {\rm div}(x) \cdot {\rm div}(\zeta_2) \nn \\
S_\delta: \quad & {\rm div}(y+z) \cdot {\rm div}(x) \cdot {\rm div}(\xi_1)
    \cdot {\rm div}(\zeta_2) \cdot {\rm div}(\phi_1) \nn \\
S_\epsilon: \quad & {\rm div}(y) \cdot {\rm div}(x) \cdot 
    {\rm div}(z\zeta_1-b\zeta_2) \cdot {\rm div}(\xi_2) \cdot
    {\rm div}(\phi_2\zeta_2-\phi_1\zeta_1 b) 
\end{align}
in the ambient space $\C^5 \times \P^1 \times \P^1 \times \P^1$, 
where ${\rm div}(a)$, common to all above, has been omitted.
Six equations in the eight-dimensional ambient space leave a complex two-dimensional variety, 
which sits within $(\tilde{X}_4)_p$ given by (\ref{oneresoflHE6}).

Similarly, two-dimensional subvarieties of $(\tilde{X}_4)_p$ exist in the fibre 
of the other branch of the matter curve, $b=0$.
\begin{align}
S_\alpha':\quad & {\rm div}(z) \cdot {\rm div}(y) \cdot {\rm div}(\xi_2) 
    \cdot {\rm div}(x+a) \cdot {\rm div}(\phi_2\zeta_2-\phi_1\zeta_1 x) \nn \\
S_\beta':\quad & {\rm div}(z) \cdot {\rm div}(y) \cdot 
    {\rm div}(\xi_2-\xi_1(x+a)) \cdot {\rm div}(x\zeta_1-\zeta_2) \cdot
    {\rm div}(\phi_2 - \phi_1 ) \nn \\
S_\gamma': \quad & {\rm div}(z) \cdot {\rm div}(y) \cdot {\rm div}(\zeta_2) 
    \cdot {\rm div}(x) \cdot {\rm div}(\phi_1\xi_2-\xi_1\phi_2 a) \nn \\
S_\delta': \quad & {\rm div}(y+z) \cdot {\rm div}(x) \cdot {\rm div}(\xi_1) 
    \cdot {\rm div}(\zeta_2) \cdot {\rm div}(\phi_1) \nn \\
S_\epsilon': \quad & {\rm div}(y) \cdot {\rm div}(x) \cdot 
    {\rm div}(z\zeta_1-a\zeta_2) \cdot {\rm div}(\xi_2) \cdot 
    {\rm div}(\phi_2), 
\end{align}
where we have omitted ${\rm div}(b)$ which is common to all above.

The surfaces over the ${\bf 5}$ matter curve $a+b=0$ are given by
\begin{align}
S_\eta:\quad &{\rm div}(z)\cdot{\rm div}(y)\cdot{\rm div}(x\xi_1-\xi_2)\cdot{\rm div}\left(x\zeta_2-(x+a)(x+b)\zeta_1\right)\cdot{\rm div}\left(x\phi_1
-\phi_2 (x+a)\right) 
\nn\\
S_\kappa:\quad &{\rm div}(z)\cdot{\rm div}(y+ab)\cdot{\rm div}\left(\zeta_2-\zeta_1x\right)\cdot{\rm div}\left(x\xi_2-(x+a)(x+b)\xi_1\right)
\cdot{\rm div}\left(\phi_2 x - \phi_1(x+b)\right) 
\nn\\
S_\iota: \quad &{\rm div}(y)\cdot{\rm div}(x)\cdot{\rm div}(z+ab)\cdot{\rm div}(\phi_1\xi_2-\xi_1\phi_2 a)\cdot{\rm div}(\phi_2\zeta_2
-\phi_1\zeta_1 b) 
\nn\\
S_\vartheta: \quad &{\rm div}(y)\cdot{\rm div}(x)\cdot{\rm div}(\xi_2)\cdot{\rm div}(\zeta_1)\cdot{\rm div}(\phi_2)  
\nn\\
S_\lambda: \quad &{\rm div}(y+ab+z)\cdot{\rm div}(x)\cdot{\rm div}(\zeta_2)\cdot{\rm div}(\xi_1)\cdot{\rm div}(\phi_1)  \, .
\end{align}
We have omitted the factor ${\rm div}(a+b)$, which is common to all above.\\
\\
The Yukawa coupling analysis of \cite{Marsano:2011hv} in our language corresponds
to proving that 
\begin{eqnarray}
 \left[ S_\alpha \right] \cdot {\rm div}(\tilde{\pi}_X^*(b)) + 
 \left[ S_{\vartheta} \right] \cdot {\rm div}(\tilde{\pi}_X^*(a)) - 
 \left[ S'_{\epsilon} \right] \cdot {\rm div}(\tilde{\pi}_X^*(a)) 
 & \equiv & 0 \, , 
  \label{eq:Yukawa-confirm-1} \\
 \left[ S_\alpha \right] \cdot {\rm div}(\tilde{\pi}_X^*(b)) - 
 \left[ S'_{\alpha} \right] \cdot {\rm div}(\tilde{\pi}_X^*(a)) + 
 S_{\tau} \cdot {\rm div}(\tilde{\pi}_X^*(a))
 & \equiv & 0 \, ,
\label{eq:Yukawa-confirm-2}
\end{eqnarray}
modulo 
$D_{1\pm, 2\pm} \cdot {\rm div}\left(\tilde{\pi}_X^*(a)\right) 
 \cdot  {\rm div}\left(\tilde{\pi}_X^*(b)\right)$.
Here, $[S_\alpha]$ is the equivalence class of $S_\alpha$ in the quotient space 
\begin{equation}
 {\rm Span}_{\Z} \left\{ S_{\alpha, \beta, \gamma, \delta, \epsilon}
		 \right\} / 
 {\rm Span}_{\Z} \left\{ D_{1\pm, 2\pm} \cdot
      {\rm div}(\tilde{\pi}_X^*(a)) \right\},
\end{equation}
and $[S'_\alpha] = - [S'_\epsilon]$ and $[S_\vartheta]$ are the
equivalence classes for the corresponding matter curves. The class
$S_{\tau}$ is given in \eqref{Stau}.
The relations above hold because 
\begin{align}
& S_\alpha \cdot {\rm div}(\tilde{\pi}_X^*(b)) = C_4, \quad 
 S'_\alpha \cdot {\rm div}(\tilde{\pi}_X^*(a)) = C_3 + C_4, \quad 
 S'_{\epsilon} \cdot {\rm div}(\tilde{\pi}_X^*(a)) = C_4 + C_5, \nn \\
& S_{\vartheta} \cdot {\rm div}(\tilde{\pi}_X^*(a)) = C_5, \quad 
 S_{\tau} \cdot {\rm div}(\tilde{\pi}_X^*(a)) = C_3. 
\end{align}
The generation of Yukawa couplings around codimension-three singularities 
can be understood at the intuitive level in the language of M2-branes 
wrapped on 2-cycles in the ALE-fibre of A-D-E type 
\cite{Tatar:2006dc, Hayashi:2008ba}, and the field theory local models 
provide a formulation of quantitative calculation of F-term Yukawa
couplings \cite{Katz:1996xe, Donagi:2008ca, Beasley:2008dc,
Hayashi:2009ge, Hayashi:2009bt}. Therefore, the confirmation of the 
topological relations among 2-cycles (\ref{eq:Yukawa-confirm-1},
\ref{eq:Yukawa-confirm-2}) does not introduce new ingredients 
to the calculation of Yukawa couplings. 
Because of localization of F-term Yukawa couplings in supersymmetric 
compactifications \cite{Cecotti:2009zf} (see also \cite{Conlon:2009qq}), 
however, we feel comfortable to repeat, in the language of algebraic geometry, 
the confirmation of the topological relations in the deformed A-D-E 
fibration \cite{Tatar:2006dc}. This comfortable confirmation can be regarded as 
a non-negative evidnece for the choice of the condition (e) \cite{Marsano:2011hv}.

\bibliographystyle{utphys}
\bibliography{res}

\end{document}

%% file: SO10bu1.pspdftex
\begin{picture}(0,0)%
\includegraphics{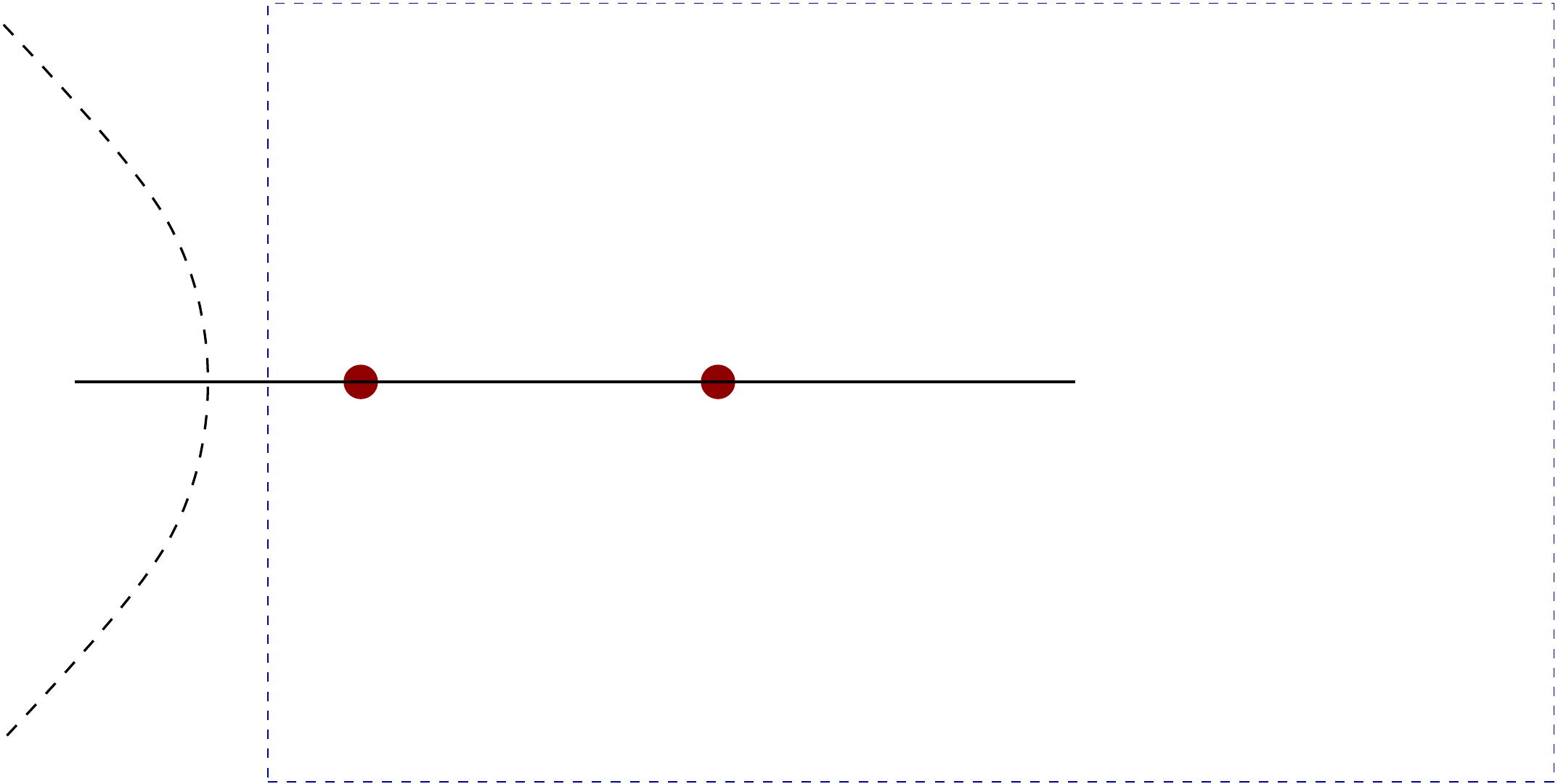}%
\end{picture}%
\setlength{\unitlength}{4144sp}%
\begingroup\makeatletter\ifx\SetFigFont\undefined%
\gdef\SetFigFont#1#2#3#4#5{%
  \reset@font\fontsize{#1}{#2pt}%
  \fontfamily{#3}\fontseries{#4}\fontshape{#5}%
  \selectfont}%
\fi\endgroup%
\begin{picture}(9799,4929)(1779,-5743)
\put(8551,-3661){\makebox(0,0)[lb]{\smash{{\SetFigFont{20}{24.0}{\rmdefault}{\mddefault}{\updefault}{\color[rgb]{0,0,0}$D_B: \{z_3=y_3=0\}$}%
}}}}
\put(2386,-5326){\makebox(0,0)[lb]{\smash{{\SetFigFont{20}{24.0}{\rmdefault}{\mddefault}{\updefault}{\color[rgb]{0,0,0}$D_\infty$}%
}}}}
\put(9721,-1276){\makebox(0,0)[lb]{\smash{{\SetFigFont{20}{24.0}{\rmdefault}{\mddefault}{\updefault}{\color[rgb]{0,0,.56}${\cal U}_3$}%
}}}}
\put(6436,-3076){\makebox(0,0)[lb]{\smash{{\SetFigFont{20}{24.0}{\rmdefault}{\mddefault}{\updefault}{\color[rgb]{0,0,0}$x_3=0$}%
}}}}
\put(4231,-3076){\makebox(0,0)[lb]{\smash{{\SetFigFont{20}{24.0}{\rmdefault}{\mddefault}{\updefault}{\color[rgb]{0,0,0}$x_3=-\beta_4$}%
}}}}
\end{picture}%

%% file: SO10bu4.pspdftex
\begin{picture}(0,0)%
\includegraphics{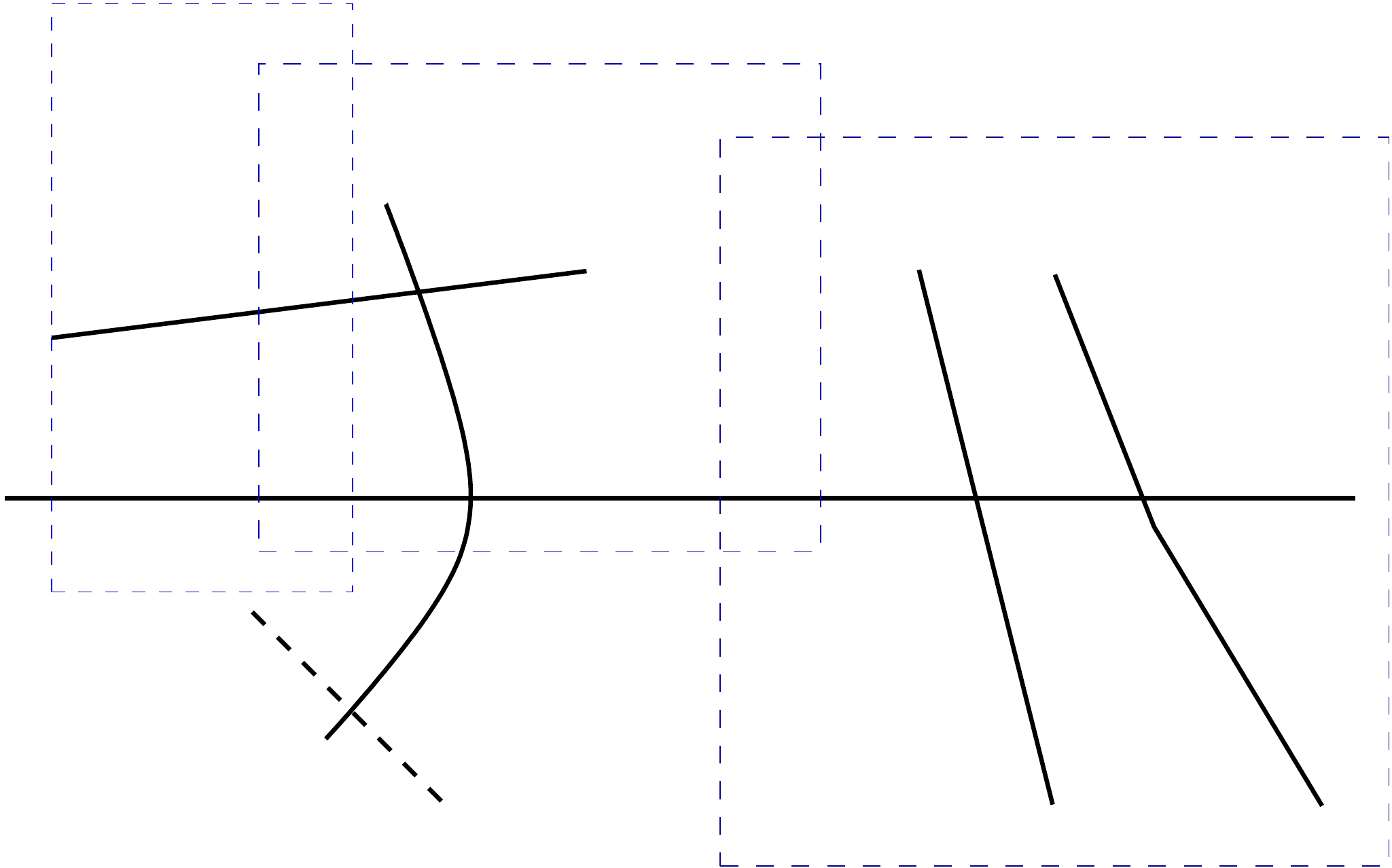}%
\end{picture}%
\setlength{\unitlength}{4144sp}%
\begingroup\makeatletter\ifx\SetFigFont\undefined%
\gdef\SetFigFont#1#2#3#4#5{%
  \reset@font\fontsize{#1}{#2pt}%
  \fontfamily{#3}\fontseries{#4}\fontshape{#5}%
  \selectfont}%
\fi\endgroup%
\begin{picture}(9360,5829)(-347,-5428)
\put(2926,-2536){\makebox(0,0)[lb]{\smash{{\SetFigFont{20}{24.0}{\familydefault}{\mddefault}{\updefault}{\color[rgb]{0,0,0}$D_B$}%
}}}}
\put(3376,-1186){\makebox(0,0)[lb]{\smash{{\SetFigFont{20}{24.0}{\familydefault}{\mddefault}{\updefault}{\color[rgb]{0,0,0}$D_A$}%
}}}}
\put(6526,-1186){\makebox(0,0)[lb]{\smash{{\SetFigFont{20}{24.0}{\familydefault}{\mddefault}{\updefault}{\color[rgb]{0,0,0}$D_-$}%
}}}}
\put(8101,-2761){\makebox(0,0)[lb]{\smash{{\SetFigFont{20}{24.0}{\familydefault}{\mddefault}{\updefault}{\color[rgb]{0,0,0}$D_C$}%
}}}}
\put(226,-61){\makebox(0,0)[lb]{\smash{{\SetFigFont{20}{24.0}{\familydefault}{\mddefault}{\updefault}{\color[rgb]{0,0,.56}${\cal U}_{311z}$}%
}}}}
\put(2476,-511){\makebox(0,0)[lb]{\smash{{\SetFigFont{20}{24.0}{\familydefault}{\mddefault}{\updefault}{\color[rgb]{0,0,.56}${\cal U}_{311\chi}$}%
}}}}
\put(7651,-961){\makebox(0,0)[lb]{\smash{{\SetFigFont{20}{24.0}{\familydefault}{\mddefault}{\updefault}{\color[rgb]{0,0,.56}${\cal U}_{313}$}%
}}}}
\put(2791,-5011){\makebox(0,0)[lb]{\smash{{\SetFigFont{20}{24.0}{\familydefault}{\mddefault}{\updefault}{\color[rgb]{0,0,0}$D_\infty$}%
}}}}
\put(5536,-1186){\makebox(0,0)[lb]{\smash{{\SetFigFont{20}{24.0}{\familydefault}{\mddefault}{\updefault}{\color[rgb]{0,0,0}$D_+$}%
}}}}
\end{picture}%

%% file: curves16matter.pspdftex
\begin{picture}(0,0)%
\includegraphics{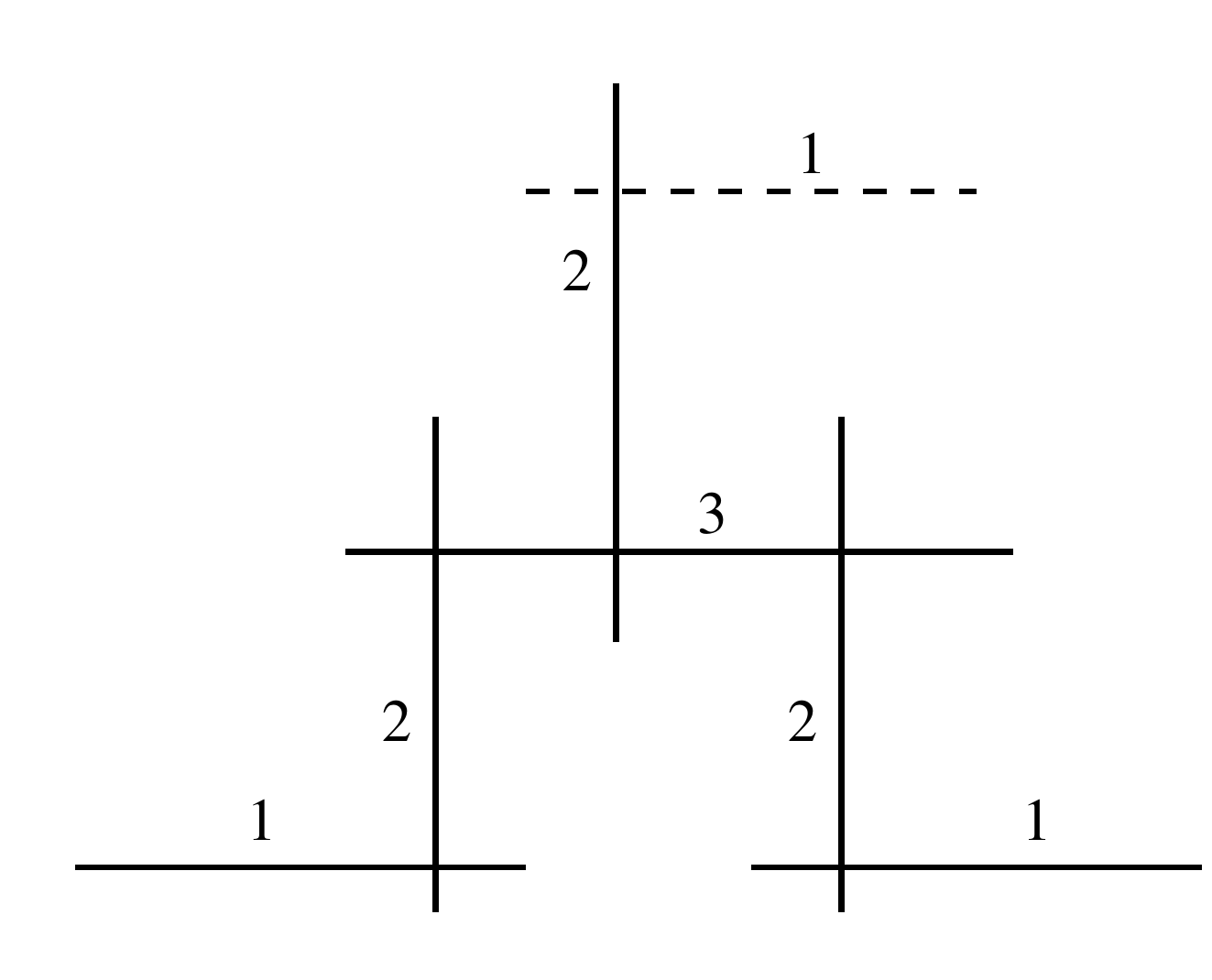}%
\end{picture}%
\setlength{\unitlength}{4144sp}%
\begingroup\makeatletter\ifx\SetFigFont\undefined%
\gdef\SetFigFont#1#2#3#4#5{%
  \reset@font\fontsize{#1}{#2pt}%
  \fontfamily{#3}\fontseries{#4}\fontshape{#5}%
  \selectfont}%
\fi\endgroup%
\begin{picture}(6033,4876)(2776,-1745)
\put(8326,-1636){\makebox(0,0)[lb]{\smash{{\SetFigFont{20}{24.0}{\rmdefault}{\mddefault}{\updefault}{\color[rgb]{0,0,0}$S_5$}%
}}}}
\put(7651,2414){\makebox(0,0)[lb]{\smash{{\SetFigFont{20}{24.0}{\rmdefault}{\mddefault}{\updefault}{\color[rgb]{0,0,0}$S_\infty$}%
}}}}
\put(5851,2864){\makebox(0,0)[lb]{\smash{{\SetFigFont{20}{24.0}{\rmdefault}{\mddefault}{\updefault}{\color[rgb]{0,0,0}$S_6$}%
}}}}
\put(7966,389){\makebox(0,0)[lb]{\smash{{\SetFigFont{20}{24.0}{\rmdefault}{\mddefault}{\updefault}{\color[rgb]{0,0,0}$S_3$}%
}}}}
\put(2791,-1591){\makebox(0,0)[lb]{\smash{{\SetFigFont{20}{24.0}{\rmdefault}{\mddefault}{\updefault}{\color[rgb]{0,0,0}$S_1$}%
}}}}
\put(4636,1199){\makebox(0,0)[lb]{\smash{{\SetFigFont{20}{24.0}{\rmdefault}{\mddefault}{\updefault}{\color[rgb]{0,0,0}$S_2$}%
}}}}
\put(6796,1154){\makebox(0,0)[lb]{\smash{{\SetFigFont{20}{24.0}{\rmdefault}{\mddefault}{\updefault}{\color[rgb]{0,0,0}$S_4$}%
}}}}
\end{picture}%

%% file: curves10matter.pspdftex
\begin{picture}(0,0)%
\includegraphics{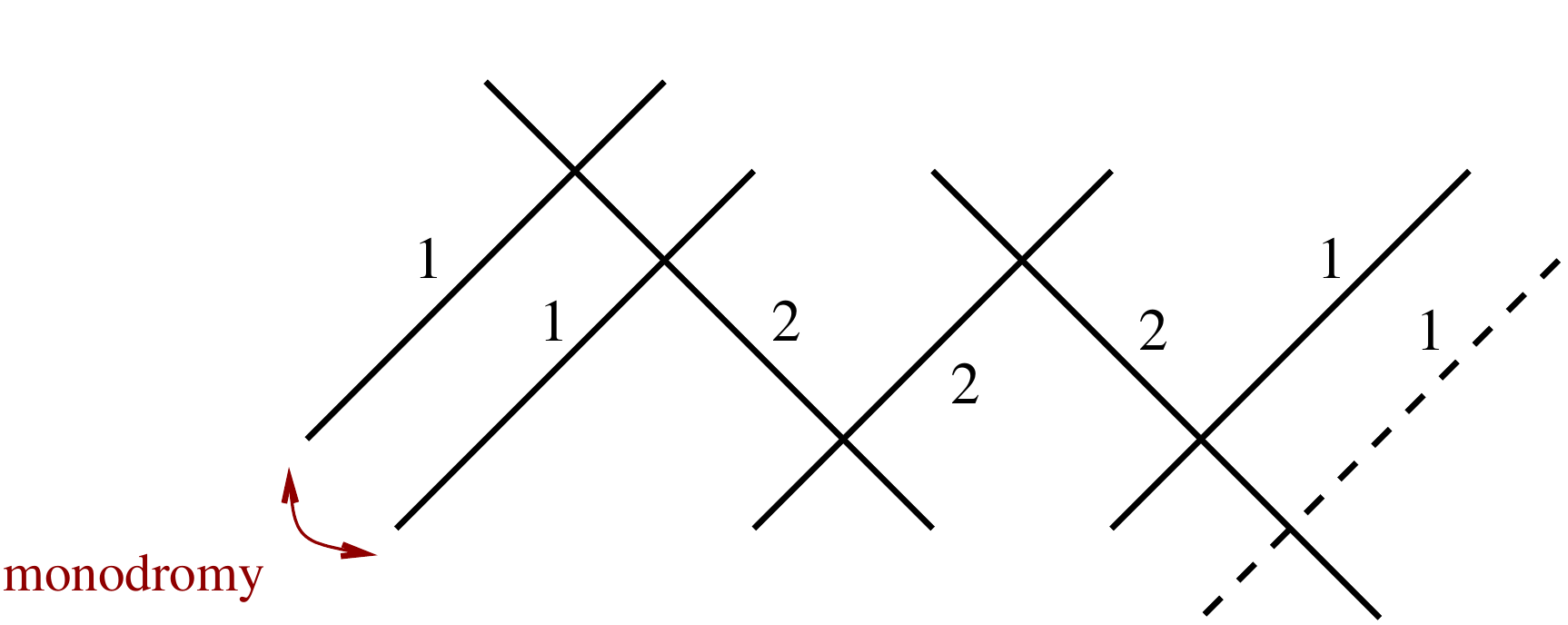}%
\end{picture}%
\setlength{\unitlength}{4144sp}%
\begingroup\makeatletter\ifx\SetFigFont\undefined%
\gdef\SetFigFont#1#2#3#4#5{%
  \reset@font\fontsize{#1}{#2pt}%
  \fontfamily{#3}\fontseries{#4}\fontshape{#5}%
  \selectfont}%
\fi\endgroup%
\begin{picture}(7878,3135)(1156,-2344)
\put(5401,119){\makebox(0,0)[lb]{\smash{{\SetFigFont{20}{24.0}{\rmdefault}{\mddefault}{\updefault}{\color[rgb]{0,0,0}$S_{ii}$}%
}}}}
\put(8281,119){\makebox(0,0)[lb]{\smash{{\SetFigFont{20}{24.0}{\rmdefault}{\mddefault}{\updefault}{\color[rgb]{0,0,0}$S_i$}%
}}}}
\put(8731,-1231){\makebox(0,0)[lb]{\smash{{\SetFigFont{20}{24.0}{\rmdefault}{\mddefault}{\updefault}{\color[rgb]{0,0,0}$S_{\infty}$}%
}}}}
\put(3151,524){\makebox(0,0)[lb]{\smash{{\SetFigFont{20}{24.0}{\rmdefault}{\mddefault}{\updefault}{\color[rgb]{0,0,0}$S_{iv}$}%
}}}}
\put(4501,-2221){\makebox(0,0)[lb]{\smash{{\SetFigFont{20}{24.0}{\rmdefault}{\mddefault}{\updefault}{\color[rgb]{0,0,0}$S_{iii}$}%
}}}}
\put(2341,-1096){\makebox(0,0)[lb]{\smash{{\SetFigFont{20}{24.0}{\rmdefault}{\mddefault}{\updefault}{\color[rgb]{0,0,0}$C_{v+}$}%
}}}}
\put(3556,-1771){\makebox(0,0)[lb]{\smash{{\SetFigFont{20}{24.0}{\rmdefault}{\mddefault}{\updefault}{\color[rgb]{0,0,0}$C_{v-}$}%
}}}}
\end{picture}%

%% file: curvesD7.pspdftex
\begin{picture}(0,0)%
\includegraphics{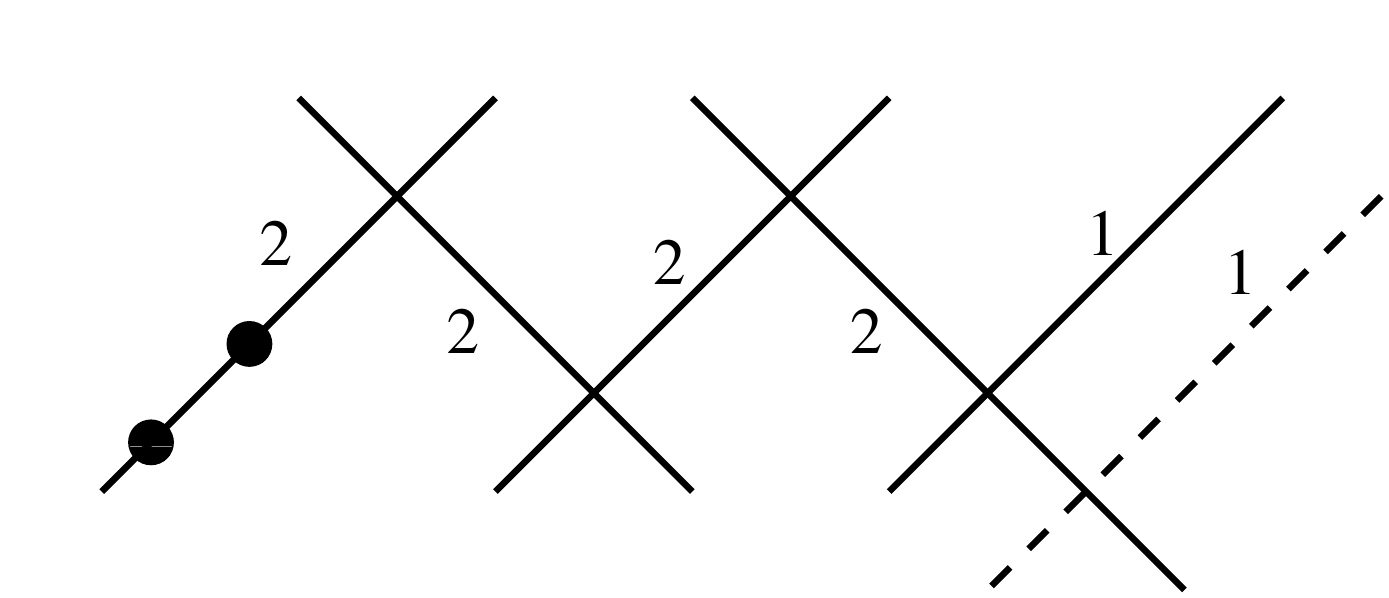}%
\end{picture}%
\setlength{\unitlength}{4144sp}%
\begingroup\makeatletter\ifx\SetFigFont\undefined%
\gdef\SetFigFont#1#2#3#4#5{%
  \reset@font\fontsize{#1}{#2pt}%
  \fontfamily{#3}\fontseries{#4}\fontshape{#5}%
  \selectfont}%
\fi\endgroup%
\begin{picture}(6348,2730)(2686,-2344)
\put(5401,119){\makebox(0,0)[lb]{\smash{{\SetFigFont{20}{24.0}{\rmdefault}{\mddefault}{\updefault}{\color[rgb]{0,0,0}$C_{ii}$}%
}}}}
\put(8281,119){\makebox(0,0)[lb]{\smash{{\SetFigFont{20}{24.0}{\rmdefault}{\mddefault}{\updefault}{\color[rgb]{0,0,0}$C_i$}%
}}}}
\put(8731,-1231){\makebox(0,0)[lb]{\smash{{\SetFigFont{20}{24.0}{\rmdefault}{\mddefault}{\updefault}{\color[rgb]{0,0,0}$C_{\infty}$}%
}}}}
\put(4501,-2221){\makebox(0,0)[lb]{\smash{{\SetFigFont{20}{24.0}{\rmdefault}{\mddefault}{\updefault}{\color[rgb]{0,0,0}$C_{iii}$}%
}}}}
\put(2701,-1996){\makebox(0,0)[lb]{\smash{{\SetFigFont{20}{24.0}{\rmdefault}{\mddefault}{\updefault}{\color[rgb]{0,0,0}$C_{v}$}%
}}}}
\put(3601,119){\makebox(0,0)[lb]{\smash{{\SetFigFont{20}{24.0}{\rmdefault}{\mddefault}{\updefault}{\color[rgb]{0,0,0}$C_{iv}$}%
}}}}
\end{picture}%

%% file: curvesE7.pspdftex
\begin{picture}(0,0)%
\includegraphics{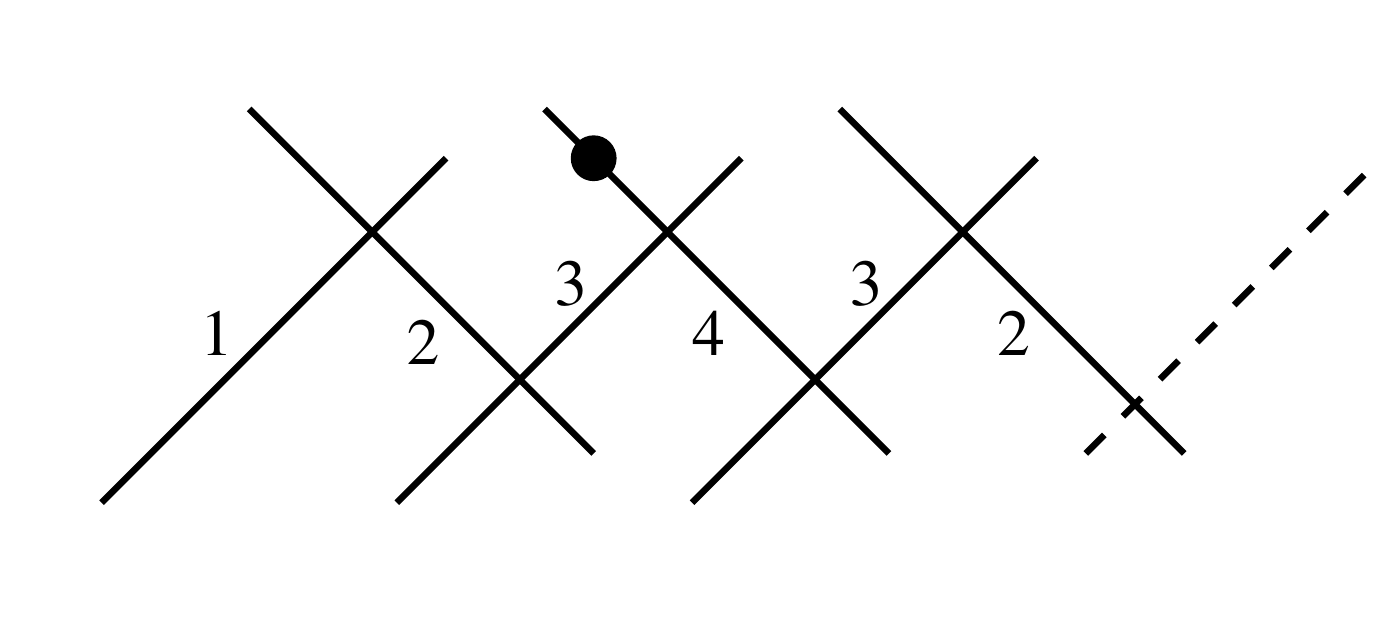}%
\end{picture}%
\setlength{\unitlength}{4144sp}%
\begingroup\makeatletter\ifx\SetFigFont\undefined%
\gdef\SetFigFont#1#2#3#4#5{%
  \reset@font\fontsize{#1}{#2pt}%
  \fontfamily{#3}\fontseries{#4}\fontshape{#5}%
  \selectfont}%
\fi\endgroup%
\begin{picture}(6348,2851)(1561,-1520)
\put(2251,1064){\makebox(0,0)[lb]{\smash{{\SetFigFont{20}{24.0}{\rmdefault}{\mddefault}{\updefault}{\color[rgb]{0,0,0}$C_b$}%
}}}}
\put(1576,-1411){\makebox(0,0)[lb]{\smash{{\SetFigFont{20}{24.0}{\rmdefault}{\mddefault}{\updefault}{\color[rgb]{0,0,0}$C_a$}%
}}}}
\put(3601,1064){\makebox(0,0)[lb]{\smash{{\SetFigFont{20}{24.0}{\rmdefault}{\mddefault}{\updefault}{\color[rgb]{0,0,0}$C_d$}%
}}}}
\put(4951,1064){\makebox(0,0)[lb]{\smash{{\SetFigFont{20}{24.0}{\rmdefault}{\mddefault}{\updefault}{\color[rgb]{0,0,0}$C_f$}%
}}}}
\put(4276,-1411){\makebox(0,0)[lb]{\smash{{\SetFigFont{20}{24.0}{\rmdefault}{\mddefault}{\updefault}{\color[rgb]{0,0,0}$C_e$}%
}}}}
\put(2926,-1411){\makebox(0,0)[lb]{\smash{{\SetFigFont{20}{24.0}{\rmdefault}{\mddefault}{\updefault}{\color[rgb]{0,0,0}$C_c$}%
}}}}
\put(7696,-16){\makebox(0,0)[lb]{\smash{{\SetFigFont{20}{24.0}{\rmdefault}{\mddefault}{\updefault}{\color[rgb]{0,0,0}$C_\infty$}%
}}}}
\end{picture}%

%% file: A4.pspdftex
\begin{picture}(0,0)%
\includegraphics{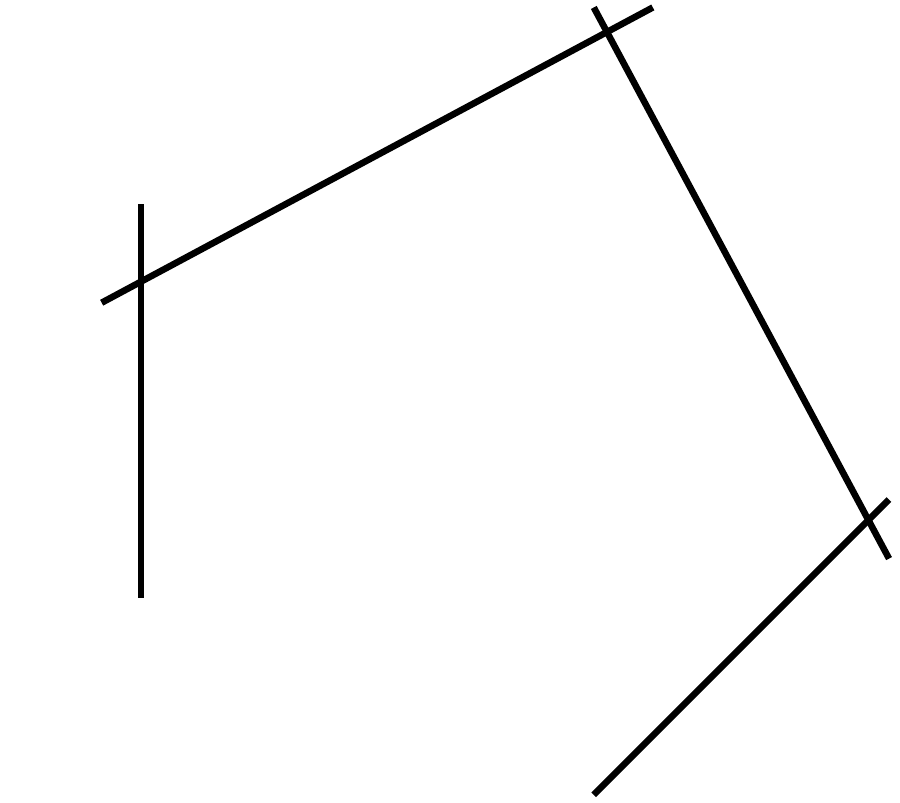}%
\end{picture}%
\setlength{\unitlength}{4144sp}%
\begingroup\makeatletter\ifx\SetFigFont\undefined%
\gdef\SetFigFont#1#2#3#4#5{%
  \reset@font\fontsize{#1}{#2pt}%
  \fontfamily{#3}\fontseries{#4}\fontshape{#5}%
  \selectfont}%
\fi\endgroup%
\begin{picture}(4098,3666)(1336,-3244)
\put(4681,-2941){\makebox(0,0)[lb]{\smash{{\SetFigFont{20}{24.0}{\rmdefault}{\mddefault}{\updefault}{\color[rgb]{0,0,0}$C_{1+}$}%
}}}}
\put(1351,-1591){\makebox(0,0)[lb]{\smash{{\SetFigFont{20}{24.0}{\rmdefault}{\mddefault}{\updefault}{\color[rgb]{0,0,0}$C_{1-}$}%
}}}}
\put(2431,-61){\makebox(0,0)[lb]{\smash{{\SetFigFont{20}{24.0}{\rmdefault}{\mddefault}{\updefault}{\color[rgb]{0,0,0}$C_{2-}$}%
}}}}
\put(4861,-691){\makebox(0,0)[lb]{\smash{{\SetFigFont{20}{24.0}{\rmdefault}{\mddefault}{\updefault}{\color[rgb]{0,0,0}$C_{2+}$}%
}}}}
\end{picture}%

%% file: A6nlH.pspdftex
\begin{picture}(0,0)%
\includegraphics{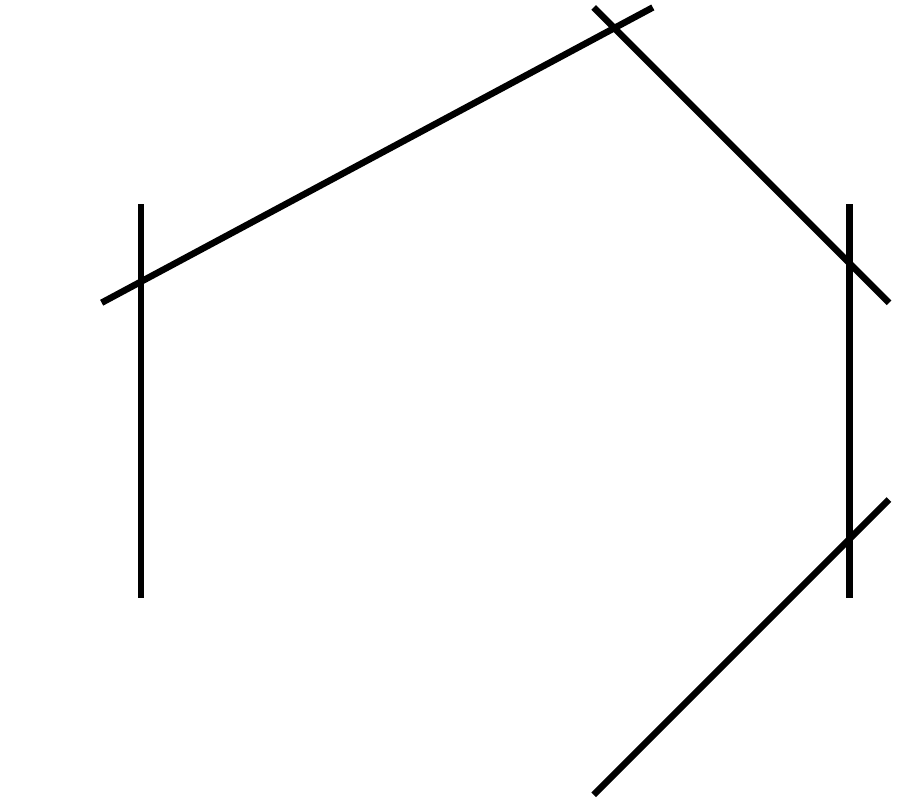}%
\end{picture}%
\setlength{\unitlength}{4144sp}%
\begingroup\makeatletter\ifx\SetFigFont\undefined%
\gdef\SetFigFont#1#2#3#4#5{%
  \reset@font\fontsize{#1}{#2pt}%
  \fontfamily{#3}\fontseries{#4}\fontshape{#5}%
  \selectfont}%
\fi\endgroup%
\begin{picture}(4098,3666)(1336,-3244)
\put(5401,-1501){\makebox(0,0)[lb]{\smash{{\SetFigFont{20}{24.0}{\rmdefault}{\mddefault}{\updefault}{\color[rgb]{0,0,0}$C_{2+}^b$}%
}}}}
\put(4681,-2941){\makebox(0,0)[lb]{\smash{{\SetFigFont{20}{24.0}{\rmdefault}{\mddefault}{\updefault}{\color[rgb]{0,0,0}$C_{1+}$}%
}}}}
\put(4681,-151){\makebox(0,0)[lb]{\smash{{\SetFigFont{20}{24.0}{\rmdefault}{\mddefault}{\updefault}{\color[rgb]{0,0,0}$C_{2+}^a$}%
}}}}
\put(1351,-1591){\makebox(0,0)[lb]{\smash{{\SetFigFont{20}{24.0}{\rmdefault}{\mddefault}{\updefault}{\color[rgb]{0,0,0}$C_{1-}$}%
}}}}
\put(2431,-61){\makebox(0,0)[lb]{\smash{{\SetFigFont{20}{24.0}{\rmdefault}{\mddefault}{\updefault}{\color[rgb]{0,0,0}$C_{2-}$}%
}}}}
\end{picture}%

%% file: A6lH.pspdftex
\begin{picture}(0,0)%
\includegraphics{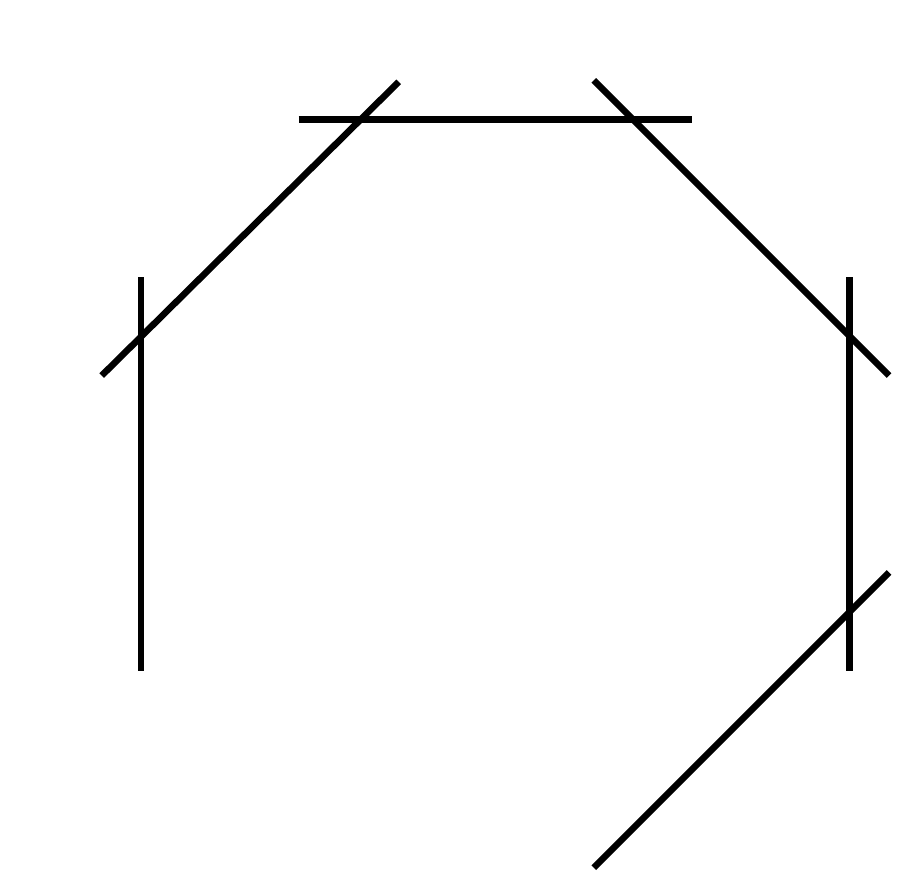}%
\end{picture}%
\setlength{\unitlength}{4144sp}%
\begingroup\makeatletter\ifx\SetFigFont\undefined%
\gdef\SetFigFont#1#2#3#4#5{%
  \reset@font\fontsize{#1}{#2pt}%
  \fontfamily{#3}\fontseries{#4}\fontshape{#5}%
  \selectfont}%
\fi\endgroup%
\begin{picture}(4098,3990)(1336,-3244)
\put(5401,-1501){\makebox(0,0)[lb]{\smash{{\SetFigFont{20}{24.0}{\rmdefault}{\mddefault}{\updefault}{\color[rgb]{0,0,0}$C_{2+}^b$}%
}}}}
\put(4681,-2941){\makebox(0,0)[lb]{\smash{{\SetFigFont{20}{24.0}{\rmdefault}{\mddefault}{\updefault}{\color[rgb]{0,0,0}$C_{1+}$}%
}}}}
\put(3241,479){\makebox(0,0)[lb]{\smash{{\SetFigFont{20}{24.0}{\rmdefault}{\mddefault}{\updefault}{\color[rgb]{0,0,0}$C_{2-}^a$}%
}}}}
\put(1351,-1591){\makebox(0,0)[lb]{\smash{{\SetFigFont{20}{24.0}{\rmdefault}{\mddefault}{\updefault}{\color[rgb]{0,0,0}$C_{1-}$}%
}}}}
\put(1891,-151){\makebox(0,0)[lb]{\smash{{\SetFigFont{20}{24.0}{\rmdefault}{\mddefault}{\updefault}{\color[rgb]{0,0,0}$C_{2-}^b$}%
}}}}
\put(4861,-151){\makebox(0,0)[lb]{\smash{{\SetFigFont{20}{24.0}{\rmdefault}{\mddefault}{\updefault}{\color[rgb]{0,0,0}$C_{2+}^a$}%
}}}}
\end{picture}%

%% file: E6ptlH.pspdftex
\begin{picture}(0,0)%
\includegraphics{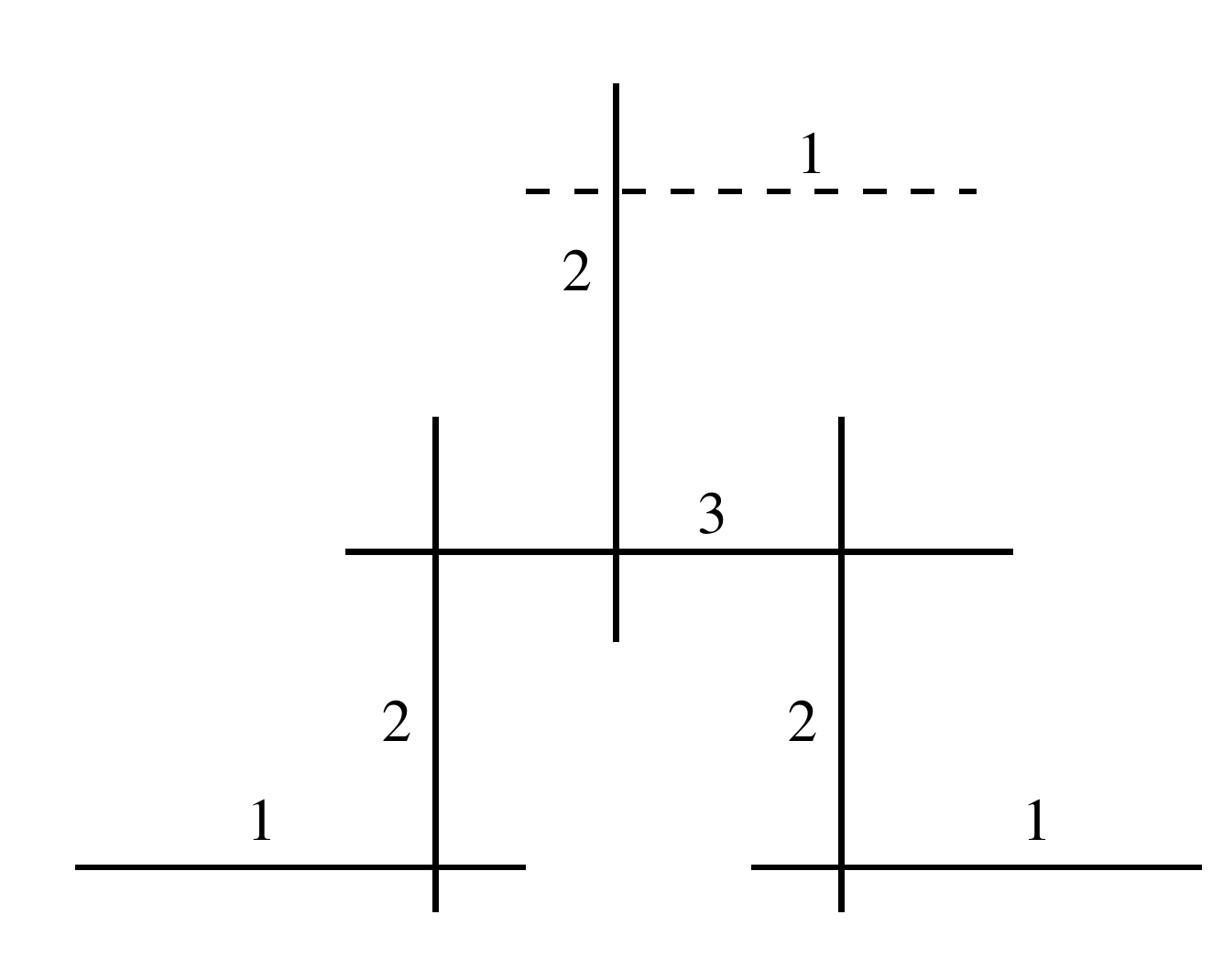}%
\end{picture}%
\setlength{\unitlength}{4144sp}%
\begingroup\makeatletter\ifx\SetFigFont\undefined%
\gdef\SetFigFont#1#2#3#4#5{%
  \reset@font\fontsize{#1}{#2pt}%
  \fontfamily{#3}\fontseries{#4}\fontshape{#5}%
  \selectfont}%
\fi\endgroup%
\begin{picture}(6033,4876)(2776,-1745)
\put(8326,-1636){\makebox(0,0)[lb]{\smash{{\SetFigFont{20}{24.0}{\rmdefault}{\mddefault}{\updefault}{\color[rgb]{0,0,0}$C_5$}%
}}}}
\put(7651,2414){\makebox(0,0)[lb]{\smash{{\SetFigFont{20}{24.0}{\rmdefault}{\mddefault}{\updefault}{\color[rgb]{0,0,0}$C_\infty$}%
}}}}
\put(5851,2864){\makebox(0,0)[lb]{\smash{{\SetFigFont{20}{24.0}{\rmdefault}{\mddefault}{\updefault}{\color[rgb]{0,0,0}$C_6$}%
}}}}
\put(7966,389){\makebox(0,0)[lb]{\smash{{\SetFigFont{20}{24.0}{\rmdefault}{\mddefault}{\updefault}{\color[rgb]{0,0,0}$C_3$}%
}}}}
\put(2791,-1591){\makebox(0,0)[lb]{\smash{{\SetFigFont{20}{24.0}{\rmdefault}{\mddefault}{\updefault}{\color[rgb]{0,0,0}$C_1$}%
}}}}
\put(4636,1199){\makebox(0,0)[lb]{\smash{{\SetFigFont{20}{24.0}{\rmdefault}{\mddefault}{\updefault}{\color[rgb]{0,0,0}$C_2$}%
}}}}
\put(6796,1154){\makebox(0,0)[lb]{\smash{{\SetFigFont{20}{24.0}{\rmdefault}{\mddefault}{\updefault}{\color[rgb]{0,0,0}$C_4$}%
}}}}
\end{picture}%